%% file: main.tex
\documentclass[11pt]{article}

\usepackage{fullpage}
\usepackage{setspace}
\usepackage{pdflscape}
\usepackage{authblk}

\usepackage{graphicx}
\usepackage[table,dvipsnames]{xcolor}
\usepackage{tcolorbox}
\usepackage{caption}
\usepackage{placeins}

\usepackage{booktabs}
\usepackage{array}
\usepackage{tabularx}
\usepackage{threeparttable}
\usepackage{threeparttablex}
\usepackage{multirow}
\usepackage{longtable}
\usepackage{ragged2e}

\usepackage{amsmath}
\usepackage{amssymb}

\usepackage{natbib}
\usepackage[linktocpage=true, colorlinks=true, linkcolor=blue, citecolor=black]{hyperref}

\newcolumntype{J}[1]{>{\justifying\arraybackslash}p{#1}}
\newcolumntype{C}[1]{>{\centering\let\newline\\\arraybackslash\hspace{0pt}}m{#1}}
\def\sym#1{\ifmmode^{#1}\else\(^{#1}\)\fi}

\begin{document}	
	
	\title{Mining Meaning: Measurement Error in AI-Assisted Literature Reviews\thanks{Corresponding author email: jdmichler@arizona.edu. Various versions (e.g., 5.6) and variants (e.g., Sol) of GPT-5 and GPT-6 were used assistively to develop visualization concepts, verify mathematical formulas, provide editorial suggestions, and draft this statement. No AI-generated figures or tables appear in this paper. All text, figures, and tables were created by the authors and all interpretations and any errors are our own. We would like to thank Chandrakant Agme, Garrett Hanson, Tim Leavy, and Aurora Seekins for research support of various forms, particularly in data labeling, associated with and leading to this paper.}}

	\author[a]{Jeffrey D. Michler}
	\author[b]{Kieran Douglas}
	\author[a]{Anna Josephson}
	\affil[a]{\small \emph{University of Arizona, Tucson, USA}}
	\affil[b]{\small \emph{University of California - Davis, Davis, USA}}
	
	\date{December 2026}
	\maketitle
	
	\thispagestyle{empty}
	
	\begin{center}\begin{abstract}
			\noindent 
			Researchers increasingly use generative AI, particularly large language models (LLMs), to automate tasks across the research pipeline. We frame LLM-assisted literature reviews as a measurement problem, linking task-level measurement error to distortion in substantive estimands. As a test case, we benchmark three implementations of GPT-4.1 against human labels and use each to characterize the literature on rainfall as an instrumental variable, drawing on a candidate corpus of 4,191 economics papers. All three implementations perform well on binary classification, but performance deteriorates as tasks demand greater contextual interpretation. Yet only 62\% of papers identified as using rainfall instruments by any implementation are identified by all three. Despite disagreement over individual papers and instrumenting relationships, broader conclusions about the literature persist across implementations and recent frontier models. Inference is thus most sensitive to the choice of measurement system at precisely the level of detail where LLMs should have a comparative advantage over human reviewers. We conclude that standard model performance metrics do not, by themselves, establish the credibility of downstream inference and that validation should match the resolution of the claim.
	\end{abstract}\end{center}
	
	{\small \noindent\emph{JEL Classification}: A11, C36, C80, D83, O33
		}
	
	\setcounter{page}{0}
	\thispagestyle{empty}
	
	\newpage
	\onehalfspacing
	
	
	\section{Introduction} \label{sec:intro}
	
	The rapid development of generative AI, particularly large language models (LLMs), has expanded the range of research tasks that can be delegated to machines. Researchers increasingly use LLMs for ideation \citep{MeinckeEtAl24, ShaerEtAl24, SiEtAl24}; data generation \citep{wang2023etal, ma24, stewart25}; coding \citep{Korinek23, PengEtAl23}; annotating and classifying text \citep{GargandFetzer25, AdamsEtAl26, BriggsEtAl26}; and writing and editing \citep{SungEtAl25, WuEtAl25}. Tasks that once required substantial amounts of human labor can now be completed quickly and at a scale that was previously infeasible. Of particular interest to empirical researchers is the use of LLMs as measurement systems that convert unstructured text into structured data at low marginal cost \citep{DellandRambachan26, LudwigEtAl26}.
	
	But how much can researchers rely on LLMs as measurement systems? Concerns exist about hallucination, context loss, operational consistency, reproducibility, and the ability to use AI-generated data to make accurate scientific inference. Yet LLMs are brittle in that we cannot reliably predict when and where a model might succeed or fail and how to interpret those successes or failures. A model may fail to find information that is present (Type II error), report information that is not present (Type I error), or correctly recognize that a concept exists while incorrectly describing what that concept is. Further, knowing that a model produces generally accurate measures provides no guarantee that estimates constructed from those measures are unbiased. Understanding how much we can rely on LLMs requires understanding the type and severity of their errors and when those errors matter for the conclusions drawn from those data.
	
	In this paper, we examine these questions by asking LLMs to read scientific papers and generate structured data for use in a literature review. When LLMs are used to generate data, the model and its implementation together function as a measurement system. We compare the performance of these systems across research tasks that vary in the amount of contextualization and interpretation they require, from relatively simple annotation to identifying and naming economic concepts from text. We then use alternative LLM measurement systems to independently construct the same literature-review dataset and compare what we learn from each. This allows us to examine both when and where LLMs succeed or fail, as well as when errors in the data they generate matter for the conclusions researchers draw from those data.
	
	Literature reviews provide a useful setting for this investigation as ``reading'' is not a single task. It can mean finding information that is explicitly stated and so relatively easy to locate, such as an article title or whether a paper contains empirical analysis. Other information requires inferring concepts from context, such as identifying the dependent variable, the source of endogeneity, or the role a particular variable plays in an econometric model. At the same time, a literature review is precisely the type of research task for which LLMs should have a comparative advantage. The low marginal cost of measurement means that models can potentially read tens of thousands of papers and produce a much more detailed, structured characterization of a literature than would be feasible with a human reviewer alone.
	
	Our test case is to understand how economists use rainfall as an instrumental variable (IV) in causal or quasi-causal studies. The rainfall-IV literature makes the reading task particularly difficult. Information on what is endogenous, how endogeneity is addressed, what the instrument is, and how rainfall is quantified can appear in equations, tables, footnotes, or different sections of a paper, can be described using inconsistent terminology, and may only be implicitly defined. Therefore, the model must do more than find words in text. It must infer the economic role those words play. The application is also substantively important. Rainfall is used to instrument for migration, health, conflict, income, agricultural production, and many other outcomes. At the same time, the validity of the exclusion restrictions underlying these applications has come under scrutiny \citep{Mellon25}. Mellon's expert human review provides a useful benchmark for assessing what we gain from measuring these relationships at greater scale and resolution.
	
	We start by developing a simple model of measurement error that distinguishes between misclassification and semantic error. We then connect these task-level errors to the substantive estimands that researchers construct from the generated data. Empirically, we build a candidate corpus of $4,191$ economics papers from which we draw a human-labeled reference sample of $439$ papers. We adopt a hierarchical prompting strategy that moves from binary classification to identifying and naming the economic concepts underlying those classifications. Using OpenAI's ChatGPT-4.1, we compare three implementation strategies: zero-shot prompting, retrieval-augmented generation (RAG), and supervised fine-tuning (SFT). To examine how well LLMs can amplify human measurement, we benchmark all three against a common human-labeled evaluation set and then use each measurement system to independently construct data from the full corpus. Rather than selecting a preferred implementation based on validation performance metrics, we carry all three datasets through the same substantive analysis of the rainfall-IV literature. To examine whether our findings are a function of our choice of LLM, we also compare the SFT implementation of GPT-4.1 with zero-shot implementations of four more recent models.
	
	Our first finding is that LLMs perform well when reading requires relatively simple annotation, but performance becomes heterogeneous as we demand greater contextualization and interpretation. Across the three implementations, accuracy and sensitivity are generally high. The models also recover relatively simple metadata, such as article titles and DOIs, with high semantic similarity. This similarity in levels of overall performance conceals different types of errors. RAG tends to generate more false positives while zero-shot tends to generate more false negatives. SFT generally produces a more balanced error profile. Thus, even for tasks on which all three implementations appear highly successful, they do not necessarily get the same papers wrong. Performance of all three implementations deteriorates as reading requires a higher degree of contextual interpretation. SFT generally improves upon these more difficult extraction tasks relative to zero-shot and RAG, but all three measurement systems produce substantial semantic errors. Correctly recognizing that a concept is present does not imply that the model can correctly identify what that concept is.
	
	Our second headline finding is that how much researchers can rely on LLM-generated data depends not just on the difficulty of the task but on what researchers ask of the generated data. Among papers classified as using a rainfall IV by at least one implementation, rainfall-IV status differs across measurement systems for more than one third of papers. Even conditional on all three systems identifying the same paper as using a rainfall IV, disagreement remains over what rainfall measure is used, what variable is instrumented, and sometimes about whether a relevant variable exists at all. As we make broader characterizations of the literature, disagreement becomes less consequential and all three measurement systems produce essentially the same picture. Economists quantify rainfall in a great many ways and use rainfall to instrument for a wide range of socioeconomic processes, raising concerns about exclusion restriction violations as well as selection on significance.
	
	The decline in measurement-system dispersion as the estimand of interest coarsens demonstrates that the importance of measurement-system choice is estimand dependent. Task-level validation metrics provide evidence of observational validity, the extent to which a measurement system implements the intended construct definition \citep{DellandRambachan26}. They do not, by themselves, determine the consequences of measurement error for a particular substantive estimand. At the paper-level, measurement-system choice can determine whether a paper is included in the rainfall-IV literature at all. At the relationship-level, measurement-system choice can change which rainfall measure instruments which endogenous variable. When the estimand is a broad characterization of the literature, many of these disagreements wash out. But broad characterizations of the literature are where LLMs add relatively little beyond what could be achieved by a traditional human review. Measurement-system choice thus matters most at precisely the level of detail where LLMs should have a comparative advantage over humans.
	
	Our study makes three contributions. First, we connect the use of LLMs as measurement systems to the economics literature on measurement error, misclassification, and generated variables. An extensive literature shows that error-contaminated measures can distort descriptive statistics, coefficient estimates, and statistical uncertainty \citep{Aigner73, HausmanEtAl98, BoundEtAl01, Schennach22, Possebom26}. Recent work extends the measurement error literature to AI-generated variables \citep{EgamiEtAl23, BattagliaEtAl24, ChristensenandHansen26}. Empirical applications examine the extent to which errors in machine-generated data distort downstream estimates \citep{TorchianaEtAl25, BrittoEtAl26}. Research on text as data and learned proxies similarly emphasizes that computational measures require construct-specific validation \citep{GrimmerandStewart13, KnoxEtAl22}. We develop a simple model of measurement error that accommodates both misclassification and semantic error and distinguishes between measurement error in the generated data and the distortion that error induces in substantive estimands. We use this framework to examine how the importance of measurement-system choice varies with the resolution of the substantive estimand.
	
	Second, we contribute to the rapidly growing body of literature that uses LLMs to generate structured data for scientific research \citep{BenoitEtAl26, ChenEtAl26, KeckEtAl26, MarwitzEtAl26}. In the social sciences, recent studies use machine-generated data to characterize causal claims made in economics \citep{GargandFetzer25} and the prevalence of publication bias in political science \citep{BriggsEtAl26}. These studies demonstrate that carefully designed LLM workflows can produce useful data at very large scale. We ask how much researchers can rely on those data. Rather than using performance validation to select a preferred measurement system, we carry data generated by alternative measurement systems through the same substantive analysis. This allows us to examine which substantive claims persist across measurement systems and how agreement across systems varies with the level of detail of those claims.
	
	Finally, we contribute to the literature that critiques the use of rainfall as an instrument \citep{RosenzweigWolpin00, Deaton10, DellEtAl14, Sarsons15}. Our paper most closely relates to \citet{Mellon25}, which documents through expert human review the large number of socioeconomic variables that rainfall has been shown to predict and argues that these relationships create widespread potential violations of the exclusion restriction. Our machine-assisted review amplifies this effort, recovering the same broad concern from a much larger candidate corpus. By documenting how economists quantify rainfall, we also show that the same types of endogenous variables are instrumented using many different measures of rainfall, raising a second potential identification concern: scope for researchers to search across alternative measures of rainfall until they obtain the desired first-stage results. The rainfall IV we observe may be the one selected not because of its theoretical fit or economic rationale but for its significance.
	
	More broadly, our paper relates to an emerging literature on how economists can make credible inference using AI-generated data. \citet{DellandRambachan26} argue that AI reduces the cost of measurement and shifts the problem researchers face from finding a scalable measure of an economic concept to choosing among many plausible measures of that concept. The consequences of that choice are evident in \citet{LudwigEtAl26}, who demonstrate that seemingly innocuous decisions about which LLM to use or how to prompt it can change the magnitude, sign, and significance of estimated coefficients. Choosing among measurement systems would be easier if we could generalize from past instances of model success or failure. A successful application demonstrates what a particular LLM is capable of, but model success is not transitive across tasks or across models. Likewise, model failure is not transitive, nor may it be evident what model failure stems from \citep{ArgyleEtAl25}. This ``jagged frontier'' of AI capabilities makes the development of new methodologies for using generative AI as a research tool difficult \citep{DellAcquaEtAl26}. Complicating matters further, what constitutes success or failure depends not only on what the model is asked to measure but also on what the researcher seeks to infer from those measurements. Measurement error that is nonignorable for one estimand may matter very little for another. And agreement across measurement systems need not imply accurate measurement but could simply be a function of common measurement errors. The promise of LLMs is not simply that they allow researchers to measure more but that they allow researchers to make claims of a degree of detail that was previously infeasible. It is the credibility of these claims that researchers ultimately need to establish.
	
	
	\section{A Simple Model of Measurement Error} \label{sec:ME-model}
	
	Valid inference begins by defining the estimand of interest and the source of uncertainty that makes the estimator stochastic \citep{ArgyleEtAl25}. Typically the uncertainty is due to either randomness in sampling from a population or randomness in assignment to treatment \citep{AbadieEtAl20}. Sampling uncertainty can be resolved if one is able to observe all units in the population. Design uncertainty generally cannot be eliminated because, for each unit, at least one of the potential outcomes remains unobserved \citep{ImbensRubin15}. In each case, uncertainty arises because the observed data do not contain all the information needed to recover the estimand.
	
	Measurement introduces an analogous source of uncertainty. When measurements are generated by algorithms whose behavior we poorly understand, it is difficult to justify assumptions about the nature and distribution of measurement errors \citep{DellandRambachan26, LudwigEtAl26}. If a measurement system were provably correct, meaning it could be guaranteed to return the correct target value for every admissible input, there would be no mismeasurement to contend with \citep{GennaroEtAl10}. Absent a proof of correctness, treating measured values as error-free amounts to assuming away the measurement problem. The relevant question is thus not only how accurately a measurement system recovers the target, but how its errors propagate into the estimand \citep{BoundEtAl01}.
	
	A quantitative literature review is itself an exercise in estimation. The population of papers and the process by which papers are selected from that population define the research context. A substantive estimand describes some feature of the population, such as the prevalence of a particular research method or the distribution of findings on a particular economic question. Estimates depend on which papers enter the review and how their contents are measured to generate the data. Random sampling introduces variation across samples of papers, while different readers can produce different measurements of the same papers. Selective inclusion and systematic errors in measurement can also bias the description of the target literature.\footnote{Selective publication creates a selection problem. Reading every published paper can characterize the published literature, but does not by itself recover the distribution of all conducted studies when publication depends on study results \citep{AndrewsKasy19, BriggsEtAl26, SnyderandZhuo26}.} 
	
	In a traditional literature review, the cost of reading and coding papers constrains both the number of papers that can be reviewed and the level of detail with which their contents can be measured. LLMs offer a potential solution to this text processing problem, allowing researchers to read far more papers than would be feasible with human reviewers alone and potentially the entire population of interest. Reading every paper in that population would eliminate sampling uncertainty. But it would not eliminate measurement uncertainty, which can persist even when extraction is deterministic. Data are made, not found, and model-generated labels are measurements of facts about papers, not the facts themselves. Our theoretical framework examines how errors in those measurements affect the conclusions drawn about the literature.
	
	To fix ideas on how measurement error might manifest in data generated by LLMs, let $T_{ij}$ be the target value for concept $i$ in paper $j$ and $Z_{ij}^{(m)}$ be the value generated by measurement system $m$. The measurement system includes the choice of model (its lineage, family, version, and variant), any additional training, the specific prompts, retrieval procedure, model settings like temperature, top-p, and seed, and post-processing of model outputs. Let $g(\cdot)$ denote the descriptive analysis of interest. Applying $g(\cdot)$ to the target values defines the substantive estimand $\theta$, while applying it to the model-generated values produces a measurement-system-specific estimate.
	
	For population $P$ and observed sample $S$, we define $\theta_{P} = g \left( \mathbf{T}_{P} \right)$, the target estimand defined on the target data, and $\widehat{\theta}_{m} = g \left( \mathbf{Z}_{S}^{(m)} \right)$, the corresponding estimate constructed from the measured data which may suffer from sampling-based and measurement-based uncertainty.\footnote{Throughout the paper we use the terms design-based, sampling-based, and measurement-based uncertainty. It is important to note that the uncertainty in the first two terms is generated by a stochastic mechanism. The uncertainty in the third term is a discrepancy that can arise either from a stochastic or deterministic process.} We can decompose the difference between the estimand and its estimate as:
	
	\begin{equation}
		\widehat{\theta}_{m} - \theta_{P} = 
		\underbrace{ g \! \left( \mathbf{Z}_{S}^{(m)} \right) - g \! \left( \mathbf{T}_{S} \right) }_{\text{measurement error}}
		+
		\underbrace{ g \! \left( \mathbf{T}_{S} \right) - g \! \left( \mathbf{T}_{P} \right) }_{\text{sampling error}}. \label{eq:mem}
	\end{equation}
	
	\noindent The second term generates sampling-based uncertainty which, if sampling is random, will itself be random. The first term is the measurement-based distortion of the estimand which need not be random. When $S = P$, as is possible when using LLMs to read an entire literature, the sampling error goes to zero. But this does not, by itself, eliminate measurement error because in using LLMs to drive $S$ as close to $P$ as possible, we rely on model-generated measures of the target values rather than the target values themselves, which, presumably, an expert human reader would be able to recover.
	
	
	\subsection{Task-Level Binary Outcomes} \label{binary-out}
	
	To characterize task-level measurement errors that give rise to measurement-induced estimand distortion, we first focus on the special case of a binary target value. The concept-level measurement error is:
	
	\begin{equation}
		Z_{ij} = T_{ij} \oplus E_{ij},
	\end{equation}
	
	\noindent where $Z_{ij}$ remains the LLM-generated measure of the (unobserved) target value $T_{ij}$ and $E_{ij}$ represents the measurement error.\footnote{To keep the notation simple, we focus on a single measurement system, $m = 1$, allowing us to omit the superscript.} In the binary case $E_{ij} = 0$ indicates no error while $E_{ij} = 1$ indicates the target value was flipped. From this accounting identity we can write the binary misclassification model for concept $i$:
	
	\begin{equation}
		\begin{pmatrix}
			\Pr[Z_{ij} = 1 \mid T_{ij} = 0] & \Pr[Z_{ij} = 0 \mid T_{ij} = 0] \\[3pt]
			\Pr[Z_{ij} = 1 \mid T_{ij} = 1] & \Pr[Z_{ij} = 0 \mid T_{ij} = 1]
		\end{pmatrix}
		=
		\begin{pmatrix}
			\alpha_{i} & 1 - \alpha_{i} \\[3pt]
			1 - \beta_{i} & \beta_{i}
		\end{pmatrix}.
	\end{equation}
	
	\noindent Here $\alpha_{i}$ is the false positive (FP) rate, $\beta_{i}$ is the false negative (FN) rate, $1 - \alpha_{i}$ is the true negative (TN) rate (specificity), and $1 - \beta_{i}$ is the true positive (TP) rate (sensitivity). FPs and FNs need not occur at the same rate.
	
	To understand how the prevalence of FPs and FNs biases the measured values, let $p_{i} = \Pr (T_{ij} = 1)$ and $q_{i} = \Pr (Z_{ij} = 1)$. Then $q_{i} = \left( 1 - \beta_{i} \right) p_{i} + \alpha_{i} \left( 1 - p_{i} \right)$. Rearranging terms, the distortion in measured prevalence is:
	
	\begin{equation}
		q_{i} - p_{i} = \alpha_{i} \left( 1 - p_{i} \right) - \beta_{i} p_{i}.
	\end{equation}
	
	\noindent Here $\alpha_{i} \left( 1 - p_{i} \right)$ is the share of units falsely classified as positive while $ \beta_{i} p_{i}$ is the share falsely classified as negative. Our estimate of the target value is biased upward when the share of FPs, $\alpha_{i} (1 - p_{i})$, exceeds the share of FNs, $\beta_{i} p_{i}$, and biased downward in the opposite case.
	
	The prevalence distortion identity clarifies why accuracy $\left( \Pr \left[ Z_{ij} = T_{ij} \right] \right)$, while being an important characteristic of a measurement system, is insufficient for capturing measurement error induced bias. Accuracy is defined as:
	
	\begin{equation}
		A_{i} = \left( 1 - \beta_{i} \right) p_{i} + \left( 1 - \alpha_{i} \right) \left( 1 - p_{i} \right). \label{eq:acc}
	\end{equation}
	
	\noindent Suppose the prevalence of the target value being $T_{ij} = 1$ is $p_{i} = 0.10$. And suppose the measurement system has high specificity $(1 - \alpha_{i} = 0.95)$ and high sensitivity $(1 - \beta_{i} = 0.90)$. This gives an accuracy rate of $A_{i} = 0.90 (0.10) + 0.95 (0.90) = 0.945$. Despite this high accuracy, the prevalence of the LLM reporting $Z_{ij} = 1$ is $q_{i} = 0.90 (0.10) + 0.05 (0.90) = 0.135$. So, the LLM-generated data have an absolute prevalence bias of $3.5$ percentage points or a relative overestimate of $35\%$ even though the measurement system classifies $94.5\%$ of observations correctly.
	
	
	\subsection{Task-Level String Outcomes} \label{sec:string-out}
	
	A second special case of the general measurement model is data generated as strings. Continue to let $T_{ij} \in \{ 0,1 \}$ be the target value, which is either present or absent, and $Z_{ij} \in \{ 0,1 \}$ be the model's binary measure of the target value. Let $N_{ij}$ be the target string (in our application a variable name) that is only defined when $T_{ij} = 1$ and let $X_{ij}$ be the LLM-generated string defined only when $Z_{ij} = 1$. In this way the binary measurement model governs the string measurement model.\footnote{This two-stage model is exactly how we design the prompts for the LLM. Variable names are generated conditional on an affirmative answer to a binary question about the presence or absence of the concept.}
	
	\paragraph{Semantic Similarity.} We first consider the quality of the generated string conditional on a true positive in the binary task. Define semantic similarity as: 
	
	\begin{equation}
		S_{ij} = s \! \left( X_{ij}, N_{ij} \right) \in[0,1] \qquad \text{if } T_{ij} = Z_{ij} = 1,
	\end{equation}
	
	\noindent with $S_{ij}$ undefined otherwise and $s(n,n) = 1$. The corresponding unit-level conditional semantic dissimilarity is simply $U_{ij} = 1 - S_{ij}$ with $U_{ij} = 0$ being perfect similarity, $U_{ij} = 1$ being no semantic similarity, and $U_{ij}$ undefined when outside the TP range.
	
	To move from the unit-level error to a system-level measure, define the mean semantic similarity score among TP observations as: 
	
	\begin{equation}
		\mu_{i} = \mathbb{E} \! \left[ S_{ij} \mid T_{ij} = Z_{ij} = 1 \right].
	\end{equation}
	
	\noindent As perfect measurement corresponds to a semantic similarity score of one, mean conditional semantic dissimilarity is: 
	
	\begin{equation}
		1 - \mu_{i} = \mathbb{E} \! \left[ U_{ij} \mid T_{ij} = Z_{ij} = 1 \right].
	\end{equation}
	
	\noindent This measure evaluates measurement error in the second-stage string-generation task conditional on the model correctly detecting that the target is present.
	
	\paragraph{Semantic Fidelity.} Although semantic similarity is defined only for TP cases, it is useful to have a measure for all observations with $T_{ij} = 1$. We define the concept of semantic fidelity as: 
	
	\begin{equation}
		F_{ij} = 
		\begin{cases}
			S_{ij}, & \text{if } T_{ij} = 1, \; Z_{ij} = 1, \\[4pt]
			0, & \text{if } T_{ij} = 1, \; Z_{ij} = 0.
		\end{cases}
	\end{equation}
	
	\noindent A TP receives fidelity equal to the semantic similarity between the target and generated strings. An FN receives zero fidelity because the model fails to produce a string when a target string exists. The corresponding unit-level error in semantic fidelity is: 
	
	\begin{equation}
		1 - F_{ij} = 
		\begin{cases}
			U_{ij}, & \text{if }T_{ij}=1, \; Z_{ij}=1, \\[4pt]
			1, & \text{if }T_{ij}=1, \; Z_{ij}=0.
		\end{cases}
	\end{equation}
	
	\noindent Thus, an FN produces maximal measurement error, while a TP produces error equal to the semantic dissimilarity between the generated and target strings.
	
	We then define the string analog of the binary measurement error model, which depends on the TP and FN rates. Taking expectations conditional on the target value being present gives us:
	
	\begin{equation}
		\mathbb{E} \left[ 1 - F_{ij} | T_{ij} = 1 \right] = \beta_{i} + \left( 1 - \beta_{i} \right) \left( 1 - \mu_{i} \right) = 1 - \left( 1 - \beta_{i} \right) \mu_{i}.
	\end{equation}
	
	\noindent The string measurement error model can be decomposed into the failure to detect $(\beta)$ plus the expected semantic error $(1 - \mu)$ weighted by the probability of detection $(1 - \beta_{i})$. Average string measurement error is thus the sum of the probability of failing to detect a target that is present and the probability-weighted expected semantic error conditional on detection.
	
	
	\subsection{Task-Level Omnibus Loss Function} \label{sec:omnibus}
	
	Finally, we define an omnibus measurement loss function that combines classification error and semantic error. For concept $i$ in paper $j$, define the unit-level loss as: 
	
	\begin{equation}
		L_{ij} =     
		\begin{cases}
			0, \; T_{ij} = 0, Z_{ij} = 0, \\[4pt]
			1, \; T_{ij} = 0, Z_{ij} = 1, \\[4pt]
			1, \; T_{ij} = 1, Z_{ij} = 0, \\[4pt]
			U_{ij}, \; T_{ij} = 1, Z_{ij} = 1.
		\end{cases}
	\end{equation}
	
	\noindent The loss is 0 for a TN, 1 for an FP or FN, and equal to the conditional semantic error for a TP. The loss function is an unweighted measure of loss, as it treats FNs and FPs as having equal maximal loss of one.\footnote{A more general loss function could assign distinct costs to each type of error (e.g., $\omega_{FP}$, $\omega_{FN}$, $\omega_{S}$).}
	
	Recalling that $p_{i} = \Pr (T_{ij} = 1)$, the task-level omnibus loss function is:
	
	\begin{equation}
		\begin{aligned}
			\mathcal{L}_{i} &= 
			\mathbb{E} \! \left[  L_{ij} \right] \\[3pt] &= 
			\alpha_{i} \left( 1 - p_{i} \right) + \beta_{i} p_{i} + \left( 1 - \beta_{i} \right) p_{i} \left( 1 - \mu_{i} \right) \\[3pt] &=
			\left( 1 - A_{i} \right) + p_{i} \left( 1 - \beta_{i} \right)  \left( 1 - \mu_{i} \right).
		\end{aligned}
	\end{equation}
	
	\noindent The first term is the population share of FPs, the second is the population share of FNs, and the third is the expected semantic error among TPs weighted by the population share of TPs. The final equality makes the hierarchical structure of the models explicit. When the task requires only binary classification, the semantic term is absent and omnibus loss reduces to $1 - A_{i}$, which is just one minus the accuracy rate.
	
	The omnibus loss function has several nice properties. It is bounded between zero and one; it equals zero under perfect measurement; it places binary and string errors on a common scale; and it decomposes loss into interpretable components. It also provides a direct basis for comparing measurement systems in identifying a single concept, $i$, for a given sample or population. 
	
	That said, it is not a perfect statistic for capturing overall measurement error. Its interpretation depends on the target prevalence, $p_{i}$; the scale of the semantic similarity function; and the maintained choice to assign maximal loss to both FPs and FNs. Additionally, while $\alpha_{i}$, $\beta_{i}$, and $\mu_{i}$ are on the same zero-to-one scale, their values do not mean the same things.\footnote{Recall that $\alpha_{i}$ and $\beta_{i}$ are percentages while $\mu_{i}$ is a semantic similarity score, not a percentage of incorrect semantic mismeasurement.} These limitations make the model suitable for comparisons across measurement systems, $m$, for a given concept and not across concepts with different prevalences. More fundamentally, $\mathcal{L}_{i}$ measures average data-level loss. It does not record where errors occur and so does not generally determine their effect on a particular substantive estimand.
	
	
	\subsection{Estimand Dependence} \label{sec:estimand}
	
	To compare measurement systems, we reintroduce the model index $m$ and define: 
	
	\begin{equation}
		\mathcal{L}_{i}^{(m)} = \alpha_{i}^{(m)} \left( 1 - p_{i} \right) + \beta_{i}^{(m)} p_{i} + p_{i} \left( 1 - \beta_{i}^{(m)} \right) \left( 1 - \mu_{i}^{(m)} \right).
	\end{equation}
	
	\noindent The target prevalence $p_{i}$ is common across measurement systems, while $\alpha_{i}^{(m)}$, $\beta_{i}^{(m)}$, and $\mu_{i}^{(m)}$ characterize system $m$'s performance on concept $i$. To connect task-level measurement error back to the substantive estimand in Equation~\eqref{eq:mem}, we define system-specific measurement-induced estimand distortion for a given set of papers as: 
	
	\begin{equation}
		\Delta_{i}^{(m)} = \widehat{\theta}_{i}^{(m)} - \theta_{i}.
	\end{equation}
	
	\noindent Thus, $\mathcal{L}_{i}^{(m)}$ measures the average quality of the model-generated data, while $\Delta_{i}^{(m)}$ measures the consequence of those errors for the particular substantive estimand. Measurement systems with similar omnibus losses may produce different estimand distortions if their errors occur for different papers or in different parts of the distribution. Conversely, one need not recover all of the target values to recover a particular estimand \citep{LudwigEtAl26}.
	
	While the target values $\mathbf{T}_{i}$ are observed in a human-labeled evaluation sample, they typically remain unobserved in the full sample or population. If they were observed for every paper, it would mean that a human had read and labeled every paper, making the use of an LLM in the review process superfluous. The implication is that estimand distortion can be calculated for a human-labeled evaluation sample but cannot be directly calculated for the full sample. Despite this limitation, we can evaluate the dependence of results on the choice of measurement system.
	
	We denote the mean estimate across the $M$ measurement systems as: 
	
	\begin{equation}
		\overline{\theta}_{i} = \frac{1}{M} \sum_{m = 1}^{M} \widehat{\theta}_{i}^{(m)}.
	\end{equation}
	
	\noindent Then, a compact measure of cross-system dependence is the measurement-system dispersion: 
	
	\begin{equation}
		D_{i} \left( \theta \right) = \left[ \frac{1}{M} \sum_{m = 1}^{M} \left( \widehat{\theta}_{i}^{(m)} - \overline{\theta}_{i} \right)^{2} \right]^{1/2}. 
	\end{equation}
	
	\noindent As the measurement systems are a fixed collection rather than a random sample from a population of LLMs, $D_{i} \left( \theta \right)$ is not a standard error or confidence interval. With that said, dispersion is a useful metric, as a larger value of $D_{i} \left( \theta \right)$ indicates that the estimate for concept $i$ is more sensitive to the measurement system used to generate the data.
	
	We can decompose the mean squared estimand distortion across measurement systems as:
	
	\begin{equation}
		\frac{1}{M} \sum_{m = 1}^{M} \left( \Delta_{i}^{(m)} \right)^{2} = D_{i} \left( \theta \right)^{2} + \left( \overline{\theta}_{i} - \theta_{i} \right)^{2}.
		\label{eq:distortion}
	\end{equation}
	
	\noindent The implication is that even when $\theta_{i}$ is unobserved, squared dispersion provides a lower bound on mean squared estimand distortion. We need not make assumptions about the nature of measurement error or its correlation across measurement systems to arrive at this result. But small dispersion does not imply small estimand distortion. Measurement systems may produce much the same estimates because they accurately measure the target values or because their errors distort the estimand in much the same way.
	
	Together, these measures distinguish three related but separate features of LLM-generated data: (1) task-level measurement error, (2) the resulting distortion in a substantive estimand, and (3) the sensitivity of that estimand to the choice of measurement system. The framework also clarifies what must be observed to evaluate each object.\footnote{Appendix~\ref{app:measures} provides a practical guide to operationalizing this framework for a variety of types of target estimands.} Target labels are required to estimate classification and semantic error rates. Without at least some labeled references one cannot determine the type and severity of measurement error in LLM-generated data. In the absence of validation data, or when examining performance on the entire corpus, system-specific estimates can be compared across measurement systems in order to determine the degree to which substantive estimates are a function of the measurement system. But this cross-system comparison does not tell us which measurement system produces the most accurate measures or which estimate is closest to the truth.
	
	
	\section{Corpus Construction and Labeling} \label{sec:data}
	
	Operationalization of this framework requires two linked data structures: a human-labeled reference sample, which provides the target values needed to estimate classification and semantic error, and a common corpus processed by each measurement system, which permits calculation of measurement-system-specific descriptive estimates and their dispersion. Corpus definition and label construction are thus not ancillary data-management choices. Rather, they determine both the population to which the descriptive claims apply and the benchmark against which measurement quality is assessed. Figure~\ref{fig:rd} in Appendix~\ref{app:pipe} presents our development pipeline from corpus construction through performance validation to substantive analysis.
	
	
	\subsection{Corpus Construction} \label{sec:corpus}
	
	The first decision we make in corpus construction is the choice of source for paper citations and metadata. While there are many searchable databases of scholarly work (e.g., Google Scholar, Web of Science, JSTOR, PubMed, Scopus), all are partial collections of the scientific record. We select the OpenAlex database \citep{Priem2022-pv} for three reasons. First, OpenAlex is the successor to the Microsoft Academic Graph (MAG), which at retirement in 2021 contained 239 million research publications. The MAG was the most complete index of scientific publications \citep{Sinha2015-zp} and was frequently used for meta-analysis of research trends in economics \citep{Jones2021-yg, Josephson2023-pl}. OpenAlex's database is a frequently updated continuation of the MAG but adds additional sources like Crossref, ORCID, PubMed, arXiv, and other institutional repositories. Because of the nearly daily updating to the bibliographic catalog, any query of OpenAlex is a single snapshot in time of what is in the catalog. At the time of our API call to OpenAlex (19 February 2025) the graph contained 248 million academic works, nearly three times the number in Scopus and Web of Science \citep{Priem2022-pv}. Second, OpenAlex is not just a search tool or database, but like the MAG, it is a catalog that disambiguates and connects scholarly work. Each catalog entry contains metadata such as title, author(s), year of publication, and DOI as well as concepts that capture what the work is about, and author and institutional relationships. This information forms a series of nodes and edges that connect the metadata from one entry to another. OpenAlex indexes 65,000 concepts allowing for greater search coverage than just querying title, abstract, and keywords. Third, OpenAlex is an open-access catalog, making research based on OpenAlex entries more reproducible. 
	
	The second decision we make is to construct a corpus of literature that is as close as possible to the universe of all papers that use rainfall as an IV. This decision means that our search criteria seek to minimize Type II errors at the expense of making many Type I errors. Therefore, our search criteria are overly broad, relative to a search criteria that seek to maximize accuracy by minimizing both Type I and Type II errors. This has the effect of increasing recall at the expense of specificity. The practical outcome of this decision is that we expect many papers in our corpus of rainfall-IV papers to not actually have a rainfall IV $(T_{ij} = 0)$. The downstream implication is that the number of TNs and FPs will increase, decreasing precision. We believe this decision is justified because it moves the most difficult error (i.e., an unretrieved paper) into the corpus. A non-rainfall-IV paper can be correctly classified by the LLM as a TN while a paper that never enters the corpus cannot be recovered. In terms of inference for the literature as a whole, the result of admitting more FPs is that we contaminate the corpus with papers that do not use rainfall IVs, potentially distorting characteristics of the rainfall-IV literature when the LLM classifies the paper as an FP. The alternative would be to reduce the coverage of the corpus, which if omissions are not random (e.g., older papers, papers in certain fields with different IV terminology) can distort conclusions about what the literature actually says.
	
	The third decision we make regards the actual search terms used to query OpenAlex. We use Boolean operators to search for works that include ``((Weather) AND (Instrumental Variable)) OR ((Rainfall) AND (Instrumental Variable))''. This results in more than 65,000 plausibly relevant academic works. We then filter these results for only English language works and works in OpenAlex's predefined domains of ``Economics'', ``Econometrics'', and ``Finance''. This narrows the search results to $4,191$ papers, comprising what is, in our view, the potential universe of economic papers that use rainfall as an IV. With the metadata on papers generated from OpenAlex we then download all works as PDFs through a combination of web scraping and manual download. 
	
	
	\subsection{Labeling}  \label{sec:label}
	
	Because our OpenAlex search prioritized recall, the candidate corpus of $4,191$ papers likely includes many papers that do not contain a rainfall IV. We select the $439$ papers ($\sim 10\%$ of the total) that have the highest OpenAlex relevance scores to form our human-labeled reference sample.\footnote{OpenAlex computes a relevance score to rank search results. The score is based on text similarity between the search terms and the metadata attached to the academic work. This semantic similarity score is then weighted by citation count, with higher cited papers receiving greater weight.} The selection of papers with the highest relevance scores increases the number of rainfall-IV papers in the human-labeled reference sample and thus the opportunities to train and evaluate string extraction. The tradeoff is that, because the human-labeled reference data are not a random subsample, validation metrics cannot be interpreted as estimates of model performance for the candidate corpus.
	
	Given that our primary goal is to characterize whether papers use a rainfall IV, the name of that rainfall IV, and the name of the endogenous variable that it instruments, we build a human-labeled reference set with a hierarchical structure and accompanying contextual information. Table~\ref{tab:info_measured} presents the extraction schema, including fields, output type, example record, and dependencies. Before formal labeling began, the principal investigators (PIs) trained two undergraduate research assistants (UGRAs) and two graduate research assistants (GRAs) to apply a common coding protocol. Training covered the study objectives, causal identification, IV estimation, and the terminology used to describe endogenous variables and instruments. Annotators reviewed worked examples and received a codebook and 40 pre-labeled reference papers. They then independently coded ten calibration papers and received oral and written feedback from the PIs before proceeding to the remaining assignments.
	
	The annotation process proceeds as follows. For each paper, annotators first record the PDF filename, article title, and DOI. If the paper is empirical, annotators assign a label of 1 and record the dependent variable(s) used in the analysis. If the paper is not empirical, annotators assign a label of 0. In this case, and whenever a dependency condition is not satisfied, the variable field is assigned \texttt{n/a}, as are all subsequent fields in the hierarchy. Conditional on the paper being empirical, annotators assign a label of 1 or 0 if the paper has an endogeneity problem or not. If there is an endogenous variable, annotators record the name of the endogenous variable(s). The same hierarchical process is applied to whether an IV is employed and if the IV is precipitation-based. For every field that contains an extracted variable name, annotators also record the section heading and the sentence or sentences where the variable name was found.
	
	Recording section headings and the sentences containing the variable names can provide LLMs with an opportunity to identify context clues for locating and correctly labeling the dependent, endogenous, and instrumental variables.\footnote{This contextual information is part of the human-labeled reference data but not part of the information the LLM is prompted to extract.} This is particularly useful in our test case because the rainfall-IV literature is extensive, interdisciplinary, and methodologically diverse. For example, explicit mention of an ``identification strategy'' is relatively rare outside of economics and even in economics it is rare prior to the Credibility Revolution. Information on the instruments varies by econometric method. Identification in dynamic panel models, selection models, or Lewbel estimators is reliant on IVs, yet papers using these methods rarely use terms like ``exclusion restriction,'' which is a common term in papers using 2SLS.\footnote{Moreover, many of these econometric models have more than one name: e.g., Blundell-Bond or Arellano-Bond for dynamic panels; endogenous switching, Heckman selection, or hazard selection for selection models; Lewbel or heteroskedasticity‑based instruments for generated instruments.} This diversity of terminology and context means that the task cannot be solved reliably through keyword matching alone. The model must therefore infer econometric roles from the surrounding context.
	
	To verify the accuracy of the human reference labels, we undertook a three-stage process. In the first stage, the two UGRAs and the two PIs were randomly assigned a subset of papers so that each paper was independently coded twice. In the second stage, the papers were reassigned to the two GRAs and the two PIs and independently coded twice again. In the third stage, the PIs compared the labels across stages and collectively reviewed disagreements so as to reconcile differences. The resulting adjudicated reference labels operationalize the target values $T_{ij}$ in our measurement error framework.
	
	
	\subsection{Prompt Engineering} \label{sec:prompt}
	
	Prompt engineering is the practice of constructing and polishing natural language inputs for LLMs to improve the quality and relevance of their outputs. Good prompt engineering can significantly enhance output quality and reduce hallucination frequency, but what makes a ``good'' prompt is not always clear, particularly if one lacks quantitative metrics to evaluate model performance. Different prompting formulations can often engender entirely different outputs even when provided to the same model with the same input information. Experimental results paired with our own exploration of the subject demonstrate that models can be highly reactionary even to the most subtle alterations in wording \citep{LeidingerEtAl23, MizrahiEtAl24, SalinasMorstatter24}. By eliciting desired model behaviors through prompt alteration, we can observe improvements to output quality without modifying model parameters \citep{sahoo2025}.
	
	Our prompt engineering follows an iterative process in which we fine-tune an independent GPT-4.1 model using a draft prompt and fine-tuning training data. After training, the model outputs its own labels for comparison with the held-out evaluation set, and we calculate performance validation metrics. The validation metrics provide information on whether errors are driven primarily by FPs or FNs. We also calculate semantic similarity scores and visually compare model-labeled strings to human-labeled strings in order to determine if model labels are conceptually aligned with the target values. Based on this information, we revise prompt text and structure and begin the next iteration on a GPT-4.1 model independent from the first.\footnote{Our decision to iterate prompts using the same training dataset does run the risk of overfitting through prompt selection. To minimize this risk, no evaluation data enter fine-tuning and every fine-tuned model is an independent run. However, the prompts are optimized to the evaluation set and so error rates may be higher when the implementations are deployed to read the entire corpus.} We had no prespecified stopping rule or target level for validation metrics and semantic similarity. Rather, we stopped iterating when successive prompt revisions produced only minor marginal improvements, with the rate of performance improvement approaching zero.
	
	Across all three implementations, we use a common document-grounded prompting framework that contains a fixed context layer and a set of 10 queries (questions). The fixed context consists of the system role (job description), extraction rules, normalized paper text, and an assistant prefill. The queries have the same hierarchical structure used by the human annotators. This hierarchical structure efficiently extracts information by reducing unnecessary queries and token use by making downstream queries conditional on preceding outputs. The queries elicit four binary classifications and six string responses from the model.
	
	Each query is uniformly structured to define the task (e.g., ``extract the article title''), recommend places to look in each PDF (e.g., ``top of first page (main heading before authors/abstract)''), specify extraction rules (e.g., ``ignore footnote markers (*, †, superscripts)''), and provide an explicit desired output format (e.g., ``ONE line: title text only. No quotes, labels, or extra words''). We then immediately repeat the text of the query because in the linear logic of an LLM, which reads left to right, any part (token) of a prompt cannot be informed by tokens that come after it. Thus, if a prompt is structured \texttt{context}-\texttt{query} the \texttt{query} is informed by the \texttt{context} but the LLM encounters the \texttt{context} without having access to the \texttt{query}. \cite{LeviathanEtAl25} shows that sending the query twice within a prompt allows each token in the query to inform every other token in the query, resulting in substantially improved performance. The full text of the queries is in Appendix~\ref{app:pipe}.
	
	
	\section{Model Implementations}  \label{sec:model}
	
	Researchers using LLMs for information extraction are faced with a set of tradeoffs, among which are task specialization, output quality, and the required data, computation, and labor. Task specialization can occur at two points, corresponding to the two distinct algorithms that constitute an LLM: the training algorithm and the text generator it produces. An off-the-shelf model can be specialized at the point of text generation with task-specific prompts alone or by supplying it with reference examples through retrieval. Alternatively, the model itself can be adapted through updating the training algorithm. We compare these approaches using the same general-purpose base models. The implementations differ in whether human-labeled reference information enters the prompt or is used to update model parameters.
	
	We use GPT-4.1 through OpenAI's API to compare three different implementations: (1) a simple zero-shot prompting strategy, (2) retrieval-augmented generation (RAG), and (3) supervised fine-tuning (SFT). Each paper is processed in a dedicated thread that becomes a single multi-turn conversation record containing the fixed context and the sequence of query/response turns for each extraction field. This allows later queries to draw on the model's response to earlier queries within the same conversation. In the final step, model output passes through some post-hoc consistency processing.
	
	We randomly divide the paper-level records into a non-held-out reference set ($80\%$) and a common held-out evaluation set ($20\%$). The non-held-out reference set supplies retrieval examples for RAG and training and development data for SFT. The common held-out evaluation set supplies the targets against which generated outputs from all three implementations are compared. The labels from this common held-out set are neither retrieved as examples nor used to update model parameters. 
	
	
	\subsection{Zero-Shot Baseline}  \label{sec:0shot}
	
	Our zero-shot implementation is a document-grounded, ``closed-book'' prompting strategy. During extraction, the model receives the normalized text of the focal paper and the prompts described in Section~\ref{sec:prompt}, but has no access to the human-labeled reference data. Zero-shot therefore isolates what the model can extract from the focal paper using the prompting framework alone and provides the baseline for evaluating the additional value of RAG and SFT.
	
\begin{landscape}
		\begin{figure}[!htbp] 
			\begin{minipage}{\linewidth}		
				\begin{center}
					\caption{Zero-Shot Implementation - Prompt Contents, Query Sequence, and Outputs  \label{fig:0sht_flo}}
					\includegraphics[width=.89\linewidth,keepaspectratio]{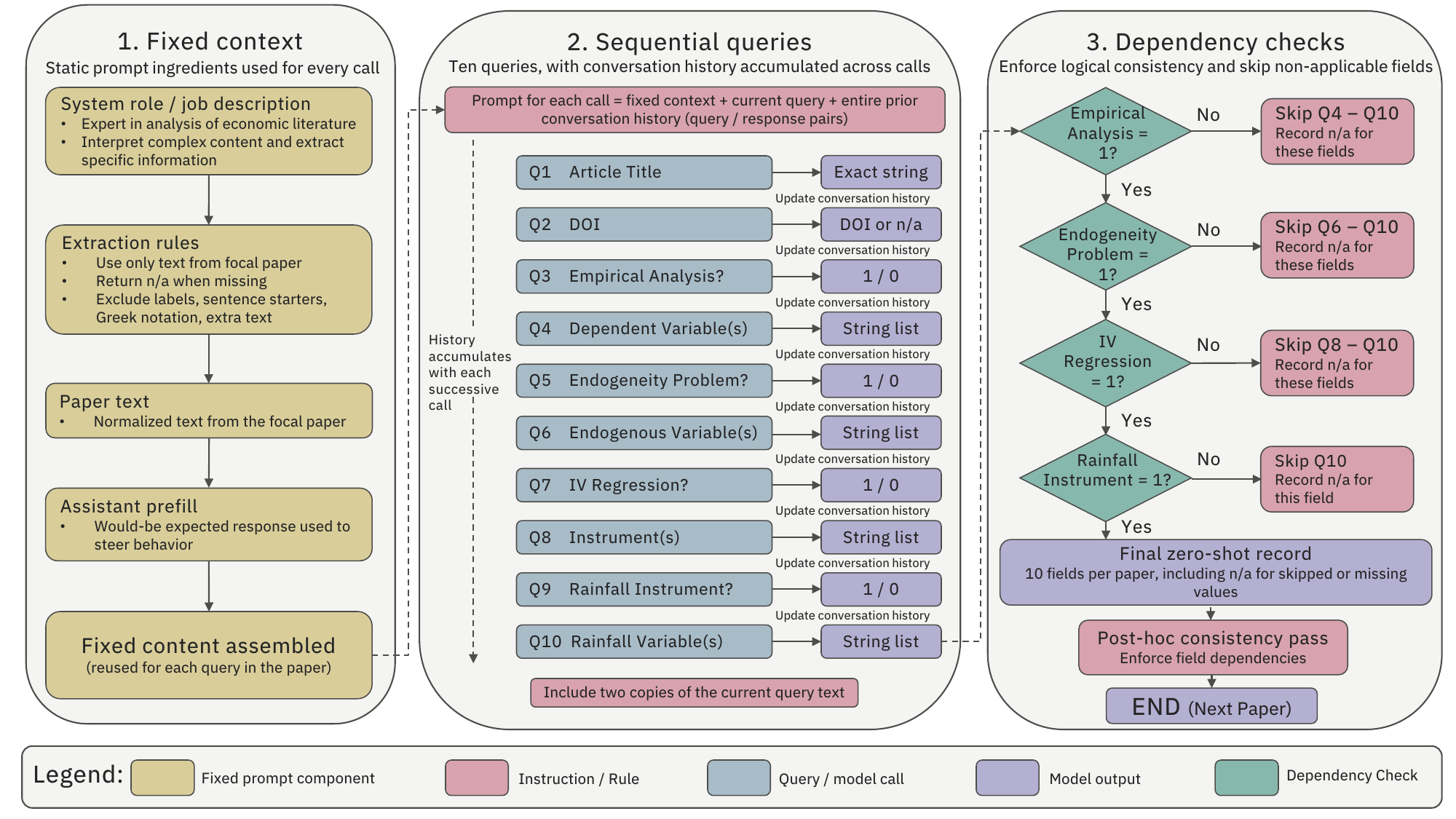}
				\end{center}
				\footnotesize  \textit{Note}: The figure shows the document-grounded, closed-book zero-shot workflow. The model receives the focal paper but no human-labeled reference examples. Panel 1 assembles the fixed context from the system role, extraction rules, normalized paper text, and assistant prefill. Panel 2 processes eligible queries sequentially. Each call includes the fixed context, two copies of the current query, and all prior query/response pairs. Panel 3 shows the dependency checks that determine which queries are skipped and the post-hoc check that enforces field dependencies. Each paper yields a ten-field record, with n/a for skipped or missing values.
			\end{minipage}
		\end{figure}   
	\end{landscape}

	Figure~\ref{fig:0sht_flo} lays out the zero-shot workflow. Each prompt supplied to the model contains a fixed context layer, the sequence of queries, and a set of dependency rules and consistency checks. Panel 1 shows the fixed context supplied to the model for each paper which consists of the system role, user rules, normalized paper text, and assistant prefill. The assistant primes the model's behavior with a fake commitment with the intention of steering later responses towards our preferred behavior. This is a would-be-expected response from a model that was fed the same system context and user queries.
	
	Panel 2 shows the sequential query/response loop. The primed model is fed each of our specified queries one at a time with the text of the current query repeated twice. After the first call, the model also receives all prior query/response turns. This creates a multi-turn sequential query/response loop, where the conversation history accumulates as the model moves through the sequence of queries.
	
	Panel 3 shows how the dependency structure governs which queries are sent to the model. Before each dependent query, the script checks the relevant response from the earlier query. If the dependency is not satisfied, the query and any downstream dependent queries are skipped, and the corresponding fields are coded as \texttt{n/a} without calling the model. After all eligible queries have been processed, a post-hoc consistency pass re-checks the four dependency gates and enforces field dependencies in the ten-field record. As zero-shot uses neither retrieved examples nor parameter updates, it is our least information-intensive implementation.
	
	
	\subsection{Retrieval-Augmented Generation (RAG)}  \label{sec:rag}
	
	RAG is a more information-intensive implementation of the same prompting strategy used in zero-shot. The focal paper, fixed context, field-specific queries, dependency structure, and creation of conversation history remain unchanged. What differs is that each eligible query is augmented with examples retrieved from the human-labeled reference data. RAG is therefore an ``open-book'' implementation. The model has access to adjudicated reference examples through the prompt, but its model weights remain unchanged \citep{sahoo2025}.
	
\begin{landscape}
		\begin{figure}[!htbp] 
			\begin{minipage}{\linewidth}		
				\begin{center}
					\caption{RAG Implementation - Reference Indexing, Retrieval, and Outputs  \label{fig:rag_flo}}
					\includegraphics[width=.89\linewidth,keepaspectratio]{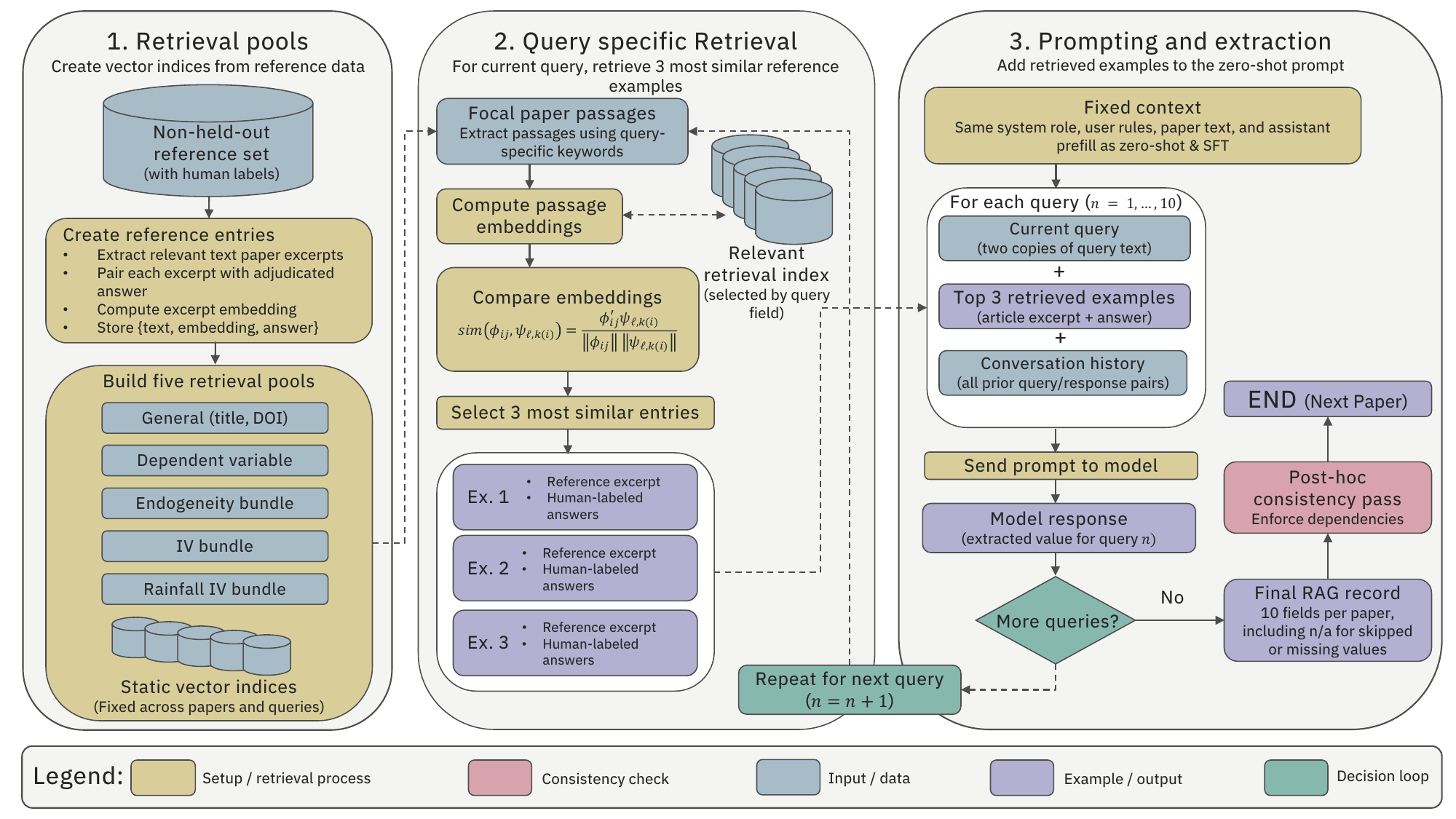}
				\end{center}
				\footnotesize  \textit{Note}: The figure shows the document-grounded, open-book RAG workflow. Panel 1 constructs five static retrieval pools from the non-held-out human-labeled reference set. Each entry contains an excerpt, its embedding, and an adjudicated answer. Panel 2 embeds query-specific focal-paper passages and uses cosine similarity to select the three most similar entries from the relevant pool. Panel 3 combines these examples with the fixed context, current query, and prior query/response pairs. Retrieval repeats for each eligible query, while model parameters remain unchanged. A post-hoc check enforces field dependencies. Each paper yields a ten-field record, with n/a for skipped or missing values.
			\end{minipage}
		\end{figure}   
	\end{landscape}

	Figure~\ref{fig:rag_flo} shows the two phases of our RAG implementation. Panel 1 covers indexing, while Panels 2 and 3 show the retrieval-conditioned generation loop. During indexing, we use the non-held-out reference set to construct five field-specific retrieval pools as vector indices. The first pool contains general information used for title and DOI extraction. The remaining four contain information on dependent variables, endogenous variables, instrumental variables, and rainfall IVs. Each reference entry contains an article excerpt, its embedding, and the corresponding adjudicated answer. The resulting indices remain fixed when the models are applied to the held-out evaluation set and the candidate corpus.
	
	Panel 2 shows the retrieval process for each eligible query. The pipeline extracts keyword-specific passages from the focal paper using a fixed set of terms related to our query concepts. The extracted focal-paper context is embedded and compared with reference-entry embeddings in the corresponding retrieval pool. To ensure we are injecting context relevant to the focal paper, we calculate cosine similarity scores between each query and all pool embeddings. Let $\boldsymbol{\phi}_{ij}$ denote the embedding of the focal-paper context for concept $i$ in paper $j$ and $\boldsymbol{\psi}_{\ell,k(i)}$ the embedding of reference entry $\ell$ in retrieval pool $k(i)$. We calculate cosine similarity as
	
	\begin{equation}
		\operatorname{sim} \! \left( \boldsymbol{\phi}_{ij}, \boldsymbol{\psi}_{\ell,k(i)} \right)
		=
		\frac{
			\boldsymbol{\phi}_{ij}^{\prime}\boldsymbol{\psi}_{\ell,k(i)}
		}{
			\left\| \boldsymbol{\phi}_{ij} \right\|
			\left\| \boldsymbol{\psi}_{\ell,k(i)} \right\|
		}.
	\end{equation}
	
	\noindent The three most similar reference entries are retrieved for each query. Each is supplied to the model as an article excerpt paired with its adjudicated answer. Panel 3 shows how these examples are combined with the fixed context, current query, and prior query/response history to form the model prompt. After each call, the query/response pair is appended to the conversation history. If eligible queries remain, the loop returns to Panel 2 to retrieve examples for the next query. Once all eligible queries have been processed, the ten-field record undergoes the same post-hoc consistency checks as in zero-shot.
	
	The principal advantage of RAG in our setting is that it gives an otherwise unchanged model access to examples of how human annotators interpret heterogeneous economic terminology. Conditioning output generation on retrieved context can improve factual grounding and reduce hallucinations \citep{LewisEtAl20}. But RAG also introduces an additional potential source of error. Performance depends not only on the model but on the quality of the reference data and whether the retrieval procedure identifies substantively relevant examples. Semantically similar text need not represent the same economic concept. Increasing the size of the retrieved context window raises the probability of surfacing substantively relevant examples, but can also introduce less relevant material that dilutes the contribution of the most informative examples. More context thus does not guarantee better performance \citep{Gu26}.
	
	
	\subsection{Supervised Fine-Tuning (SFT)}  \label{sec:sft}
	
	SFT uses the human-labeled reference data to adapt model parameters before extraction. Unlike RAG, which supplies examples within each prompt, SFT uses adjudicated responses to modify how the model responds to subsequent queries. In our setting, the aim is to improve interpretation of the indirect phrasing, qualifications, and domain-specific terminology used to describe endogeneity and IV strategies. We fine-tune the same base models used in zero-shot and RAG. The tuning mechanism is represented by Low-Rank Adaptation (LoRA), a parameter-efficient fine-tuning method \citep{HuEtAl21}.\footnote{OpenAI does not disclose its parameter-updating method but the generative AI community is in general agreement that GPT fine-tuning uses LoRA as all other major fining-tuning platforms use LoRA. That said, GPT's use of LoRA is an assumption not a verified description of the platform's implementation.}
	
\begin{landscape}
		\begin{figure}[!htbp] 
			\begin{minipage}{\linewidth}		
				\begin{center}
					\caption{SFT Implementation - Training, Prompting, and Outputs  \label{fig:sft_flo}}
					\includegraphics[width=.89\linewidth,keepaspectratio]{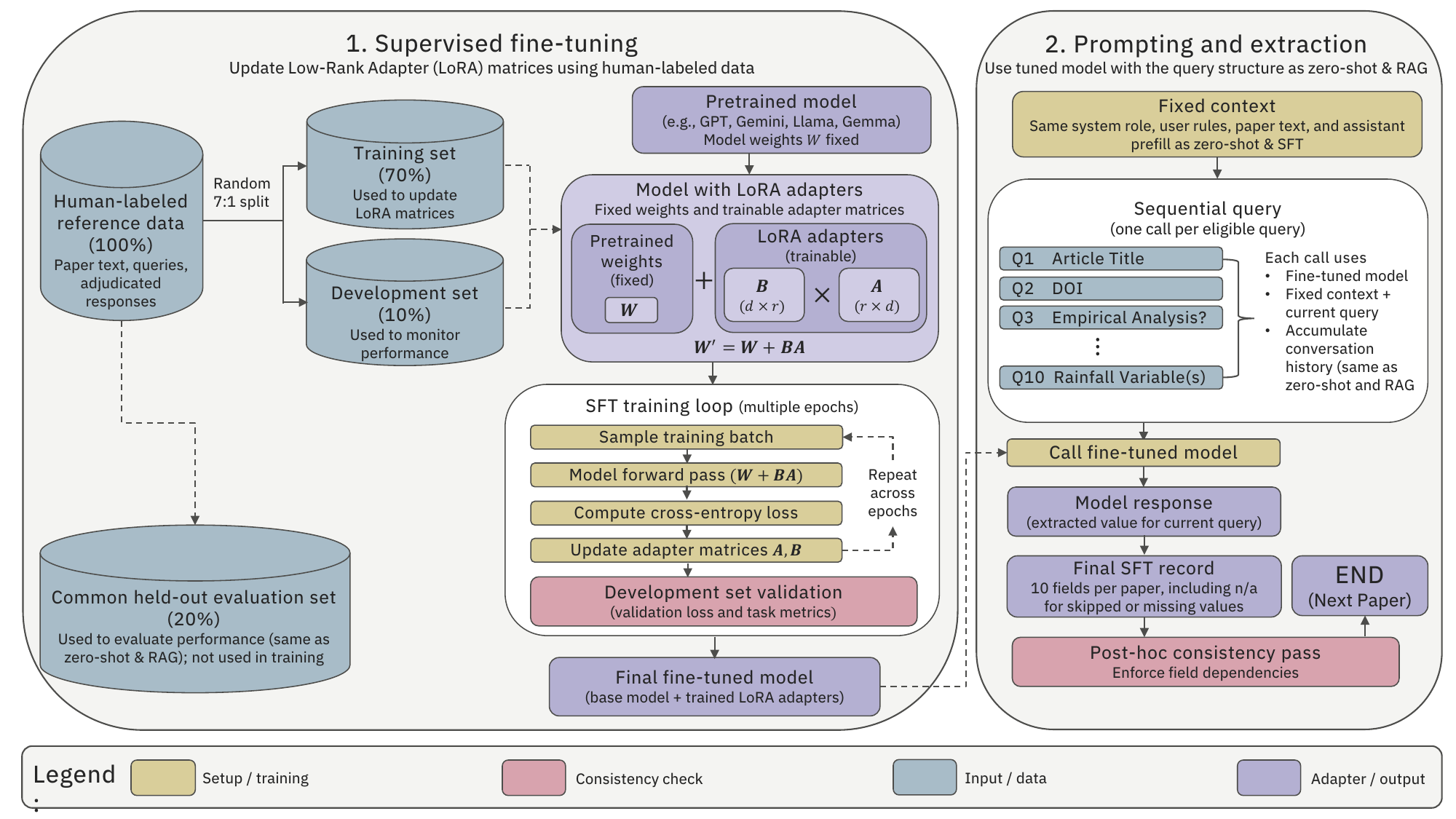}
				\end{center}
				\footnotesize  \textit{Note}: The figure shows supervised fine-tuning followed by paper-level extraction. Panel 1 partitions the human-labeled reference sample into training (70\%), development (10\%), and common held-out evaluation (20\%) sets. Training minimizes cross-entropy loss by updating low-rank adaptation (LoRA) matrices \(A\) and \(B\), while pretrained weights \(W\) remain fixed. The development set monitors performance without updating parameters. The held-out set is used for performance evaluation, not training. Panel 2 applies the final fine-tuned model to each paper using the fixed context, current query, and accumulated query/response history. Extraction uses no retrieved examples or further parameter updates. A post-hoc check enforces field dependencies in the ten-field record, with n/a for skipped or missing values.
			\end{minipage}
		\end{figure}   
	\end{landscape}

	Figure~\ref{fig:sft_flo} separates fine-tuning from subsequent extraction. Panel 1 shows the allocation of reference data and the training process. We divide the non-held-out reference set 7:1 into a training and development set. The development set is called the validation set by fine-tuning platforms. The pretrained model enters fine-tuning with its base weights fixed.\footnote{This differs from full fine-tuning, in which elements in a series of matrices that make up a base model’s architecture (e.g., receiving, embedding, predicting, etc.) are made adjustable, whereby the underlying model is fundamentally modified to fit a specific task.} For each selected weight matrix $\mathbf{W}$, LoRA introduces two smaller trainable matrices, $\mathbf{A}$ and $\mathbf{B}$. Their product provides a low-rank adjustment to the original weights such that: 
	
	\begin{equation}
		\mathbf{W}' = \mathbf{W} + \mathbf{B}\mathbf{A}.
	\end{equation}
	
	\noindent Only the adapter matrices are updated during training. This reduces the number of trainable parameters relative to full fine-tuning while allowing the model's responses to adapt to the extraction task.
	
	The training loop in Panel 1 processes batches of labeled conversation records. At each step, the model computes token probabilities for the adjudicated responses and the cross-entropy loss associated with those responses. The loss is used to update $\mathbf{A}$ and $\mathbf{B}$, while $\mathbf{W}$ remains fixed. The updated adapter matrices are used in the next training step, and the process repeats across batches and epochs. Within the training records, later queries are conditioned on the adjudicated responses to earlier queries. During extraction, by contrast, later queries are conditioned on the model's own prior responses. The fine-tuning development set is evaluated during training and at completion. Development-set metrics allow us to monitor performance beyond the training examples without using those records for parameter updates. We deploy the final model returned by the completed job rather than selecting an earlier checkpoint based on development-set performance.
	
	Panel 2 shows extraction using the final fine-tuned model. Each paper begins a new thread with the same fixed context and field-specific queries used in zero-shot. Eligible queries are processed sequentially, and the same dependency rules and post-hoc consistency checks produce a ten-field record for each paper. No reference examples are retrieved and no parameters are updated during extraction. Relative to zero-shot and RAG, SFT changes the model used to process the prompt rather than the information supplied about the focal paper or relevant examples. Despite requiring a higher upfront investment in computation and labor, SFT can scale more efficiently than zero-shot or RAG approaches \citep{liu2024lost, dunn2022structured}. 
	
	
	\section{Performance Validation} \label{sec:valid}
	
	Prior to deploying each implementation on the entire corpus, we evaluate performance based on a set of metrics designed to capture accuracy and to quantify the types of successes and failures.
	
	
	\subsection{Validation Criteria} \label{sec:valid-crit}
	
	We evaluate binary classification and string extraction performance using the validation metrics defined in Section~\ref{sec:ME-model}. Following a modified version of the answer evaluation rubric in \citet{balaguer2024rag}, we also conduct a targeted manual review of model-implementation outputs to characterize common failure modes and ensure that apparent quantitative gains correspond to substantively correct extraction.
	
	Confusion matrices summarize agreement between the LLM-generated measures $\left( Z_{ij} \right)$ and target values $\left( T_{ij} \right)$ by cross-tabulating the target values (rows) against the measured values (columns). Each confusion matrix contains four cells: true positives ($\mathrm{TP}$), cases where $T_{ij} = Z_{ij} = 1 $; true negatives ($\mathrm{TN}$), $T_{ij} = Z_{ij} = 0 $; false positives ($\mathrm{FP}$), cases where $T_{ij} = 0$ and $Z_{ij} = 1 $; and false negatives ($\mathrm{FN}$), cases where $T_{ij} = 1$ and $Z_{ij} = 0 $. Table~\ref{tab:confusion_matrix} visualizes these quantities. Together, they characterize the measurement system's error profile by distinguishing false detections from missed positive cases while also recording correct detections and correct rejections.
	
	From the confusion matrix cells we report accuracy, sensitivity, specificity, precision, $\operatorname{F1}$-score, and balanced accuracy. Accuracy is the share of correct classifications: $(\mathrm{TP} + \mathrm{TN})/(\mathrm{TP} + \mathrm{TN} + \mathrm{FP} + \mathrm{FN})$. Sensitivity (the true positive rate) is calculated as $\mathrm{TP}/(\mathrm{TP} + \mathrm{FN})$ and measures how often target-positive cases are detected. Specificity (the true negative rate) is calculated as $\mathrm{TN}/(\mathrm{TN} + \mathrm{FP})$ and measures how often target-negative cases are correctly rejected. Precision (the positive predictive value) is calculated as $\mathrm{TP}/(\mathrm{TP} + \mathrm{FP})$ and captures the reliability of positive predictions. The $\operatorname{F1}$-score is the harmonic mean of precision and sensitivity and summarizes positive-class performance when both FPs and FNs matter. Balanced accuracy is the arithmetic mean of sensitivity and specificity and is particularly informative when class frequencies are uneven, because it weights performance on the positive and negative classes equally.
	
	For non-binary responses, we evaluate semantic agreement between the model-generated string $\left( X_{ij} \right)$ and the target string $\left( N_{ij} \right)$ using sentence embeddings and cosine similarity. We encode each string using all-MiniLM-L6-v2, a SentenceTransformer model built on a Bidirectional Encoder Representations from Transformers (BERT) family architecture \citep{ReimersGurevych19}. SentenceTransformers are transformer-based models that are fine-tuned to produce sentence-level embeddings for semantic similarity. For TP cases, semantic similarity is $S_{ij} = s \! \left( X_{ij}, N_{ij} \right)$ where $s$ is the cosine similarity between the two embeddings. This allows us to measure semantic alignment beyond exact string matching and to summarize performance using both the mean and distribution of semantic similarity across the evaluation set.
	
	
	\subsection{Performance Validation} \label{sec:valid-perf}
	
\begin{figure}[!htbp] 
		\begin{minipage}{\linewidth}		
			\begin{center}
				\caption{GPT: Evidence, Performance, and Semantic Stability  \label{fig:gpt_eval}}
				\includegraphics[width=.85\linewidth,keepaspectratio]{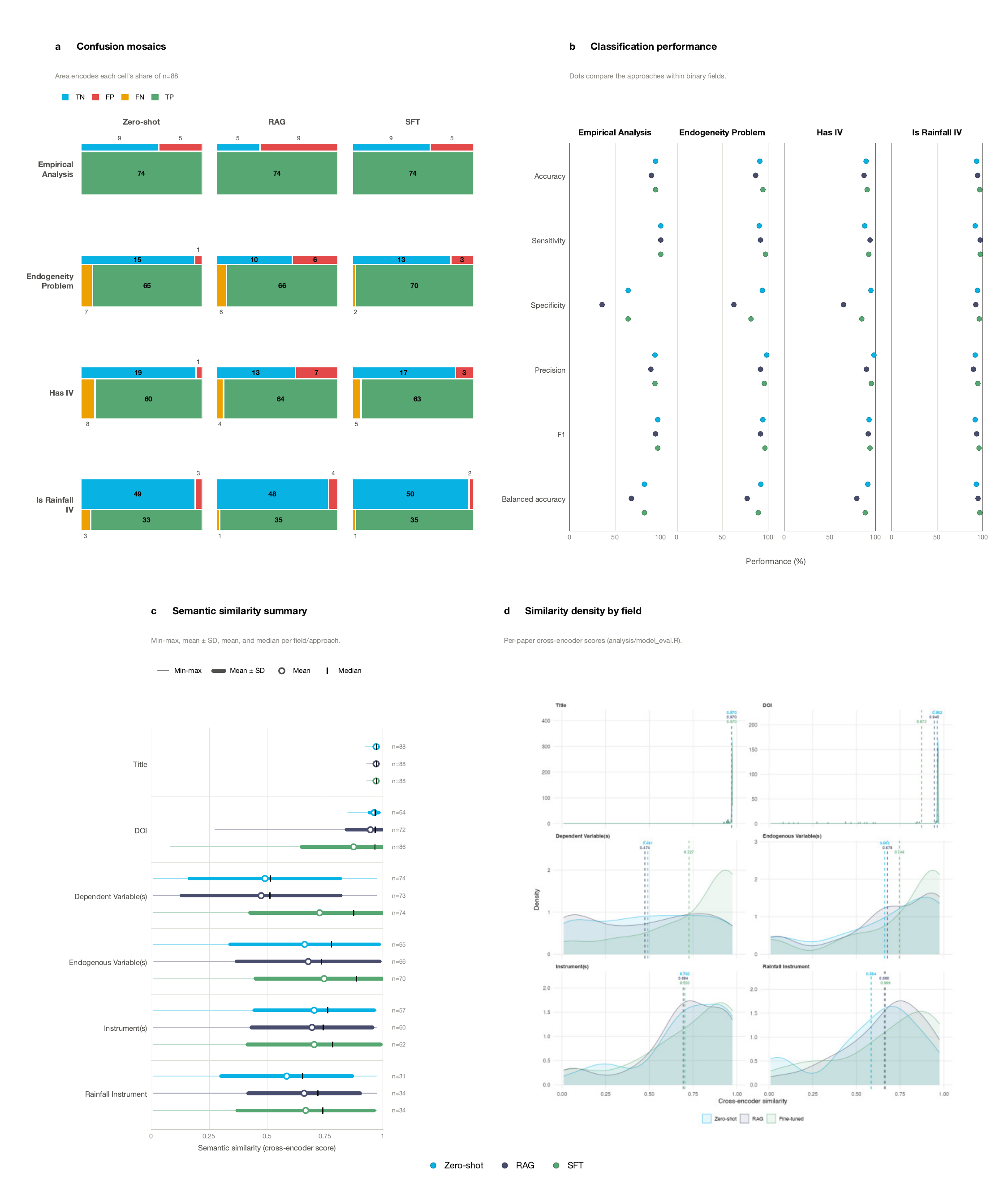}
			\end{center}
			\footnotesize  \textit{Note}: The figure compares zero-shot, retrieval-augmented generation (RAG), and supervised fine-tuning (SFT) on a common held-out evaluation set ($n = 88$ papers). Panel A presents confusion mosaics for four binary extraction tasks, with cell areas and labels indicating true negative (TN), false positive (FP), false negative (FN), and true positive (TP) counts. Panel B reports accuracy, sensitivity, specificity, precision, F1 score, and balanced accuracy by implementation and task. Panels C and D evaluate string extraction using semantic similarity between generated and target strings. Panel C reports the minimum--maximum range, mean $\pm$ one standard deviation, mean, and median by field and implementation, with sample sizes at right. Panel D plots score distributions, with dashed vertical lines indicating implementation-specific means. For conditional string-extraction fields, scores are defined only for TP observations, so sample sizes vary across fields and implementations.
		\end{minipage}
	\end{figure}

	Figure~\ref{fig:gpt_eval} summarizes task-level evaluation metrics on the common held-out evaluation set for the three implementations of GPT-4.1. Panel A presents the confusion mosaics for each binary outcome. Panel B uses the raw confusion matrices to calculate binary task-level validation metrics. Panel C presents range, mean, and standard deviation of semantic similarity scores while Panel D draws the distribution of these scores.
	
	What is immediately obvious is that all implementations perform very well on binary classification tasks. Of the $88$ papers in our held-out evaluation sample, each implementation misclassifies (FP or FN) $12$ or fewer papers for every binary query. Consequently, all implementations have high accuracy (share of correct answers) and precision (positive predictive value). In fact, accuracy for the three implementations on the empirical analysis, IV use, and rainfall-IV tasks varies by only three to four percentage points. The difference in accuracy is larger for endogeneity, driven by RAG's relatively poor performance, but even on this task the largest difference is only eight percentage points.
	
	Despite this high accuracy and precision, when implementations make errors the types of errors differ substantially. Measures of specificity (the TN rate) and sensitivity (the TP rate) capture these differences. The RAG implementation produces noticeably more FPs than either zero-shot or SFT for empirical analysis, endogeneity, and IV use. These errors reduce specificity and balanced accuracy. As an example, for empirical analysis RAG achieves accuracy of $90\%$ but specificity of only $36\%$. In other words, RAG is misidentifying papers as being empirical, having an endogeneity problem, and using an IV at a higher rate than zero-shot or SFT.
	
	The zero-shot implementation makes almost as many errors as RAG but exhibits a different error profile. It has very few FPs but many more FNs for endogeneity, IV, and rainfall IV, resulting in higher specificity but lower sensitivity. Thus, zero-shot is missing papers that actually do have an endogeneity problem, use an IV, and use a rainfall IV. Empirical analysis is the exception where zero-shot has sensitivity levels that match those of the other implementations. In fact, all three implementations correctly identify all $74$ papers that are empirical, yielding perfect sensitivity, with performance differences arising entirely from FPs among the remaining $14$ papers.
	
	SFT generally has the strongest and most balanced performance of the three implementations. It produces relatively few FPs and FNs across all four tasks and performs particularly well in detecting the presence of a rainfall IV, where it only misses one positive and two negative cases. Its performance is nearly as strong for IV use and endogeneity, with only a small number of errors of either type. As a result, SFT combines sensitivity and specificity more consistently than RAG and avoids the stronger FN tendency of zero-shot. This produces high balanced accuracy in addition to high accuracy, precision, and F1 scores. Taken together, the results reinforce why accuracy alone is insufficient for evaluating LLM-generated data. While all three implementations are highly accurate, there are meaningful differences in when and how implementations get it wrong, and these different patterns of measurement error can translate into downstream estimand distortion.
	
	Panels C and D show a different error pattern for string extraction tasks. Semantic similarity is strongest when the string extraction tasks involves relatively simple metadata, like title ($\approx 0.97$) and DOI ($>0.87$). Performance declines and diverges across implementations when the task requires greater contextual interpretation to identify substantive economic concepts. Similarity scores for dependent variables, endogenous variables, instruments, and rainfall instruments are markedly lower and considerably more variable. The zero-shot and RAG implementations perform similarly on these more demanding fields, with mean semantic similarity generally between $0.50$ and $0.70$. SFT consistently produces higher similarity scores, with means roughly between $0.65$ and $0.75$. The largest differences in semantic similarity arise in dependent-variable extraction. Zero-shot and RAG scores spread broadly across the scale, with considerable density near zero, while SFT places more density on high-scoring outputs. Mean similarity is $0.73$ for SFT, compared with $0.49$ for zero-shot and $0.47$ for RAG.
	
	For the other variable names, SFT always outperforms zero-shot and RAG in terms of mean score, though sometimes these differences can be very minor. For instrument names, all three produce nearly identical means: zero-shot ($0.70$), RAG ($0.69$), and SFT ($0.70$). Nevertheless, the similarity in means masks considerable variation in the quality of individual extractions. SFT concentrates more density at high scores but its advantage is not uniform. RAG has higher means than zero-shot, yet typically has larger mass near zero, meaning RAG produces some very poorly matched strings. These comparisons cover only evaluated strings and do not necessarily include the same papers across implementations. They therefore describe the quality of recovered strings rather than whether every required string was recovered. Higher average similarity does not establish consistently accurate or complete extraction.
	
	Taken together, the panels distinguish successful classification from accurate extraction. An implementation can correctly detect that a concept is present while producing a poorly matched description of that concept. Neither high classification accuracy nor a high mean similarity score fully characterizes measurement error in the generated data.
	
	
	\section{Results} \label{sec:results}
	
	
	We next examine whether differences in task-level measurement performance translate into substantive differences in the description of the rainfall-IV literature.\footnote{Prior to conducting this analysis, we remove 19 duplicate records corresponding to papers stored under filenames that differ only in DOI punctuation, reducing the common model-labeled corpus from 3,840 to 3,821 unique papers.} As we do not have target labels for the candidate corpus, we cannot directly calculate estimand distortion. Instead, we compare the labels generated by the three implementations paper-by-paper to characterize how strongly measured facts depend on implementation. Then we use each dataset separately to construct Sankey diagrams that map the relationships between rainfall IVs and the endogenous variables that are being instrumented. By visualizing these nodes and their connecting edges, we can see how economists use rainfall IVs and determine the extent to which our conclusions about the literature are a function of the measurement system used to summarize the literature.
	
	Differences in the strings generated by the LLMs do not necessarily imply differences in what the variables measure. Similarly, two distinct variables may measure the same underlying economic concept. To examine the extent to which measurement-system dependence is a function of the resolution of the estimand, we develop two taxonomies. Drawing on the JEL classification system, we define 23 economic concepts into which we can group the endogenous variables identified by the LLMs. For the rainfall instruments, we define 21 categories that capture differences in how economists quantify rainfall. Both taxonomies include two residual categories. For classification, we use Qwen3-32B, an open-weight LLM developed by Alibaba and selected following an evaluation of alternative classification procedures. We use the same taxonomies and classification procedure across all three implementations. Appendix~\ref{app:class} details the construction of the taxonomies, the process by which we classify variables, and the performance of the classification model.
	
	\subsection{Paper-Level Comparison} \label{sec:paper}
	
	Each implementation outputs a dataset in which the unit record is a single paper and the generated labels are the responses to the queries in the prompt. We focus on rainfall identification because it determines which papers enter the substantive analysis. As can be seen in Table~\ref{tab:gpt_agree}, of the $3,821$ papers in the common model-labeled corpus, the zero-shot implementation identifies $151$ papers as using a rainfall IV while RAG identifies $160$ papers. SFT identifies $180$ papers, or about $13\%$ more than RAG and $19\%$ more than zero-shot. The union of these sets contains $209$ papers that are identified as using a rainfall IV by at least one implementation. Of these, all three implementations agree on $130$ papers ($62\%$), exactly two agree on $22$ papers ($11\%$), and $57$ papers are identified by exactly one implementation ($27\%$). Thus, rainfall-IV status differs across measurement systems for $38\%$ of papers identified by at least one system. For a more precise metric of measurement-system dependence, we calculate the pairwise Jaccard index, which is the intersection of two sets divided by their union. For zero-shot and RAG, the Jaccard overlap is $0.79$, for zero-shot and SFT the overlap is $0.70$, and for RAG and SFT the overlap is $0.69$. Thus, there is meaningful measurement-system dependence, at least on the extensive margin. Anywhere from one fifth to nearly one third of papers identified by either of two implementations are not identified by both.
	
	Conditional on the $130$ papers that all three implementations agree on as having a rainfall IV, there is still substantial disagreement on what the actual rainfall IV is and what it instruments. After a basic normalization, all three implementations give exactly the same endogenous variable strings for $35$ papers ($27\%$), the same rainfall-IV strings for $23$ papers ($18\%$), and agreement on both sets of strings for just $8$ papers ($6\%$). The corresponding disagreement rates represent an upper bound on substantive disagreement because we are enforcing exact string matches. When variables are assigned to broader categories, all three implementations agree on the endogenous variable categories for $84$ papers ($65\%$), the rainfall-IV categories for $52$ papers ($40\%$), and both sets of categories for $40$ papers ($31\%$). Thus, while some disagreements reflect differences in terminology, there are still real and meaningful disagreements between implementations. For example, the same paper is characterized as instrumenting opium production by zero-shot and RAG but as instrumenting conflict by SFT. In another paper, all three implementations identify agricultural production as the endogenous variable, but zero-shot and RAG identify two measures of mean monsoon rainfall as instruments while SFT identifies total annual rainfall. These types of disagreements change which nodes are connected in the network rather than merely change the label of a node.
	
	\subsection{Relationship-Level Comparison} \label{sec:rel}
	
	For each implementation, we construct a Sankey diagram in which the nodes represent categories of rainfall IVs and endogenous variables, the edges represent instrumenting relationships, and the width of the edges indicates the frequency with which we observe those relationships. The Sankeys therefore allow us to move beyond the paper-level comparison of measurement-system dependence and ask whether those differences change the substantive inference we draw from the literature. Following \citet{Mellon25}, the relevant feature is not simply how often particular rainfall measures or endogenous variables appear, but the relationships between them. Evidence that the same or closely related rainfall measures are used to instrument multiple endogenous variables identifies potential alternative causal pathways through which rainfall may affect socioeconomic outcomes and therefore raises concerns about the exclusion restriction in rainfall-IV applications.
	
	The zero-shot implementation identifies $188$ rainfall-IV/endogenous-variable relationships grouped into $17$ rainfall-IV categories and $23$ endogenous variable categories. Total Rainfall ($68$) and Two-Sided Rainfall Extremes ($18$) both connect to Agricultural Production, Inputs, \& Land ($23$), Income \& Income Shocks ($32$), Conflict, Crime, \& Social Unrest ($7$), and Macroeconomic Growth, Output, \& Volatility ($19$). Rainfall Deviation from Normal ($19$) connects to three of these four categories. Thus, distinct measures of rainfall (levels, deviations, extremes) are used to instrument the same types of endogenous variables. Looking at the graph from the other direction, Agricultural Production, Inputs, \& Land receives links from six different measures of rainfall, Conflict, Crime, \& Social Unrest from three, and Macroeconomic Growth, Output, \& Volatility from seven. Here, a single socioeconomic concept is instrumented using numerous different ways to quantify rainfall.
	
	RAG identifies $204$ rainfall-IV/endogenous-variable relationships, grouped into $15$ rainfall categories and $23$ endogenous categories. Total Rainfall ($77$) is somewhat more common than its zero-shot counterpart, while Rainfall Deviation from Normal ($18$) and Two-Sided Rainfall Extremes ($18$) change little. Total Rainfall and Two-Sided Rainfall Extremes again connect to Agricultural Production, Inputs, \& Land ($27$), Income \& Income Shocks ($25$), Conflict, Crime, \& Social Unrest ($9$), and Macroeconomic Growth, Output, \& Volatility ($32$). Similar to zero-shot, many different rainfall metrics feed into a single endogenous variable category. Agricultural Production, Inputs, \& Land receives links from seven different measures of rainfall, Conflict, Crime, \& Social Unrest from five, and Macroeconomic Growth, Output, \& Volatility from eight. Despite these compositional differences in the variables and their relationships to each other, the broad network structure remains similar to zero-shot.
	
	SFT identifies $201$ rainfall-IV/endogenous-variable relationships, grouped into $19$ rainfall-IV categories and $24$ endogenous variable categories. Total Rainfall ($75$) remains prominent, while Rainfall Deviation from Normal ($16$) remains similar across implementations and Two-Sided Rainfall Extremes ($12$) declines relative to the other two implementations. Total Rainfall and Rainfall Deviation from Normal both connect to Agricultural Production, Inputs, \& Land ($15$), Income \& Income Shocks ($21$), Conflict, Crime, \& Social Unrest ($8$), and Macroeconomic Growth, Output, \& Volatility ($19$). These categories receive links from four, eight, five, and seven different measures of rainfall, respectively. Relative to RAG, SFT identifies fewer relationships involving agricultural and macroeconomic variables but more involving Labor Markets \& Employment ($16$ versus $3$) and Poverty, Inequality, \& Welfare ($11$ versus $5$). SFT therefore alters the relative importance of particular nodes and relationships while preserving the broad network structure produced by zero-shot and RAG.
	
	Read from left to right, the diagrams identify potential alternative pathways that rainfall-IV applications need to address. For example, a plausible causal pathway may connect agricultural production to income to conflict to macroeconomic performance. Yet total rainfall is used to instrument all four across implementations. A study that uses a rainfall shock to instrument agricultural production in a regression of conflict must justify why rainfall does not affect conflict through income independently of agricultural production. This concern is particularly salient if a different study uses a rainfall shock to instrument income in a regression of conflict. The more variables that rainfall instruments for that could affect the outcome independently of the endogenous variable of interest, the harder it is to defend the exclusion restriction in any particular application. The Sankeys make this concern visible. The relevance argument for one endogenous variable identifies a potential competing pathway for another application.
	
	Read from right to left, the same diagrams raise a different concern. Agricultural variables receive links from total rainfall, mean rainfall, deviations from normal, and rainfall variability across all three implementations. Macroeconomic and household-welfare variables likewise receive numerous different types of rainfall instruments. Differences in setting and theory may justify using one or more measures of rainfall to instrument for a specific endogenous variable. But these measures cannot simply be assumed to be interchangeable. The range of alternatives used in the literature creates scope for choosing from the many plausible instruments the rainfall metric that produces the desired first-stage results. The rainfall IV we observe may be the one that was selected not because of its theoretical fit or economic rationale but for its significance.
	
	The Sankey diagrams reveal measurement-system dispersion in the composition and relative importance of individual rainfall-instrument relationships, but considerably less dispersion in the broader conclusions we draw from the literature. Across implementations, the network retains the same basic structure. Many measures of rainfall are used to instrument a wide range of distinct economic and social processes. Thus, while measurement-system choice affects which relationships appear and how prominent they are, it does not materially change the substantive concerns raised by the literature. The overall variety of relationships, rather than implementation differences in any one relationship, raises concerns about exclusion restriction violations and selection on significance.
	
	
	\subsection{Sensitivity to Model Choice} \label{sec:frontier}
	
	Given the rapid pace at which LLMs are improving, a natural question is whether the degree and types of measurement error we document are a function of our choice to use GPT-4.1. In Appendix~\ref{app:alt-mod}, we replicate our performance validation and substantive analysis using zero-shot implementations of four frontier LLMs and compare the results to those of our SFT implementation. Despite differences in model capabilities, we observe much the same pattern of measurement error. All four models correctly classify between 84 and 86 of the 88 papers on the rainfall-IV task, though they differ in the types of errors they make. And, as with the three implementations of GPT-4.1, performance deteriorates when we ask the models to identify and name the economic concepts underlying those classifications.
	
	Interestingly, SFT outperforms all four frontier models on the more demanding string extraction tasks. For example, mean semantic similarity for dependent variables is $0.73$ for SFT compared with values no greater than $0.52$ for the newer models (Figure~\ref{fig:frnt_eval}). We observe similar differences for endogenous variables and rainfall IVs. Thus, adapting an older model to a specific research task can yield greater semantic similarity than using a more recent general-purpose model without fine-tuning.
	
	When we use the data generated by the frontier models to characterize the rainfall-IV literature, we observe considerable variation in both the number of papers and the relationships identified. For example, Gemma identifies fewer papers as using a rainfall IV than SFT (137 versus 180) but identifies $66\%$ more rainfall-IV/endogenous-variable relationships. There are similarly large differences in the relative importance of particular relationships. Yet we recover the same broad characterization of the literature and, consequently, the same concerns about the exclusion restriction and selection on significance. Thus, the estimand dependence of measurement-system dispersion that we document when varying how we implement an LLM persists when we vary the LLM itself.
	
	
	\section{Discussion} \label{sec:discuss}
	The empirical results illustrate an implication of our theoretical framework that measurement-system dispersion, $D_{i} \left( \theta \right)$, is a function of the estimand rather than being an intrinsic property of the generated data. That we recover much the same pattern of results using frontier models suggests that the measurement problems we document are not simply a function of the current state of the technology. As new generations of LLMs are developed, the amount of measurement error in the data they generate may decline. But absent a proof of correctness, researchers will continue to have to determine when and under what conditions that measurement error matters for substantive inference.
	
	The decline in measurement-system dependence as the estimand coarsens highlights an important tension in the use of LLMs for literature reviews, and in the research process more broadly. Measurement-system choice matters little when one wants to establish broad descriptions of a literature. Yet this is where LLMs add relatively little. A competent traditional literature review can already establish conclusions like ``the rainfall-IV literature has an exclusion restriction problem,'' ``trade liberalization tends to raise income,'' or ``climate shocks affect agricultural production.''  The comparative advantage of LLMs lies instead in the potential to produce fine-grained characterization at the scale of an entire literature. But this is precisely the setting where our results show measurement-system dependence to be most pronounced. 
	
	\cite{Mellon25} provides a useful benchmark for this tradeoff between scale and measurement-system dependence. He reads 289 papers and manually documents relationships between weather and socioeconomic variables, using those paper-level relationships to argue that weather instruments face widespread potential exclusion restriction violations. We recover the same broad conclusion regardless of which measurement system generates our data. But the fine-grained evidence differs. Starting from a candidate corpus of $4,191$ papers, we use LLMs to amplify this type of expert human review, documenting not only what rainfall is used to instrument but also how economists quantify rainfall. But the papers included in the review, the rainfall IVs, the endogenous variables, and the relationships between them all vary across measurement systems. We recover the broad conclusion that a human review has already established, but we are considerably less certain about the finer details of the literature that our machine-assisted review adds.
	
	The second concern we identify in the rainfall-IV literature, selection on significance, applies equally to the use of LLMs in conducting our review. Just as an economist can search across alternative measures of rainfall until they obtain the desired first-stage results, a researcher using an LLM can search across alternative measurement systems until they obtain the desired substantive results. Even when the economic concept to be measured is well defined, the choices of model, prompt, and implementation are free parameters of the measurement process. The low marginal cost of generating alternative measures shifts opportunities for specification search upstream, from how researchers model the data to how they generate the data \citep{DellandRambachan26, LudwigEtAl26}. The choice of measurement system should therefore be informed by what researchers seek to measure rather than by what they hope to find.
	
	The importance of estimand resolution also changes what performance validation metrics tell us in applications that use LLMs to generate data from the scientific literature. \cite{BriggsEtAl26} provide a useful example. They show that carefully validated LLMs can recover structured information from nearly 100,000 academic papers at a fraction of the cost of human coders. They report accuracy, precision, sensitivity, and specificity typically ranging from $0.75$ to $1.00$. These metrics establish that the generated data closely reproduce their validation labels, but they do not establish how residual measurement error affects a particular estimand constructed from those data. The principal results in \cite{BriggsEtAl26} are highly aggregated across a very large corpus, a setting in which residual paper-level disagreement may have relatively little effect on the resulting conclusions. The same degree of residual error may be much more consequential when the estimand concerns an individual paper, concept, or relationship. Because \cite{BriggsEtAl26} select a preferred model and do not carry alternative measurement systems through the analysis, their validation exercise cannot reveal how sensitive their substantive results are to model choice. High task-level performance is not, by itself, evidence that a substantive estimand is robust to the choice of measurement system.
	
	\cite{GargandFetzer25} provide a complementary example. They use LLM-generated descriptions of causes, effects, and exogenous variation to construct fine-grained claim graphs for 44,852 academic papers, which they then use to characterize causal structure and novelty in the economics literature. Their validation results are substantially stronger for coarse method classification than for the semantic content from which claim-graph edges are constructed. They report accuracy, precision, sensitivity, and specificity with values typically ranging from $0.51$ to $0.94$, while mean semantic similarity with expert-labeled causes, effects, and sources of exogenous variation ranges from only $0.14$ to $0.27$. These metrics, together with repeated extraction, snippet-based checks, external benchmarking, and sensitivity to alternative aggregation thresholds, establish stability and agreement with external labels along important dimensions of their chosen extraction system, but they do not establish how residual measurement error affects a particular estimand constructed from those data. Unlike the highly aggregated setting in \cite{BriggsEtAl26}, the distinctive contribution of \cite{GargandFetzer25} depends directly on fine-grained paper-level measurement, precisely the level of estimand resolution at which both their semantic-validation results indicate considerable error and our results show measurement-system dependence to be most acute. Further, as \cite{GargandFetzer25} do not carry alternative measurement systems through the analysis, their robustness exercises cannot establish whether the paper-level claim graphs, or the conclusions drawn from those graphs, would persist under a different measurement system. Their results thus illustrate that even extensive task-level validation and robustness checks within a chosen extraction system do not, by themselves, establish the robustness of downstream substantive estimands to measurement-system choice.
	
	The implication of Equation~\eqref{eq:distortion} is that comparing estimates across measurement systems and validating those estimates are complementary exercises. While squared dispersion provides a lower bound on mean squared estimand distortion, agreement across measurement systems does not establish observational validity. To establish validity, we need to anchor the measurements to an explicit definition of the concept rather than to the outputs of other LLMs \citep{DellandRambachan26}. Further, the design of the validation exercise should be informed by the substantive estimand of interest. If, for example, we are interested in how frequently a particular measure of rainfall is used to instrument a particular economic concept, it is the distortion in this estimated frequency that we need to assess rather than simply the accuracy with which the LLM identifies papers that use a rainfall IV. Validation should match the resolution of the claim.
	
	
	\section{Conclusions} \label{sec:conc}
	
	Literature reviews are exercises in measurement. Researchers define a population of papers, read them, code information, and use those data to characterize what a literature says. LLMs dramatically increase the number of papers and the level of detail that can be coded. But the data they produce are still measurements. When those measurements are imperfect, the relevant question is not simply how often the model succeeds or fails but whether the errors change what a researcher concludes.
	
	In this paper, we develop a measurement error framework for LLM-generated data and apply it to a review of how economists use rainfall as an instrumental variable. We start with a model of measurement error that distinguishes between types of tasks as well as types of errors. The model highlights that standard validation metrics reveal little about how task-level measurement errors distort inference on substantive estimands further downstream. We then compare zero-shot, RAG, and SFT implementations of GPT-4.1. We validate each against a common held-out evaluation set and use each measurement system to independently construct the data used in the review. This allows us to compare task-level performance with the stability of the resulting substantive conclusions.
	
	Our results are encouraging for simple tasks but less so as we demand more from the model in terms of contextualization and interpretation. All three implementations perform well when asked to classify whether a paper is empirical, has an endogeneity problem, uses an IV, or uses rainfall as an IV. Performance is more heterogeneous when we ask the implementations to identify and name the economic concepts underlying those classifications. SFT generally improves on these more difficult extraction tasks relative to zero-shot and RAG, but substantial semantic errors remain across all three measurement systems. The same pattern of measurement error persists among the more recent LLMs, with zero-shot implementations of all four frontier models performing worse than SFT on the more demanding string extraction tasks. Applied across the common model-labeled corpus, the implementations frequently disagree on actual variable names, what the relationship is between IV and endogenous variables, and even which papers are part of the rainfall-IV literature. High task-level performance therefore does not imply that the datasets produced by different measurement systems are interchangeable.
	
	Yet these differences in the generated data do not always change what we infer about the literature. Across implementations and model families, the same broad picture of the rainfall-IV literature emerges. Rainfall instruments a wide range of socioeconomic processes, while the same types of endogenous variables are instrumented using many different measures of rainfall. Read in one direction, these relationships identify potential alternative causal pathways that threaten the exclusion restriction. Read in the other direction, they reveal substantial researcher degrees of freedom over how to quantify rainfall, creating scope for selection on significance. These broad conclusions survive the choice of measurement system but could also be produced by a human review of a much smaller number of papers. The finer details of the literature, which is where LLMs appear to have a comparative advantage, do not.
	
	Selecting a preferred measurement system based on accuracy, precision, sensitivity, specificity, or semantic similarity is not sufficient to ensure credible inference. These metrics tell us how well a measurement system reproduces human labels on an evaluation sample. They do not tell us whether the remaining measurement error matters for the inference those data are meant to support. Nor does agreement across measurement systems establish that estimates are close to the substantive estimand, as the systems may make similar errors.
	
	Taken together, our results suggest a simple rule: validation should match the resolution of the claim. Absent that, researchers should demonstrate that inference on substantive estimands is robust to the choice of measurement system. When LLM-generated data are used only to characterize a literature in broad terms, measurement-system choice may have little effect on the conclusion. But broad conclusions are where LLMs offer the least advantage over traditional literature reviews. The comparative advantage of LLMs lies in producing fine-grained information across a literature at a scale human readers cannot feasibly achieve. But that advantage comes with a risk. The finer the estimand, the more inference can depend on the measurement system that generated the data. The promise of LLMs is therefore not simply that they allow researchers to measure more. It is that they allow researchers to make more detailed claims from those measurements. Realizing this promise requires establishing not just that LLMs measure well, but that they measure well enough to support those claims.
	
	\newpage
	\singlespacing
	\bibliographystyle{chicago}
	\bibliography{bibl}

	
	\newpage 
	\FloatBarrier
	
	\begin{figure}[!htbp] 
		\begin{minipage}{\linewidth}		
			\begin{center}
				\caption{Zero-Shot: Rainfall Instruments and Endogenous Variables \label{fig:gpt_0st}}
				\includegraphics[width=\linewidth,keepaspectratio]{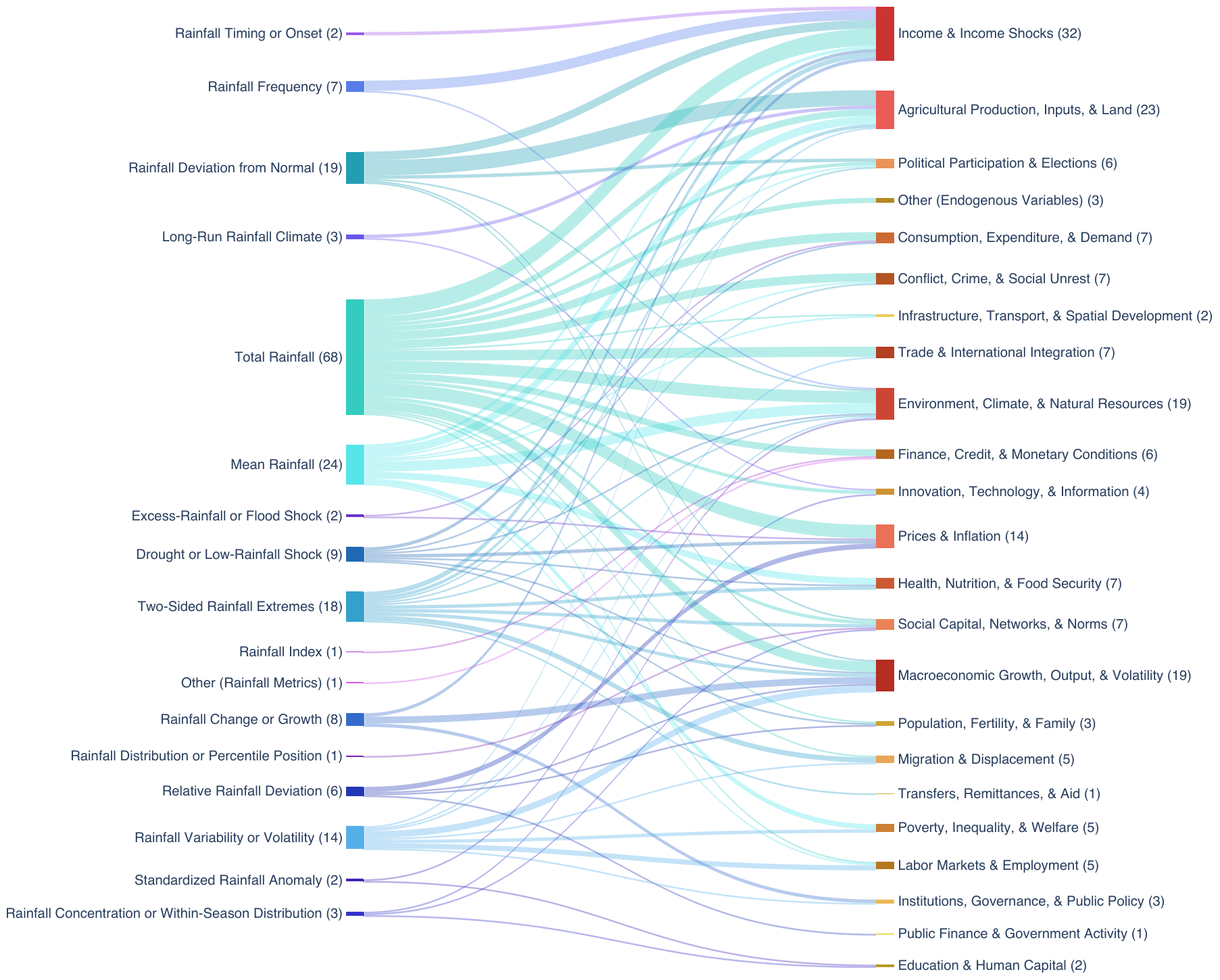}
			\end{center}
			\footnotesize  \textit{Note}: The Sankey diagram maps harmonized categories of rainfall instruments identified by the zero-shot implementation (left) to the corresponding harmonized categories of endogenous variables (right). Flow width is proportional to the number of extracted rainfall instrument–endogenous variable linkages, with flows colored according to the rainfall-instrument category. Values in parentheses report the total number of extracted linkages associated with each category. Counts refer to model-generated linkages and need not correspond to unique papers.
		\end{minipage}
	\end{figure}
	
	\begin{figure}[!htbp] 
		\begin{minipage}{\linewidth}		
			\begin{center}
				\caption{RAG: Rainfall Instruments and Endogenous Variables \label{fig:gpt_rag}}
				\includegraphics[width=\linewidth,keepaspectratio]{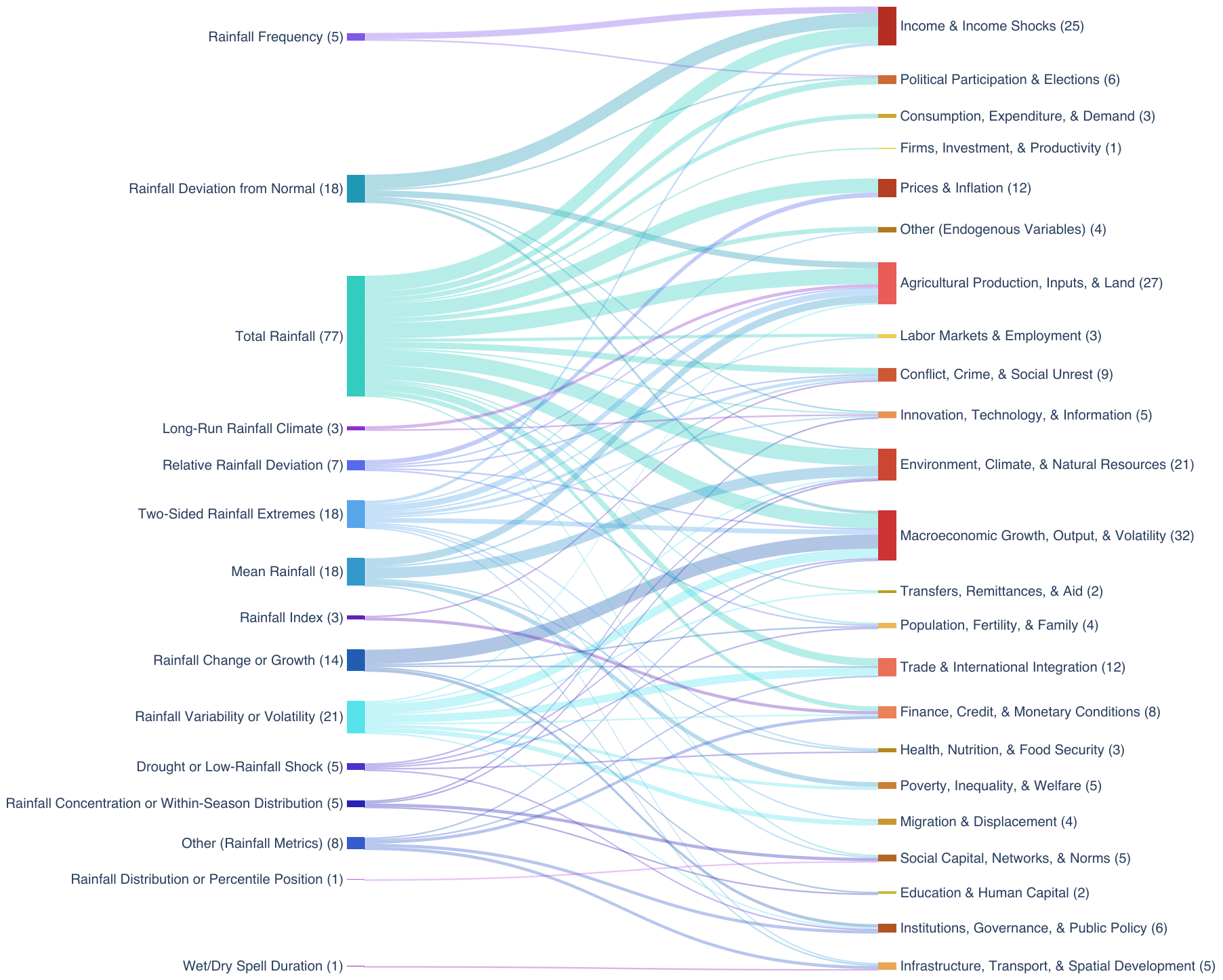}
			\end{center}
			\footnotesize  \textit{Note}: The Sankey diagram maps harmonized categories of rainfall instruments identified by the retrieval-augmented generation (RAG) implementation (left) to the corresponding harmonized categories of endogenous variables (right). Flow width is proportional to the number of extracted rainfall instrument–endogenous variable linkages, with flows colored according to the rainfall-instrument category. Values in parentheses report the total number of extracted linkages associated with each category. Counts refer to model-generated linkages and need not correspond to unique papers.
		\end{minipage}
	\end{figure}

	\begin{figure}[!htbp] 
		\begin{minipage}{\linewidth}		
			\begin{center}
				\caption{SFT: Rainfall Instruments and Endogenous Variables \label{fig:gpt_sft}}
				\includegraphics[width=\linewidth,keepaspectratio]{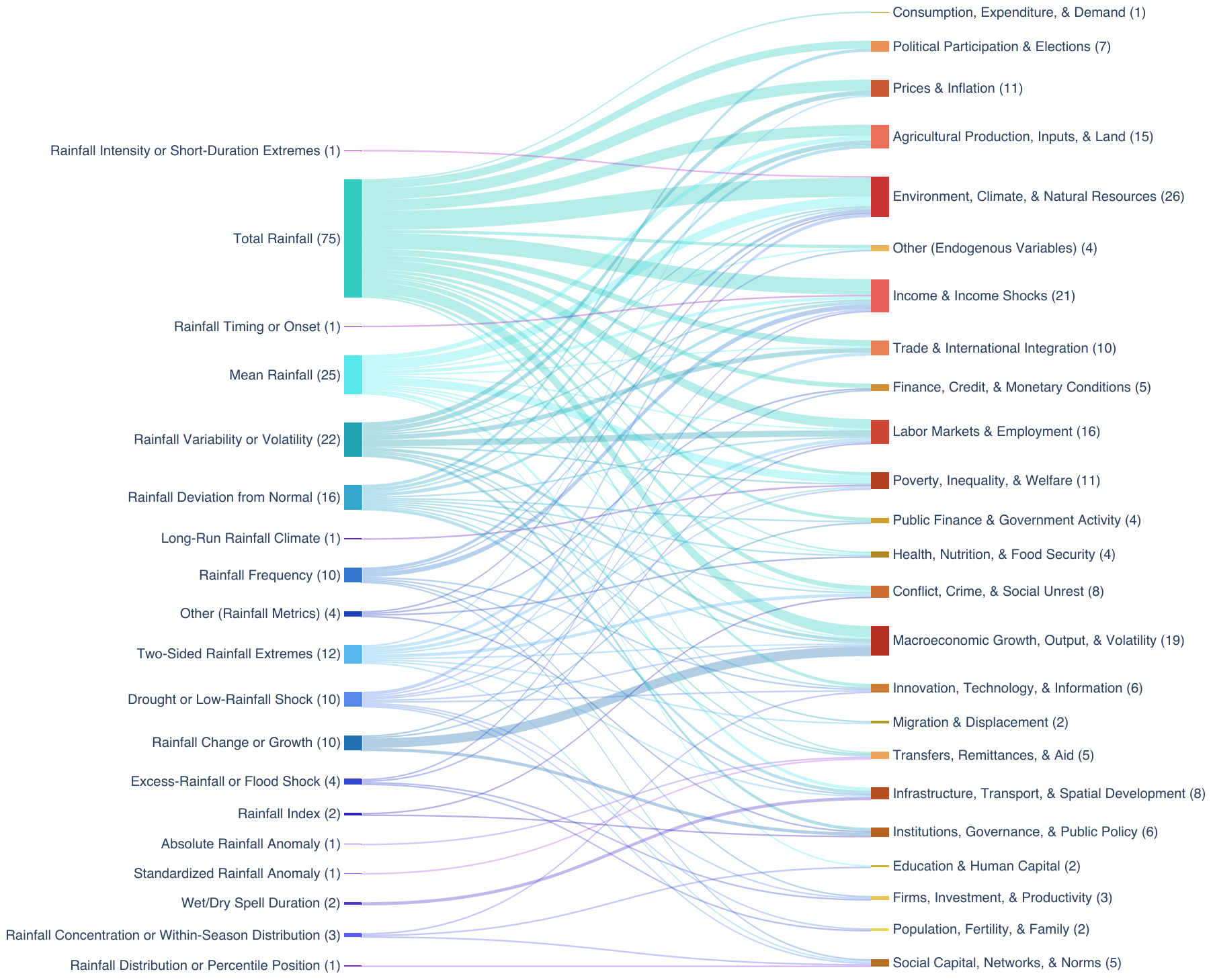}
			\end{center}
			\footnotesize  \textit{Note}: The Sankey diagram maps harmonized categories of rainfall instruments identified by the supervised fine-tuning (SFT) implementation (left) to the corresponding harmonized categories of endogenous variables (right). Flow width is proportional to the number of extracted rainfall instrument–endogenous variable linkages, with flows colored according to the rainfall-instrument category. Values in parentheses report the total number of extracted linkages associated with each category. Counts refer to model-generated linkages and need not correspond to unique papers.
		\end{minipage}
	\end{figure}
	
	
	
	\FloatBarrier
	
	
	\begin{table}[!htbp]
		\centering
		\caption{Information Extracted from Each Paper and Prompt Dependencies}
		\label{tab:info_measured}
		\begin{threeparttable}
			\footnotesize \setlength{\tabcolsep}{4pt} \renewcommand{\arraystretch}{1.18}
			\begin{tabularx}{\textwidth}{
					@{} > {\raggedright\arraybackslash}p{0.23\textwidth} >{\raggedright\arraybackslash}X >{\centering\arraybackslash}p{0.11\textwidth} >{\raggedright\arraybackslash}p{0.25\textwidth} @{} }
				\\[-1.8ex]\hline 
				\hline \\[-1.8ex]
				Measure & Information captured & Output & Dependency \\
				\midrule
				Article Title & Article title, including subtitle & Exact string &       None \\
				DOI & Version-of-record DOI & DOI or \texttt{n/a} & None \\
				Empirical Analysis & Presence of empirical statistical estimation &      1/0 & None \\
				Dependent Variable(s) & Name(s) of main regression outcome(s) & String   list & \texttt{Empirical Analysis} \(=1\) \\
				Endogeneity Problem  & Main analysis has an endogenous regressor & 1/0   & \texttt{Empirical Analysis} \(=1\) \\
				Endogenous Variable(s) & Name(s) of endogenous regressor(s)
				& String list & \texttt{Endogeneity Problem} \(=1\) \\
				Instrumental Variable Regression & Use of an IV-type estimator with      excluded instrument & 1/0 & \texttt{Endogeneity Problem} \(=1\) \\
				Instrumental Variable(s) & Name(s) of excluded instrument(s) & String    list & \texttt{Instrumental Variable Regression}      \(=1\) \\
				Rainfall Instrument & An excluded instrument is precipitation-based &    1/0 & \texttt{Instrumental Variable Regression} \(=1\) \\
				Rainfall Variable(s) & Name of specific precipitation-based              instrument(s) & String list & \texttt{Rainfall Instrument} \(=1\) \\
				\\[-1.8ex]\hline 
				\hline \\[-1.8ex]
			\end{tabularx}	
			\begin{tablenotes}[flushleft]
				\footnotesize
				\item \textit{Note:} Article title, DOI, and empirical analysis are evaluated independently. Dependent variables and the presence of an endogeneity problem are evaluated only for papers containing empirical analysis. Subsequent fields follow the conditional structure shown in the dependency column. A field is not evaluated when its dependency condition is not satisfied and it and all subsequent fields are coded \texttt{n/a}. String lists preserve the variable names used in the paper and separate multiple values with semicolons. Complete prompt instructions and output rules appear in Appendix~\ref{app:prompt}.
			\end{tablenotes}
		\end{threeparttable}
	\end{table}

	
	\begin{table}[htbp] \centering
		\caption{Binary Classification and Validation Metrics} \label{tab:confusion_matrix}
		\renewcommand{\arraystretch}{1.6}
		\setlength{\tabcolsep}{10pt}
		\begin{tabular}{cc|cc|l}
			&  & \multicolumn{2}{c|}{\textbf{LLM-Generated Value}} & \\
			&  & $Z_{ij} = 1$ & $Z_{ij} = 0$ & \\
			\cline{2-5}
			\multirow{-2}{*}{\rotatebox[origin=c]{90}{\textbf{Target Value}}}
			& $T_{ij} = 0$ & $\text{FP}: \alpha_i$ & $\text{TN}: 1 - \alpha_i$ & $\displaystyle \text{Specificity} = \Pr( Z_{ij} = 0\mid T_{ij} = 0 ) = 1-\alpha_i$ \\
			& $T_{ij} = 1$ & $\text{TP}: 1-\beta_i$ & $\text{FN}: \beta_i$ & $\displaystyle \text{Sensitivity} = \Pr( Z_{ij} = 1\mid T_{ij} = 1 ) = 1-\beta_i$ \\
			\cline{2-5}
			\multicolumn{5}{c}{$\displaystyle \text{Accuracy} = \Pr( Z_{ij} = T_{ij} ) = ( 1 - \beta_i ) p_i + ( 1 - \alpha_i )( 1 - p_i ), \qquad p_i = \Pr( T_{ij} = 1 )$}
		\end{tabular}
		\begin{tablenotes}[flushleft]
			\footnotesize
			\item \textit{Note:} $i$ indexes the extraction task and $j$ indexes papers. $\alpha_i$ is the false-positive rate, $\beta_i$ is the false-negative rate, and $p_i = \Pr( T_{ij} = 1 )$ is the prevalence of the target value. Cell entries report the corresponding classification type and conditional probability.
		\end{tablenotes}
	\end{table}

	
	\input{tables/gpt_agree}

	
	\FloatBarrier
	\clearpage
	\newpage
	\appendix
	\onehalfspacing
	
	\begin{center}
		\section*{Online-Only Appendix to ``Mining Meaning: Measurement Error in AI-Assisted Literature Reviews''} \label{sec:app}
	\end{center}
	
	
	\setcounter{table}{0}
	\renewcommand{\thetable}{A.\arabic{table}}
	\setcounter{figure}{0}
	\renewcommand{\thefigure}{A.\arabic{figure}}
	\setcounter{equation}{0}
	\renewcommand{\theequation}{A.\arabic{equation}}
	
	\section{Operationalizing the Measurement Error Model} \label{app:measures}
	This appendix describes how to move from task-level validation of an LLM measurement system to an assessment of how measurement-system choice affects the substantive results of a literature review or any research-related activity tasked to an LLM. The key distinction is that validation metrics characterize the quality of the model-generated data, whereas $\widehat{\theta}_{i}^{(m)}$ is the result of applying a substantive analysis, $g_i$, to the data generated by model $m$. The measurement error metrics are the primitives while the substantive estimands capture the effects of measurement error on inference. There is thus no unique value of $\theta$ until the researcher specifies the descriptive quantity or statistical relationship of interest.
	
	
	\subsection{Task-Level Validation Metrics}	\label{app:task-val}
	Let $\mathcal{E}$ denote the evaluation sample, with $j \in \mathcal{E}$ a paper (or other object) in the evaluation sample. Additionally, let $A_{i}^{(m)}$, $R_{i}^{(m)}$, $C_{i}^{(m)}$, $\Pi_{i}^{(m)}$, $\operatorname{F1}_{i}^{(m)}$, and $\operatorname{BA}_{i}^{(m)}$ denote the population accuracy, sensitivity, specificity, precision, F1 score, and balanced accuracy and  denote their evaluation-sample analogs as $\widehat{A}_{i}^{(m)}$, $\widehat{R}_{i}^{(m)}$, $\widehat{C}_{i}^{(m)}$, $\widehat{\Pi}_{i}^{(m)}$, $\widehat{\operatorname{F1}}_{i}^{(m)}$, and $\widehat{\operatorname{BA}}_{i}^{(m)}$. Then, for concept $i$ and measurement system $m$, define
	
	\begin{align}
		\mathrm{TP}_{i}^{(m)} &= \sum_{j \in \mathcal{E}} \mathbf{1} \left\{         T_{ij} = 1,\, Z_{ij}^{(m)} = 1 \right\}, \\[3pt]
		\mathrm{FN}_{i}^{(m)} &= \sum_{j \in \mathcal{E}} \mathbf{1} \left\{        T_{ij} = 1,\, Z_{ij}^{(m)} = 0 \right\}, \\[3pt]
		\mathrm{FP}_{i}^{(m)} &= \sum_{j \in \mathcal{E}} \mathbf{1} \left\{         T_{ij} = 0,\, Z_{ij}^{(m)} = 1 \right\}, \\[3pt]
		\mathrm{TN}_{i}^{(m)} &= \sum_{j \in \mathcal{E}} \mathbf{1} \left\{        T_{ij} = 0,\, Z_{ij}^{(m)} = 0 \right\}.
	\end{align}
	
	\noindent The corresponding binary validation measures are as follows.
	
	\paragraph{Accuracy.} Accuracy is the proportion of observations classified correctly:
	
	\begin{equation}
		\widehat{A}_{i}^{(m)} = \frac{\mathrm{TP}_{i}^{(m)} + \mathrm{TN}_{i}^{(m)}}{\mathrm{TP}_{i}^{(m)} + \mathrm{TN}_{i}^{(m)} + \mathrm{FP}_{i}^{(m)} + \mathrm{FN}_{i}^{(m)}}.
	\end{equation}
	
	\noindent Accuracy summarizes overall classification performance, but it is prevalence dependent and does not distinguish FPs from FNs.
	
	\paragraph{Sensitivity.} Sensitivity, also called the TP rate or recall, is
	
	\begin{equation}
		\widehat{R}_{i}^{(m)} = \frac{\mathrm{TP}_{i}^{(m)}}{\mathrm{TP}_{i}^{(m)} + \mathrm{FN}_{i}^{(m)}}.
	\end{equation}
	
	\noindent Sensitivity measures the proportion of target-positive observations correctly identified by the measurement system.
	
	\paragraph{Specificity.} Specificity, or the TN rate, is
	
	\begin{equation}
		\widehat{C}_{i}^{(m)} = \frac{\mathrm{TN}_{i}^{(m)}}{\mathrm{TN}_{i}^{(m)} + \mathrm{FP}_{i}^{(m)}}.
	\end{equation}
	
	\noindent Specificity measures the proportion of target-negative observations correctly identified.
	
	\paragraph{Precision.} Precision, or the positive predictive value, is
	
	\begin{equation}
		\widehat{\Pi}_{i}^{(m)} = \frac{\mathrm{TP}_{i}^{(m)}}{\mathrm{TP}_{i}^{(m)} + \mathrm{FP}_{i}^{(m)}}.
	\end{equation}
	
	\noindent Precision measures the proportion of model-generated positive classifications that are target positive.
	
	\paragraph{F1 score.} The F1 score is the harmonic mean of precision and sensitivity:
	
	\begin{equation}
		\widehat{\operatorname{F1}}_{i}^{(m)} = \frac{2 \widehat{\Pi}_{i}^{(m)} \widehat{R}_{i}^{(m)}}{\widehat{\Pi}_{i}^{(m)} + \widehat{R}_{i}^{(m)}}.
	\end{equation}
	
	\noindent The F1 score balances positive predictive performance and positive case recovery but does not incorporate true negatives.
	
	\paragraph{Balanced accuracy.} Balanced accuracy gives equal weight to positive and negative case performance:
	
	\begin{equation}
		\widehat{\operatorname{BA}}_{i}^{(m)} = \frac{1}{2} \left( \widehat{R}_{i}^{(m)} + \widehat{C}_{i}^{(m)} \right).
	\end{equation}
	
	\noindent Balanced accuracy is useful when the positive and negative classes are highly unequal in size.
	
	\paragraph{Semantic similarity.} For a binary TP, let $N_{ij}$ denote the target string and $X_{ij}^{(m)}$ the string generated by measurement system $m$. Semantic similarity is
	
	\begin{equation}
		S_{ij}^{(m)} = s \! \left( X_{ij}^{(m)}, N_{ij} \right) \in[0,1]     \qquad \text{if } T_{ij} = Z_{ij}^{(m)} = 1.
	\end{equation}
	
	\noindent The validation-sample mean semantic similarity among true positives is
	
	\begin{equation}
		\widehat{\mu}_{i,\mathcal{E}}^{(m)} = \frac{1}{\mathrm{TP}_{i}^{(m)}} \sum_{j \in \mathcal{E}} \mathbf{1} \left\{ T_{ij} = Z_{ij}^{(m)} = 1 \right\} S_{ij}^{(m)}. \label{eq:msemantic}
	\end{equation}
	
	\noindent Semantic similarity evaluates the quality of the generated string conditional on a correct positive classification. It does not penalize FNs, because no generated string exists in those cases, and it does not evaluate FPs, because no target string exists.
	
	\paragraph{Omnibus Loss Function.} The sample analog of the task-level omnibus loss defined in the main text is
	
	\begin{equation}
		\begin{aligned}
			\widehat{\mathcal{L}}_{i,\mathcal{E}}^{(m)} 
			&= \frac{ \mathrm{FP}_{i}^{(m)} + \mathrm{FN}_{i}^{(m)} + \displaystyle \sum_{j \in \mathcal{E}} \mathbf{1} \left\{ T_{ij} = Z_{ij}^{(m)} = 1 \right\} \left( 1 -S_{ij}^{(m)} \right)}{J_\mathcal{E}} \\[3pt]
			&= \left( 1 - \widehat{A}_{i}^{(m)} \right) + \frac{\mathrm{TP}_{i}^{(m)}}{J_\mathcal{E}} \left(1 - \widehat{\mu}_{i,\mathcal{E}}^{(m)} \right).
		\end{aligned}   
	\end{equation}
	
	\noindent These statistics characterize task-level measurement performance. They do not, by themselves, determine the consequences of measurement error for the substantive quantities to be estimated. The omnibus loss cannot be interpreted as an error rate because it combines classification errors with semantic dissimilarity scores.
	
	
	\subsection{From Task-Level Binary Validation to Substantive Estimand} 	\label{app:bi-est}
	Let $r \in \{ \mathcal{E}, P \}$ index either the evaluation sample or the full analysis sample or population, and let $J_r$ denote the number of papers in $r$. Additionally, let $\mathbf{T}_{i,r}$ collect the target values for task $i$ and let $\mathbf{Z}_{i,r}^{(m)}$ collect the corresponding values generated by measurement system $m$. Then define the target estimand and its model-generated analog as
	
	\begin{equation*}
		\theta_{i,r} = g_i \left( \mathbf{T}_{i,r} \right), \qquad
		\widehat{\theta}_{i,r}^{(m)} = g_i \! \left( \mathbf{Z}_{i,r}^{(m)} \right),
	\end{equation*}
	
	\noindent where $g_i$ is the substantive analysis applied to concept $i$. Depending on the research question, $g_i(\cdot)$ may return a prevalence, mean, difference between groups, time trend, correlation, regression coefficient or even a string value like a variable name. The system-specific measurement-induced distortion is
	
	\begin{equation}
		\Delta_{i,r}^{(m)} = \widehat{\theta}_{i,r}^{(m)} - \theta_{i,r}
	\end{equation}
	
	\noindent This quantity is defined for either $r = \mathcal{E}$ or $r = P$ but it can only be directly calculated when the target values are observed, which will ordinarily be the case for only the evaluation sample.
	
	The mean estimate across the $M$ measurement systems is
	
	\begin{equation}
		\overline{\theta}_{i,r} = \frac{1}{M} \sum_{m=1}^{M} \widehat{\theta}_{i,r}^{(m)}.
	\end{equation}
	
	\noindent And measurement-system dispersion is
	
	\begin{equation}
		D_{i,r}(\theta) = \left[ \frac{1}{M} \sum_{m=1}^{M} \left( \widehat{\theta}_{i,r}^{(m)} - \overline{\theta}_{i,r} \right)^2 \right]^{1/2}.
	\end{equation}
	
	\noindent Because the measurement systems constitute a fixed, prespecified collection of models and implementations rather than a random sample from a population of systems, the denominator is $M$. Similarly, $\overline{\theta}_{i,r}$ is the center of the estimates produced by the selected measurement systems. It is not necessarily closer to the target estimand than any particular system-specific estimate and should not be interpreted as an unbiased measurement-system average estimator without additional assumptions. Nor is the quantity $D_{i,r}(\theta)$ a standard error or confidence interval. It is a descriptive measure of the sensitivity of the substantive result to choice of measurement system. 
	
	
	\paragraph{Example 1: Prevalence.} Suppose the substantive estimand for concept $i$ is the proportion of papers in which that concept is present. In this case, the function $g_i$ is the sample mean, so that
	
	\begin{equation*}
		\theta_{r} = \frac{1}{J_r} \sum_{j \in r} T_{j}, \qquad
		\widehat{\theta}_{r}^{(m)} = \frac{1}{J_r} \sum_{j \in r} Z_{j}^{(m)}.
	\end{equation*}
	
	\noindent In the notation of the binary measurement model, $\theta_{r}$ is the target prevalence $p$, $\widehat{\theta}_{r}^{(m)}$ is the measured prevalence $q^{(m)}$, and $\Delta_{r}^{(m)} = q^{(m)} - p$.
	
	For example, if the target prevalence in the evaluation sample was $\theta_{\mathcal{E}} = 0.10$ and four measurement systems produced prevalence estimates
	
	\begin{equation*}
		0.10, \qquad 0.12, \qquad 0.11, \qquad 0.15.
	\end{equation*}
	
	\noindent Then, the measurement systems produce a mean estimate of $12\%$, with absolute bias of $2$ percentage points and relative bias of $20\%$, with cross-system dispersion of approximately $1.9$ percentage points.
	
	
	\paragraph{Example 2: Numerical Measures and Regression Relationships.} 
	The same procedure applies when binary outputs are combined into a numerical summary. Examples include the number of hypotheses tested in a paper, the number of variables assigned a particular econometric role, or the share of a prespecified set of concepts, like JEL codes, detected in each paper. Let $Y_j^{(m)}$ denote the numerical summary constructed from the data generated by measurement system $m$.
	
	Suppose, for example, that the researcher is interested in the relationship between this numerical summary and a paper characteristic, $W_j$, such as year of publication.  For each measurement system, estimate the same regression:
	
	\begin{equation}
		Y_{j}^{(m)} = a^{(m)} + \theta^{(m)} W_j + \mathbf{B}_{j}^{\prime} \boldsymbol{\gamma}^{(m)} + \varepsilon_{j}^{(m)},
	\end{equation}
	
	\noindent where $a^{(m)}$ is the intercept and $\mathbf{B}_j$ contains any additional paper characteristics included in the analysis. The estimated coefficient on $W_j$ is the system-specific substantive estimate $\widehat{\theta}^{(m)}$. The researcher then can calculate the cross-system mean and measurement-system dispersion.
	
	The same logic applies if the LLM-generated variable enters on the right-hand side of a regression, or if $g_i(\cdot)$ is a nonlinear estimator. The relevant requirement is that the same specification, sample restrictions, covariates, transformations, and estimation procedure be applied to every model-generated dataset. Each $\widehat{\theta}_{i,r}^{(m)}$ may also have a conventional sampling standard error when such an interpretation is supported by the sampling design. That standard error measures within-system uncertainty, whereas $D_{i,r}(\theta)$ measures between-system sensitivity.
	
	
	\subsection{From Task-Level String Validation to Substantive Estimand} \label{app:string-est}
	Let $\mathbf{N}_{i,r}$ collect the target strings for task $i$ and let $\mathbf{X}_{i,r}^{(m)}$ collect the corresponding strings generated by measurement system $m$. Because strings do not have a natural arithmetic structure, a string-based analysis must first map the strings into a scalar descriptive quantity. Then define the target estimand and the model-generated estimate as
	
	\begin{equation*}
		\theta_{i,r} = g_i \left( \mathbf{T}_{i,r}, \mathbf{N}_{i,r} \right), \qquad
		\widehat{\theta}_{i,r}^{(m)} = g_i \! \left( \mathbf{Z}_{i,r}^{(m)}, \mathbf{X}_{i,r}^{(m)} \right),
	\end{equation*}
	
	\noindent The target indicators and strings appear together because a target string is only defined when $T_{i,r} = 1$ and a generated string is only defined when $Z_{i,r}^{(m)} = 1$.
	
	The system-specific measurement-induced distortion, the mean estimate across the measurement systems, and the measurement-system dispersion are all defined analogously to the binary outcome case. The system-specific estimate, cross-system mean, and measurement-system dispersion can be calculated for the full sample once the generated strings have been transformed into a common scalar representation.
	
	
	\paragraph{Example 1: Categorization.} Raw strings can be mapped onto a common analytic representation using a fixed harmonization rule. Let
	
	\begin{equation}
		h(\cdot): \mathcal{X} \rightarrow \{ 1, \ldots, K \}
	\end{equation}
	
	\noindent map each target or generated string into one of $K$ standardized concepts or categories.
	
	For example, suppose task $i$ extracts the name of an instrumental variable and $K$ is a set of standardized economic concepts, such as JEL codes. Define the target and generated category indicators as
	
	\begin{equation*}  
		\theta_{k,r} = \frac{1}{J_r} \sum_{j \in r} \mathbf{1} \left\{ T_{j} = 1, \, h(N_{j}) = k \right\}, \qquad
		\widehat{\theta}_{k,r}^{(m)} = \frac{1}{J_r} \sum_{j \in r} \mathbf{1} \left\{ Z_{j}^{(m)} = 1, \, h(X_{j}^{(m)}) = k \right\}.
	\end{equation*}
	
	\noindent Thus, $\theta_{k,r}$ is the share of papers whose target string belongs to category $k$, while $\widehat{\theta}_{k,r}^{(m)}$ is the corresponding share measured using the strings generated by measurement system $m$.
	
	The denominator should be chosen to match the estimand. Dividing by all papers estimates the share of papers using a variable in category $k$. Dividing by the number of model-positive observations instead estimates the share of extracted variables assigned to category $k$. These are distinct descriptive quantities.
	
	The harmonization rule $h(\cdot)$ must be held fixed across measurement systems. If harmonization is itself performed by an LLM or another learned model, that procedure forms part of the measurement system and should be included in the definition of $m$.
	
	
	\paragraph{Example 2: Semantic Proximity to a Category.} A researcher may be interested not only in whether an extracted string belongs to a discrete category but also how closely a string maps to a substantive concept. Let $b_k \in \mathcal{X}$ denote a prespecified reference string representing concept $k$. For example, if task $i$ extracts the name of an instrumental variable, $b_k$ might be the reference string ``weather shock.''
	
	Using the same normalized semantic similarity function, $s( \cdot, \cdot)$, define the target mean semantic proximity and its model-generated analog as
	
	\begin{equation*}
		\theta_{k,r} = \frac{1}{J_r} \sum_{j \in r \\ T_{j} = 1} s \! \left( N_{j}, b_{k} \right), \qquad
		\widehat{\theta}_{k,r}^{(m)} = \frac{1}{J_r} \sum_{j \in r \\ Z_{j}^{(m)} = 1} s \! \left( X_{j}^{(m)}, b_{k} \right).
	\end{equation*}
	
	\noindent A paper contributes zero when no target or generated string is present and contributes a value between zero and one when a string is present. Thus, $\theta_{k,r}$ is the average semantic proximity of the target strings to concept $k$, while $\widehat{\theta}_{k,r}^{(m)}$ is the corresponding quantity generated by measurement system $m$.
	
	Alternatively, the estimands can be written as
	
	\begin{equation*}
		\theta_{k,r} = p_{r} \overline{s}_{k,r} \qquad
		\widehat{\theta}_{k,r}^{(m)} = \widehat{q}_{r}^{(m)} \overline{s}_{k,r}^{(m)},
	\end{equation*}
	
	\noindent where $p_{r}$ is the share of papers for which the target string exists, $\widehat{q}_{r}^{(m)}$ is the share of papers for which measurement system $m$ produces a string, and $\overline{s}_{k,r}$ is the mean semantic proximity to $b_k$ conditional on a string being produced. This estimand combines an extensive margin (whether a string exists) with an intensive margin (how closely that string relates to the reference concept).
	
	However, care must be taken in interpreting this extensive/intensive margin estimand. For example, if the target strings exist in $40\%$ of papers in the evaluation sample with mean semantic proximity to $b_{k} = 0.60$, then the resulting estimand is $\theta_{k,\mathcal{E}} = 0.24$. Suppose measurement system $m$ produces a string in $30\%$ of papers in the evaluation sample and with mean semantic proximity to $b_{k} = 0.80$. Then $\widehat{\theta}_{k,\mathcal{E}} = 0.24$, the same overall estimate despite different rates of prevalence and semantic proximity. Therefore, researchers should report the prevalence and conditional semantic proximity components alongside the substantive estimands.
	
	As in all the examples, the system-specific estimates $\left( \widehat{\theta}_{i,r}^{(m)} \right)$, cross-system means $\left( \overline{\theta}_{i,r} \right)$, and measurement-system dispersion $\left( D_{i,r} (\theta) \right)$ can be calculated for both the evaluation sample $(\mathcal{E})$ and the full sample or population $(P)$. The absolute bias $\left( q_{i,\mathcal{E}}^{(m)} - p_{i,\mathcal{E}} \right)$, relative bias $\left( \left[ q_{i,\mathcal{E}}^{(m)} - p_{i,\mathcal{E}} \right] / p_{i,\mathcal{E}} \right)$, target estimand $\left( \theta_{i,\mathcal{E}} \right)$, and system-specific distortion $\left( \Delta_{i,\mathcal{E}}^{(m)} \right)$ can only be calculated where the corresponding target values are observed, which is typically only in the evaluation sample.
	
	
	\subsection{Evaluation Sample and Full Sample} \label{app:full}
	The evaluation sample and full sample support different empirical quantities.
	
	\paragraph{Evaluation Sample.} Because both target and generated values are observed, the evaluation sample supports calculation of
	
	\begin{itemize}
		\item confusion-matrix quantities, including accuracy, sensitivity, specificity, precision, F1, and balanced accuracy;
		\item semantic similarity and the omnibus measurement loss;
		\item the target-data estimand, $\theta_{i,\mathcal{E}}$;
		\item each system-specific estimate, $\widehat{\theta}_{i,\mathcal{E}}^{(m)}$;
		\item direct estimand distortion, $\Delta_{i,\mathcal{E}}^{(m)}$;
		\item the cross-system mean, $\overline{\theta}_{i,\mathcal{E}}$, and measurement-system dispersion, $D_{i,\mathcal{E}}(\theta)$ .
	\end{itemize}
	
	\noindent The evaluation sample therefore reveals both how accurately each task is performed and how the resulting errors affect the particular descriptive estimand.
	
	\paragraph{Full sample.} In the full sample, the target values generally remain unobserved, meaning the performance metrics cannot be directly calculated. Researchers can nevertheless calculate $\widehat{\theta}_{i,P}^{(m)}$ for every measurement system along with $\overline{\theta}_{i,P}$ and $D_{i,P}(\theta)$. Without target labels, the full-sample estimand distortion $\Delta_{i,P}^{(m)}$ cannot be calculated.
	
	The interpretation of full-sample dispersion is therefore narrower: it measures dependence on measurement-system choice, not absolute error. Low dispersion does not establish validity because all systems may make similar errors. High dispersion indicates that substantive conclusions are sensitive to the system used to generate the data but does not, by itself, identify which estimate is closest to the target.
	
	When the evaluation sample is not randomly drawn from the population of interest, using the evaluation data to correct full-sample estimates requires additional assumptions. \citet{DellandRambachan26} distinguish between two such assumptions: measurement stability and outcome stability. Measurement stability requires that the distribution of measured values conditional on target values be the same in the evaluation sample and the full sample. Outcome stability requires that the distribution of target values conditional on measured values be the same in the two samples.
	
	The distinction matters for how researchers use the evaluation data to correct full-sample estimates. Consider the case where the substantive estimand is prevalence. Under measurement stability, researchers can use evaluation-sample estimates of the FP and FN rates to recover target prevalence from measured prevalence using the binary measurement error model in Section~\ref{binary-out}, provided that $1 - \alpha_i - \beta_i \neq 0$. Under outcome stability, researchers can instead estimate target prevalence by weighting the evaluation-sample probabilities of a target positive conditional on positive and negative classifications by the prevalence of those classifications in the full sample.
	
	In our study, we select papers for human labeling based on their OpenAlex relevance scores precisely because we want to increase the prevalence of rainfall-IV papers in the reference sample. Even if sensitivity and specificity are the same across samples, differences in target prevalence will generally result in differences in precision. Moreover, papers with different relevance scores may differ in the difficulty with which the relevant information can be extracted. Our sampling procedure provides little basis for assuming either outcome stability or measurement stability. Consequently, we do not use the evaluation data to correct full-sample estimates but instead examine the extent to which those estimates depend on the choice of measurement system.
	
	
	\subsection{Practical Workflow} \label{app:workflow}
	A researcher implementing this framework can proceed as follows.
	
	\begin{enumerate}
		\item \textbf{Define the target and estimand.} Specify each target concept $T_{ij}$ and the substantive quantity $\theta_i = g_i(\mathbf{T}_i)$ before evaluating measurement-system performance. Examples include prevalence, a difference between periods, a time trend, or a regression coefficient.
		
		\item \textbf{Fix the measurement systems.} Define each measurement system $m$ to include the model, model version, prompts, retrieval procedure, decoding settings, parser, and any string-harmonization procedure.
		
		\item \textbf{Use a common set of papers.} Apply every measurement system to the same validation and analysis observations. Otherwise, cross-system differences may reflect differences in sample composition rather than measurement-system choice.
		
		\item \textbf{Construct an expert-labeled evaluation sample.} Record the target binary values and, when applicable, target strings using a fixed coding and adjudication protocol. If the evaluation data are to be used to correct full-sample estimates, draw the evaluation sample using a probability sampling design or justify the assumptions under which such a correction can be made.
		
		\item \textbf{Calculate task-level validation measures.} For each concept and measurement system, calculate the confusion matrix, accuracy, sensitivity, specificity, precision, F1, balanced accuracy, mean semantic similarity, and omnibus measurement loss.
		
		\item \textbf{Apply the substantive analysis to the validation data.} Calculate $\theta_{i,\mathcal{E}}$ and $\widehat{\theta}_{i,\mathcal{E}}^{(m)}$ for every measurement system. Report the system-specific distortions, $\Delta_{i,\mathcal{E}}^{(m)}$.
		
		\item \textbf{Apply the same analysis to the full sample.} Holding the specification fixed, calculate $\widehat{\theta}_{i,P}^{(m)}$ for every measurement system.
		
		\item \textbf{Summarize measurement-system sensitivity.} Calculate and report each system-specific estimate, the cross-system mean, $\overline{\theta}_{i,P}$, and measurement-system dispersion, $D_{i,P}(\theta)$. Do not interpret the cross-system mean as the target value or the dispersion as a confidence interval.
		
		\item \textbf{Keep sampling uncertainty and measurement-system sensitivity separate.} Report conventional standard errors or confidence intervals for each system-specific estimate when justified by the sampling design. Report $D_i(\theta)$ separately as a measure of sensitivity to the measurement system.
		
		\item \textbf{Preserve dependence when quantifying uncertainty.} If uncertainty intervals are constructed for validation metrics, estimand distortion, or measurement-system dispersion, resample papers jointly across all measurement systems. Models should not be resampled, because they constitute the fixed set being compared.
	\end{enumerate}
	
	\noindent The central operational principle is simple: define the substantive analysis once, apply it unchanged to every model-generated dataset, and compare the resulting estimates. Task-level validation shows how accurately the data are measured; estimand distortion and measurement-system dispersion show whether those errors matter for the conclusions drawn from model outputs.
	
	
	\newpage
	
	\setcounter{table}{0}
	\renewcommand{\thetable}{B.\arabic{table}}
	\setcounter{figure}{0}
	\renewcommand{\thefigure}{B.\arabic{figure}}
	\setcounter{equation}{0}
	\renewcommand{\theequation}{B.\arabic{equation}}
	
	\section{Additional Details on Implementation Pipeline}\label{app:pipe}
\begin{landscape}
		\begin{figure}[!htbp] 
			\begin{minipage}{\linewidth}		
				\begin{center}
					\caption{Overall Research Design  \label{fig:rd}}
					\includegraphics[width=.89\linewidth,keepaspectratio]{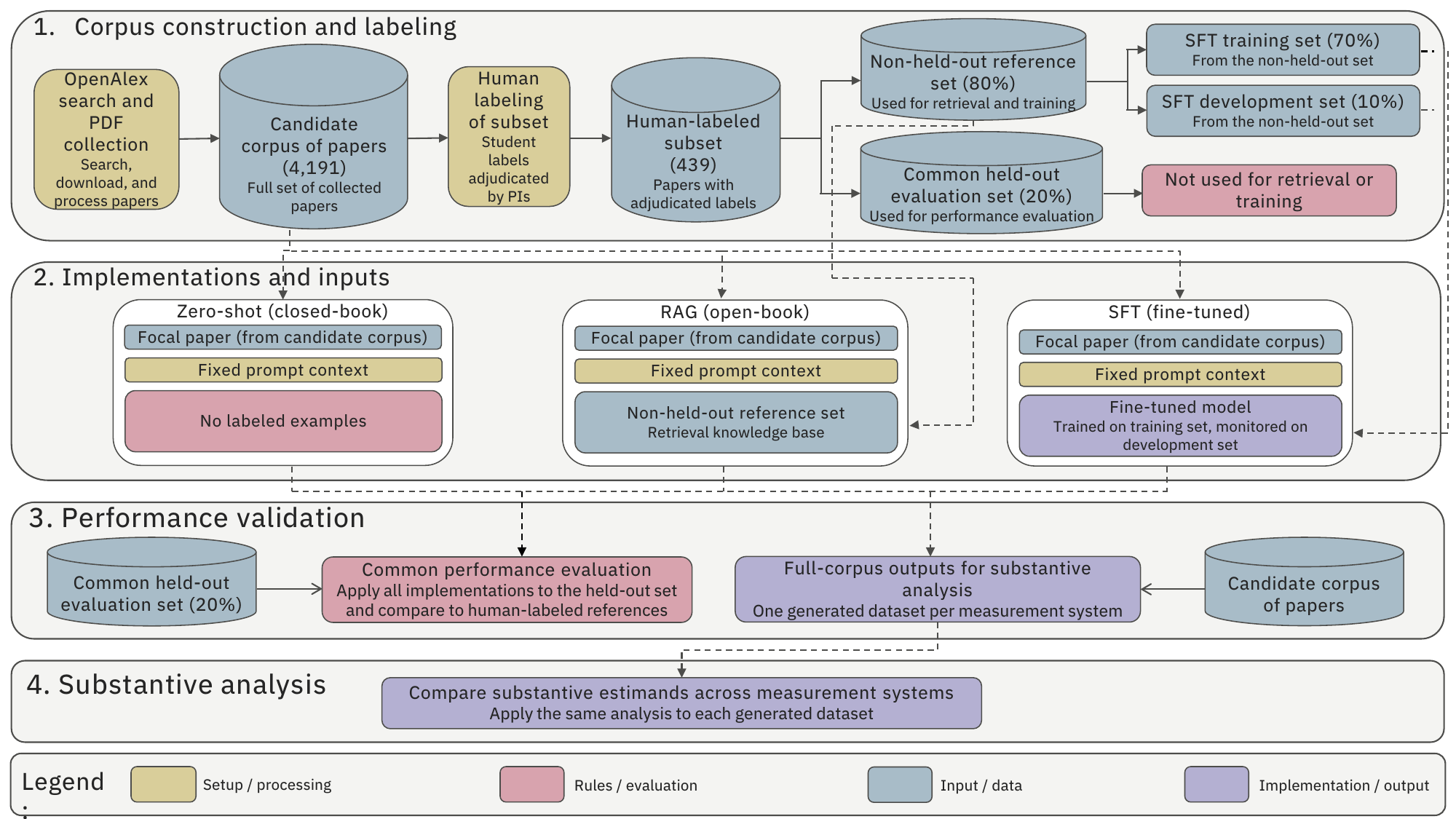}
				\end{center}
				\footnotesize  \textit{Note}: The figure summarizes corpus construction, implementation inputs, and evaluation. Panel 1 shows the collection of $4,191$ candidate papers and human labeling of a subset of 439 papers. Of the labeled sample, 20\% forms the common held-out evaluation set and 80\% supplies reference data for retrieval and fine-tuning. Panel 2 distinguishes implementation inputs. Zero-shot receives no labeled examples, RAG retrieves reference examples, and SFT uses labeled data for training and development monitoring. All implementations receive the focal paper and fixed prompt context. Panel 3 compares outputs with adjudicated labels on the common held-out set and produces a full-corpus dataset for each measurement system. Panel 4 applies the same substantive analysis to each generated dataset and compares the resulting estimates.
			\end{minipage}
		\end{figure}   
	\end{landscape}

	Figure~\ref{fig:rd} provides a visual description of the research design pipeline. Panel 1 shows the collection of $4,191$ candidate papers and human labeling of a subset of 439 papers. Of the labeled sample, 20\% forms the common held-out evaluation set and 80\% supplies reference data for retrieval and fine-tuning. Panel 2 distinguishes implementation inputs. Zero-shot receives no labeled examples, RAG retrieves reference examples, and SFT uses labeled data for training and development monitoring. All implementations receive the focal paper and fixed prompt context. Panel 3 compares outputs with adjudicated labels on the common held-out set and produces a candidate corpus output dataset for each measurement system. Panel 4 applies the same substantive analysis to each generated dataset and compares the resulting estimates.
	
	
	\subsection{Fixed Context}
	To promote consistency for our repetitive metadata-extraction task across a sequence of queries and model implementations, we make use of a fixed prompting structure which appears at the beginning of each API call to condition model behavior. A model is first provided a system role, which establishes its task role and defines output constraints. This is followed by a user role, which defines the specific extraction task and provides the model with the exact context from which it is expected to identify the requested metadata. Finally, an assistant role (prefill) acknowledges these defined constraints and initializes the interaction. This prompting strategy does not eliminate per-paper model discretion, but helps constrain its judgment to supplied evidence and an expected response schema which can be repurposed across all implementations to improve comparability. 
	
	\subsubsection*{System Role} 
	
	\begin{tcolorbox}[colback=gray!5!white, colframe=gray!75!black]
		\{
		\hspace*{0.2cm}``role'': ``system'', ``content'': (\\
		\hspace*{0.5cm}``You are an AI assistant that is an expert in analysis of economic literature.''\\``You interpret complex content and extract specific information, especially about''\\
		``empirical methods, endogeneity problems, and instrumental variable strategies.''\\
		``Rules: Use only information from the provided text to answer each query.''\\
		``If the requested information is not available, answer exactly n/a.''\\
		``Follow the requested output format exactly.''\\
		``Exclusions: Do not include the user queries, any labels, greek letters, or additional''\\
		``text beyond what is requested. Do not respond with symbolic notation, only words.''\\
		``For binary questions with justification requests, always start with 0 or 1''
		\},
	\end{tcolorbox}
	
	
	\subsubsection*{User Role} 
	
	\begin{tcolorbox}[colback=gray!5!white, colframe=gray!75!black]
		\{
		\hspace*{0.2cm}``role'': ``User'', ``content'': (\\
		\hspace*{0.5cm}``You will be asked a sequence of extraction questions about the same academic article. ''\\``I am specifically interested in how researchers address endogeneity problems, particularly ''\\
		``using instrumental variables and especially rainfall-based instruments. ''\\
		``Use answers you have already given as context for later questions when helpful, ''\\
		``but always ground your answers in the provided text.''\\
		``Here are the relevant sections from the article:''\\
		``Rainfall Shocks and Smallholder Market Participation in Rural Malawi ...''\\
		``We estimate the effect of market participation on household income using a two-stage''\\
		``least squares strategy. Because market participation is plausibly endogenous to income...''
		\},
	\end{tcolorbox}
	
	\subsubsection*{Assistant Role (Prefill)} 
	
	\begin{tcolorbox}[colback=gray!5!white, colframe=gray!75!black]
		\{
		\hspace*{0.2cm}``role'': ``Assistant'', ``content'': (\\
		\hspace*{0.5cm}``Understood. I will answer each extraction question using only the provided article text.''
		\},
	\end{tcolorbox}

	\subsection{Model Queries}\label{app:prompt}
	
	
	\subsubsection*{Article Title} 
	
	\begin{tcolorbox}[colback=gray!5!white, colframe=gray!75!black]
		\{
		\hspace*{0.2cm}`key': `Article Title',\\
		\hspace*{0.5cm}`question': 'Task: Extract the article title. Look: top of first page (main heading before authors/abstract). Rules: include subtitle if in the same heading; exclude authors/affiliations/journal headers/footers/section headers; ignore footnote markers (*, \dag, superscripts). Output: ONE line: title text only. No quotes, labels, or extra words. --- Task: Extract the article title. Look: top of first page (main heading before authors/abstract). Rules: include subtitle if in the same heading; exclude authors/affiliations/journal headers/footers/section headers; ignore footnote markers (*, \dag, superscripts). Output: ONE line: title text only. No quotes, labels, or extra words.'
		\},
	\end{tcolorbox}
	
	
	\subsubsection*{DOI}
	\begin{tcolorbox}[colback=gray!5!white, colframe=gray!75!black]
		\{
		\hspace*{0.2cm}`key': `DOI',\\
		\hspace*{0.5cm}`question': `Task: Extract the DOI of THIS article (version of record), not DOIs in references. Look: first page/front matter for doi:, DOI, https://doi.org/. Rules: remove URL/prefix (doi:, https://doi.org/); remove spaces/line breaks; strip trailing punctuation; output lowercase. Output: ONE token = normalized DOI (10.xxxx/xxxx) OR exactly n/a. --- Task: Extract the DOI of THIS article (version of record), not DOIs in references. Look: first page/front matter for doi:, DOI, https://doi.org/. Rules: remove URL/prefix (doi:, https://doi.org/); remove spaces/line breaks; strip trailing punctuation; output lowercase. Output: ONE token = normalized DOI (10.xxxx/xxxx) OR exactly n/a.'
		\}
	\end{tcolorbox}
	
	
	\subsubsection*{Empirical Analysis}
	\begin{tcolorbox}[colback=gray!5!white, colframe=gray!75!black]
		\{
		\hspace*{0.2cm}`key': `Empirical Analysis',\\
		\hspace*{0.5cm}`question': `Task: Does the article contain empirical quantitative statistical estimation (e.g., regressions/econometrics with estimated coefficients/SEs/p-values)? Look: `we estimate/regress', model equations with error terms, regression tables with coefficients. Rules: do NOT infer from topic/title/abstract. Count only analysis done in THIS article (not summaries or reviews of other papers). Exclude review articles, articles in hydrology, and articles that contain purely theoretical work only, qualitative/descriptive only, or numerical simulation or model calibration only. Output: 1 if yes, 0 if no. Output exactly one character. --- Task: Does the article contain empirical quantitative statistical estimation (e.g., regressions/econometrics with estimated coefficients/SEs/p-values)? Look: `we estimate/regress', model equations with error terms, regression tables with coefficients. Rules: do NOT infer from topic/title/abstract. Count only analysis done in THIS article (not summaries or reviews of other papers). Exclude review articles, articles in hydrology, and articles that contain purely theoretical work only, qualitative/descriptive only, or numerical simulation or model calibration only. Output: 1 if yes, 0 if no. Output exactly one character.'
		\},
	\end{tcolorbox}
	
	
	\subsubsection*{Dependent Variable(s)}
	\begin{tcolorbox}[colback=gray!5!white, colframe=gray!75!black]
		\{
		\hspace*{0.2cm}`key': `Dependent Variable(s)',\\
		\hspace*{0.5cm}`dependency': {`key': `Empirical Analysis', `value': `1'}, \\
		\hspace*{0.5cm}`question': `Task: List the main dependent/outcome variable(s) used in the primary regression/econometric results. Look: LHS of main equations; column headers of main regression tables; text describing the main empirical model. Rules: exclude first-stage outcomes, RHS variables (treatments/endogenous regressors/instruments/controls/covariates/fixed effects), mediators/moderators. Keep names exactly as written in the paper/table (including any log/ln/differences/units if shown). Output: ONE line: variable name(s) only; separate multiple with `; '. Output the name of the dependent variable, not the symbolic notation representing the name of the variable in an equation (e.g., $y_{it}$). --- Task: List the main dependent/outcome variable(s) used in the primary regression/econometric results. Look: LHS of main equations; column headers of main regression tables; text describing the main empirical model. Rules: exclude first-stage outcomes, RHS variables (treatments/endogenous regressors/instruments/controls/covariates/fixed effects), mediators/moderators. Keep names exactly as written in the paper/table (including any log/ln/differences/units if shown). Output: ONE line: variable name(s) only; separate multiple with `; '. Output the name of the dependent variable, not the symbolic notation representing the name of the variable in an equation (e.g., $y_{it}$).'
		\},
	\end{tcolorbox}
	
	
	\subsubsection*{Endogeneity Problem}
	\begin{tcolorbox}[colback=gray!5!white, colframe=gray!75!black]
		\{
		\hspace*{0.2cm}`key': `Endogeneity Problem',\\
		\hspace*{0.5cm}`dependency': {`key': `Empirical Analysis', `value': `1'}, \\
		\hspace*{0.5cm}`question': `Task: In the MAIN empirical analysis, do the authors treat any RHS variable as endogenous (correlated with the error term) and address it explicitly? Look: statements that a regressor is endogenous + a method to address it (IV/2SLS/3SLS/LIML, control function/2SRI, GMM with instruments, first stage, weak-IV tests, etc.). Rules: do NOT count generic mentions of `endogeneity' or references to endogeneity issues in other papers without an actual endogenous regressor in the main specs of this paper. Output: 1 if yes, 0 if no. Output exactly one character. --- Task: In the MAIN empirical analysis, do the authors treat any RHS variable as endogenous (correlated with the error term) and address it explicitly? Look: statements that a regressor is endogenous + a method to address it (IV/2SLS/3SLS/LIML, control function/2SRI, GMM with instruments, first stage, weak-IV tests, etc.). Rules: do NOT count generic mentions of `endogeneity' or references to endogeneity issues in other papers without an actual endogenous regressor in the main specs of this paper. Output: 1 if yes, 0 if no. Output exactly one character.'
		\},
	\end{tcolorbox}
	
	
	\subsubsection*{Endogenous Variable(s)}
	\begin{tcolorbox}[colback=gray!5!white, colframe=gray!75!black]
		\{
		\hspace*{0.2cm}`key': `Endogenous Variable(s)',\\
		\hspace*{0.5cm}`dependency': {`key': `Endogeneity Problem', `value': `1'}, \\
		\hspace*{0.5cm}`question': `Task: List the explanatory variable(s) explicitly treated as endogenous in the main analysis. Look: `we instrument X', `X is endogenous', first-stage descriptions/tables, reduced form, weak-IV/overid tests. Rules: exclude instruments, dependent variables, and ordinary controls. Keep names exactly as written in the paper/table (including any log/ln/differences/units if shown). Output: ONE line: variable name(s) only; separate multiple with `; '. Output the name of the endogenous variable, not the symbolic notation representing the name of the variable in an equation (e.g., $x_{it}$). --- Task: List the explanatory variable(s) explicitly treated as endogenous in the main analysis. Look: `we instrument X', `X is endogenous', first-stage descriptions/tables, reduced form, weak-IV/overid tests. Rules: exclude instruments, dependent variables, and ordinary controls. Keep names exactly as written in the paper/table (including any log/ln/differences/units if shown). Output: ONE line: variable name(s) only; separate multiple with `; '. Output the name of the endogenous variable, not the symbolic notation representing the name of the variable in an equation (e.g., $x_{it}$).'
		\},
	\end{tcolorbox}
	
	
	\subsubsection*{Instrumental Variable Regression}
	\begin{tcolorbox}[colback=gray!5!white, colframe=gray!75!black]
		\{
		\hspace*{0.2cm}`key': `Instrumental Variable Regression',\\
		\hspace*{0.5cm}`dependency': {`key': `Endogeneity Problem', `value': `1'}, \\
		\hspace*{0.5cm}`question': `Task: In the MAIN empirical analysis, do the authors implement an IV-type estimator with excluded instruments to address endogeneity? Look: an actual excluded instrument set used in a first stage/reduced form; 2SLS/IV/3SLS/LIML; IV-Probit; control function/2SRI; GMM with instruments; exclusion restriction; first-stage equations/tables. Rules: do NOT count non-statistical uses of 'instrument' (survey instrument, measurement instrument), references to other papers that use IVs, arguments as to why IVs are not necessary for this paper, or papers that only use RCT/RDD/DiD/event study without an IV first stage. Output: 1 if yes, 0 if no. Output exactly one character. --- Task: In the MAIN empirical analysis, do the authors implement an IV-type estimator with excluded instruments to address endogeneity? Look: an actual excluded instrument set used in a first stage/reduced form; 2SLS/IV/3SLS/LIML; IV-Probit; control function/2SRI; GMM with instruments; exclusion restriction; first-stage equations/tables. Rules: do NOT count non-statistical uses of 'instrument' (survey instrument, measurement instrument), references to other papers that use IVs, arguments as to why IVs are not necessary for this paper, or papers that only use RCT/RDD/DiD/event study without an IV first stage. Output: 1 if yes, 0 if no. Output exactly one character.'
		\},
	\end{tcolorbox}
	
	
	\subsubsection*{Instrumental Variable(s)}
	\begin{tcolorbox}[colback=gray!5!white, colframe=gray!75!black]
		\{
		\hspace*{0.2cm}`key': `Instrumental Variable(s)',\\
		\hspace*{0.5cm}`dependency': {`key': `Instrumental Variable Regression', `value': `1'},\\
		\hspace*{0.5cm}`question': `Task: List the excluded instrument(s) used in the main IV analysis. Look: `we instrument X with Z', `Z is our instrument', `excluded instrument', first-stage/reduced-form equations or tables. Rules: exclude endogenous regressors themselves, dependent variables, and regular controls. Keep names exactly as written in the paper/table (including any log/ln/differences/units if shown). Output: ONE line: instrument name(s) only; separate multiple with `; '. Output the name of the instrumental variable, not the symbolic notation representing the name of the variable in an equation (e.g., $z_{it}$). --- Task: List the excluded instrument(s) used in the main IV analysis. Look: `we instrument X with Z', `Z is our instrument', `excluded instrument', first-stage/reduced-form equations or tables. Rules: exclude endogenous regressors themselves, dependent variables, and regular controls. Keep names exactly as written in the paper/table (including any log/ln/differences/units if shown). Output: ONE line: instrument name(s) only; separate multiple with `; '. Output the name of the instrumental variable, not the symbolic notation representing the name of the variable in an equation (e.g., $z_{it}$).'
		\},
	\end{tcolorbox}
	
	
	\subsubsection*{Rainfall Instrument}
	\begin{tcolorbox}[colback=gray!5!white, colframe=gray!75!black]
		\{
		\hspace*{0.2cm}`key': `Rainfall Instrument',\\
		\hspace*{0.5cm}`dependency': {`key': `Instrumental Variable Regression', `value': `1'}, \\
		\hspace*{0.5cm}`question': `Task: Is any excluded instrument based on rainfall/precipitation? Look: rainfall, precipitation, drought, monsoon rainfall/onset, wet-day counts, SPI/SPEI/PDSI, precipitation-derived indices used as the EXCLUDED instrument. Rules: count only if the variable is precipitation-based and used as an excluded instrument in an IV first stage/reduced form. Do NOT count precipitation used only as regressor/control/interaction/exposure/outcome or if its relationship to the dependent variable is discussed. Do NOT count ENSO or other climate indices unless explicitly stated to be precipitation-based AND used as the excluded instrument. Ignore mentions of rainfall as an instrument in other papers. Output: 1 if yes, 0 if no. Output exactly one character. --- Task: Is any excluded instrument based on rainfall/precipitation? Look: rainfall, precipitation, drought, monsoon rainfall/onset, wet-day counts, SPI/SPEI/PDSI, precipitation-derived indices used as the EXCLUDED instrument. Rules: count only if the variable is precipitation-based and used as an excluded instrument in an IV first stage/reduced form. Do NOT count precipitation used only as regressor/control/interaction/exposure/outcome or if its relationship to the dependent variable is discussed. Do NOT count ENSO or other climate indices unless explicitly stated to be precipitation-based AND used as the excluded instrument. Ignore mentions of rainfall as an instrument in other papers. Output: 1 if yes, 0 if no. Output exactly one character.'
		\},
	\end{tcolorbox}
	
	
	\subsubsection*{Rainfall Variable(s)}
	\begin{tcolorbox}[colback=gray!5!white, colframe=gray!75!black]
		\{
		\hspace*{0.2cm}`key': `Rainfall Variable(s)',\\
		\hspace*{0.5cm}`dependency': {`key': `Rainfall Instrument', `value': `1'}, \\
		\hspace*{0.5cm}`question': `Task: List the specific rainfall/precipitation-based excluded instrument(s) used in the main IV analysis. Look: first-stage/reduced-form equations/tables for the precipitation-based instrument name(s) (e.g., total/mean rainfall over a window, rainfall deviations/shocks, monsoon onset, rainfall index, coefficient of variation of rainfall). Rules: keep names exactly as written in the paper/table (including window/statistic/units/logs if shown). Output: ONE line: rainfall instrument name(s) only; separate multiple with `; '. Output the name of the rainfall instrument, not the symbolic notation representing the name of the variable in an equation (e.g., $r_{it}$). --- Task: List the specific rainfall/precipitation-based excluded instrument(s) used in the main IV analysis. Look: first-stage/reduced-form equations/tables for the precipitation-based instrument name(s) (e.g., total/mean rainfall over a window, rainfall deviations/shocks, monsoon onset, rainfall index, coefficient of variation of rainfall). Rules: keep names exactly as written in the paper/table (including window/statistic/units/logs if shown). Output: ONE line: rainfall instrument name(s) only; separate multiple with `; '. Output the name of the rainfall instrument, not the symbolic notation representing the name of the variable in an equation (e.g., $r_{it}$).'
		\},
	\end{tcolorbox}
	
	
	\newpage
	
	\setcounter{table}{0}
	\renewcommand{\thetable}{C.\arabic{table}}
	\setcounter{figure}{0}
	\renewcommand{\thefigure}{C.\arabic{figure}}
	\setcounter{equation}{0}
	\renewcommand{\theequation}{C.\arabic{equation}}
	
	
	\section{Classification Method and Matching} \label{app:class}
	
	In order to make apples-to-apples comparisons for the information extracted by different implementations, we need to translate the specific variables identified in individual papers into a common set of categories. We develop two taxonomies, one for the economic concepts represented by the endogenous variables and another for the rainfall measures represented in the instruments. This allows us to aggregate results across papers while preserving substantively meaningful differences in how variables are defined. We then test a set of LLM-classification pipelines to determine how best to sort variable names into the common categories.
	
	
	\subsection{Matching Endogenous Variables to Economic Concepts}
	
	We develop a taxonomy of economic concepts based on the American Economic Association's Journal of Economic Literature (JEL) classification system. Rather than mapping variables directly to individual JEL codes, we group related codes into 23 broader economic concepts. These categories cover the major areas of economic activity, including income, consumption, agricultural production, labor markets, health, migration, governance, and finance. Our objective is to create categories sufficiently broad to aggregate related variables across papers but sufficiently narrow to retain economically meaningful distinctions. The resulting categories and their associated JEL codes are reported in Table \ref{tab:econ_tax}. The JEL codes provide conceptual anchors for each category rather than serving as the classification output.
	
	An important distinction is that we classify the economic object represented by the endogenous variable rather than the broader topic of the paper, the sector being studied, or the population of interest. For example, farm income is classified as income, agricultural employment as labor markets, agricultural exports as trade, and government health expenditure as public finance. Similarly, variables representing the same economic concept remain in the same category regardless of their units, functional form, or empirical transformation. We include two residual categories, Other for variables that fall outside the 23 economic concepts and Unclassifiable for variables for which the extracted description provides insufficient information to determine what is being measured.
	
	
	\subsection{Matching Instrumental Variables to Rainfall Categories}
	
	Unlike the endogenous variables, which span the breadth of economics, the instrumental variables all measure some aspect of precipitation. However, economists quantify rainfall in numerous ways, and these differences can be substantively important. We develop a taxonomy of 21 rainfall categories that distinguish between measures of rainfall levels, changes, deviations from historical norms, extreme events, and other characteristics of precipitation. For example, total rainfall, standardized rainfall anomalies, drought indicators, and the duration of dry spells capture different features of rainfall and are therefore assigned to separate categories. Table \ref{tab:rain_tax} reports the categories, their definitions, and examples of rainfall measures that belong to each.
	
	In constructing these categories, we distinguish between differences that change the underlying rainfall measure and those that simply reflect how the measure is implemented. Rainfall totals measured over a month, growing season, or year belong to the same category, as do contemporaneous and lagged versions of the same measure. By contrast, deviations from historical rainfall, standardized anomalies, and one-sided rainfall shocks are treated separately because they capture different aspects of the rainfall distribution. We also distinguish named rainfall indices, such as the Standardized Precipitation Index (SPI), from conventional standardized anomalies, and the frequency of dry days from the duration of consecutive dry spells. Similarly, cumulative or recurrent shock exposure represents a history of rainfall events rather than a single lagged realization. We include two residual categories, Other for measures that do not fit the 21 substantive categories and Unclassifiable for measures whose meaning or construction cannot be determined from the extracted description.
	
	
	\subsection{Matching Extracted Variables to Categories}
	
	While the task of classifying a variable into one of a finite set of categories is seemingly less complex than that of identifying the variable in a paper, there is still considerable scope for measurement error. One could develop a set of rules based on the words that appear in a variable name, though this would require the researcher to anticipate the myriad ways economists refer to the same concept. Alternatively, one could rely on models that quantify semantic similarity. The issue here is that semantic similarity need not correspond to the distinctions that are of interest to the researcher. Agricultural employment and agricultural income, for example, are semantically similar but represent distinct economic concepts. An advantage of using an LLM to classify variables is that one can provide the model with instructions as to which differences between variables are and are not substantively meaningful. Whether there are gains to supplementing the LLM with information on semantic similarity is an empirical question. To examine this question, we use the Qwen3 family of open-weight models, which includes both LLMs of varying sizes and models designed specifically to quantify (embedding models) and rank (reranking models) the semantic similarity of strings.
	
	We implement six different approaches to classifying variables. The simplest is to provide the variable name and the relevant taxonomy to Qwen3-32B (32B Direct). We then augment the prompt with instructions regarding the application of the taxonomy and allow the model a second attempt at classification if it returns a category that does not exist in the taxonomy (32B Solo). A more involved approach is to use an embedding model to identify categories that are semantically similar to the variable name and a reranking model to refine the ordering of these categories. The relative similarity of each category to the variable name is then provided as additional information to the LLM tasked with making the classification decision. We implement this approach using two versions of Qwen3 that differ in the number of parameters (8B Pipeline and 32B Pipeline). Finally, there may be gains to providing the LLM with information on the context in which the variable appears in the literature. We augment the 32B Solo and 32B Pipeline implementations with information that the GPT models generated about the paper from which the variable comes (32B Solo + Context and 32B Pipeline + Context). While this additional information may allow the model to make inference about otherwise ambiguous variable names, there is also scope for measurement error in the contextual information to induce measurement error in the classification of variables.
	
	We assess the performance of each implementation against human labels for 42 rainfall IVs and 85 endogenous variables. Increasing the size of the LLM from 8B to 32B in the pipeline increases accuracy from 50\% to 67\% for rainfall IVs and from 84\% to 87\% for endogenous variables. Interestingly, there are no similar gains to increasing the complexity of the classification procedure. In fact, if we dispense with the embedding and reranking models altogether, accuracy increases to 71\% for rainfall IVs and 89\% for endogenous variables for both 32B Direct and 32B Solo. Providing the LLM with additional contextual information does little to improve the classification of rainfall IVs and reduces accuracy in the classification of endogenous variables. Much of this latter decline in accuracy comes from the LLM assigning variables to substantive categories when the human coders did not. Comparisons across the other performance metrics yield much the same conclusion. Given that the additional computational costs associated with the more complex implementations are not accompanied by systematic improvements in classification accuracy, we rely on Qwen3-32B as a stand-alone classifier.
	

	\newpage
	
	\input{tables/econ_tax}

	
	\input{tables/rain_tax}

	
	\begin{landscape}
		\begin{figure}[!htbp] 
			\begin{minipage}{\linewidth}		
				\begin{center}
					\caption{Qwen3 Classifiers: Evidence, Performance, and Semantic Stability  \label{fig:qwen3_eval}}
					\includegraphics[width=\linewidth,keepaspectratio]{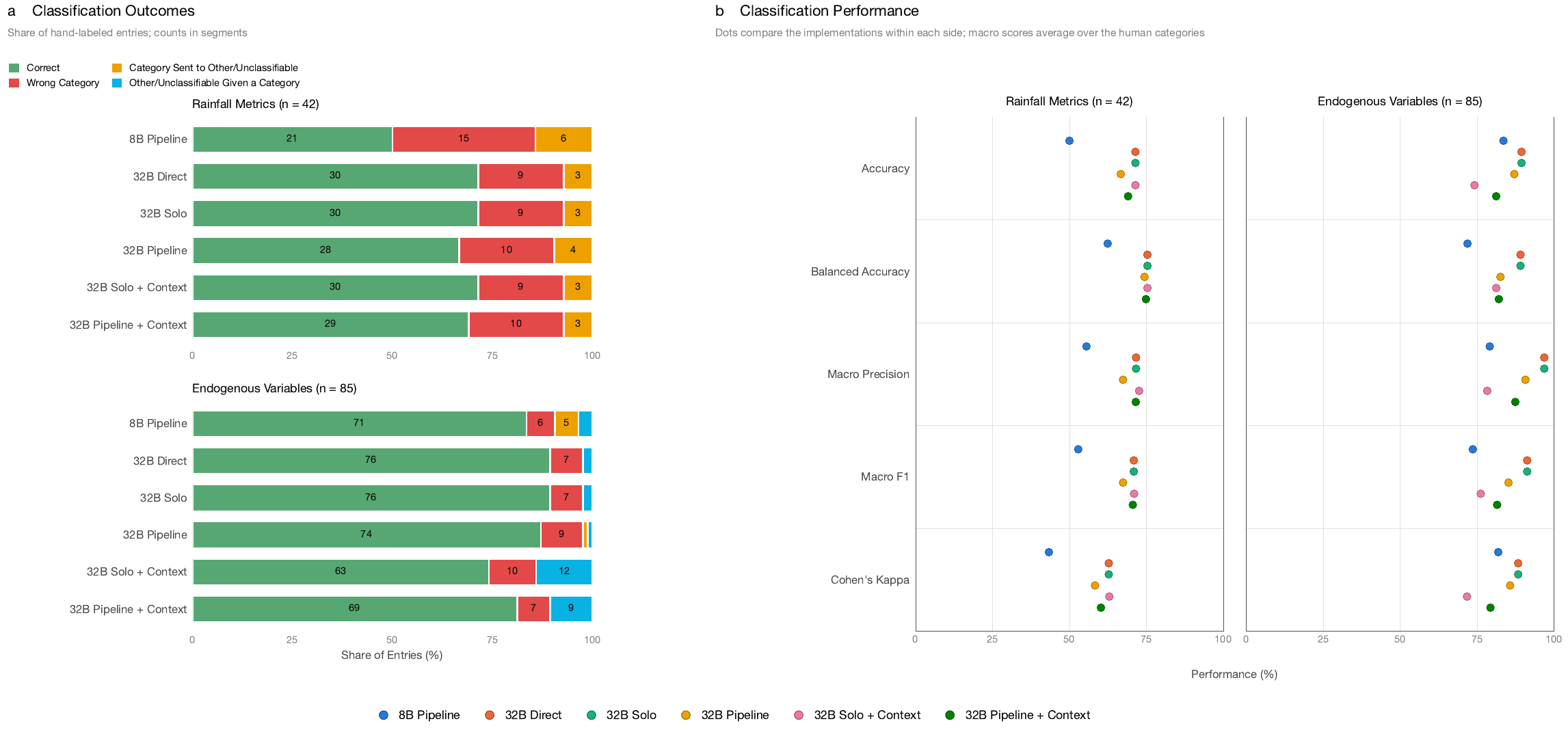}
				\end{center}
				\footnotesize  \textit{Note}: The figure compares the performance of six implementations of Qwen3 in classifying rainfall IVs and endogenous variables against human labels. Panel (a) reports the share of variables that are correctly classified, assigned to the wrong category, or incorrectly assigned to or from one of the residual categories. The values within the bars are counts of variables. Panel (b) reports five different metrics of classification performance. Macro-averaged metrics weight each category represented in the human-labeled data equally, regardless of the frequency with which variables belong to that category.
			\end{minipage}
		\end{figure}    
	\end{landscape}
	

	\newpage
	
	\setcounter{table}{0}
	\renewcommand{\thetable}{D.\arabic{table}}
	\setcounter{figure}{0}
	\renewcommand{\thefigure}{D.\arabic{figure}}
	\setcounter{equation}{0}
	\renewcommand{\theequation}{D.\arabic{equation}}
	
	
	\section{Comparison to Frontier Models} \label{app:alt-mod}
	
	\subsection{Validation of Frontier Models} \label{app:alt-val}
	
	To examine whether our findings reflect limitations of an older generation of LLMs, we compare our SFT implementation of GPT-4.1 with zero-shot implementations of four newer models: GPT-6.1 Sol, Gemma 4 26B, Llama 4 Maverick, and Claude Opus 5.5.\footnote{This comparison does not isolate the effect of model vintage because GPT-4.1 is fine-tuned while the newer models use zero-shot prompting. A like-for-like comparison could instead use our zero-shot GPT-4.1 implementation alongside the newer zero-shot models.} Figure~\ref{fig:frnt_eval} presents the same four-panel validation exercise using the common held-out sample of 88 papers. The results suggest that improvements in general model capabilities have not eliminated the measurement problems documented in our main analysis.
	
	Panels A and B show that binary classification performance remains high across models. For rainfall-IV identification, all five measurement systems correctly classify between 84 and 86 of the 88 papers, yielding accuracy rates of approximately 95\% to 98\%. However, similar accuracy conceals differences in the composition of errors. SFT produces two false positives and one false negative, while GPT-6.1 produces no false positives but misses two positive cases, and Claude produces no false positives but misses three. Consequently, the number of papers classified as using a rainfall IV ranges from 33 to 37, compared with 36 in the human-labeled data. Differences are more pronounced for earlier classification tasks. SFT misses only two of the 72 papers with an endogeneity problem, while each of the four newer models misses at least five. Conversely, SFT produces three false positives on this task compared with none for GPT-6.1. For empirical analysis, nearly all positive cases are correctly identified, but the models incorrectly classify several of the 14 non-empirical papers as empirical. Thus, even the newest models exhibit the same tension between sensitivity and specificity documented in our main analysis. High overall accuracy does not imply that measurement systems produce the same prevalence estimates or introduce the same direction of error.
	
	Panels C and D reveal substantially greater differences when models must identify and name economic concepts. All five systems perform well on relatively simple metadata, with mean semantic similarity for article titles of approximately 0.97. The newer models also achieve higher mean similarity for DOIs, although these comparisons are conditional on considerably fewer evaluated observations than SFT. Performance deteriorates markedly for substantive economic variables, and the newer models do not consistently improve upon our older, fine-tuned model. For dependent variables, SFT achieves mean semantic similarity of 0.73, compared with values no greater than 0.52 across the four frontier models. The corresponding comparisons are 0.75 versus at most 0.59 for endogenous variables and 0.67 versus at most 0.59 for rainfall instruments. Differences are smaller for instrumental variables generally, but SFT continues to outperform all four newer models. Notably, dependent-variable extraction exhibits the weakest performance among the frontier models, indicating that semantic mismeasurement is not confined to identifying specialized rainfall instruments.
	
	The distributions in Panel D further demonstrate that differences in mean similarity do not fully characterize the quality of extracted information. SFT concentrates considerably more density at high similarity scores, particularly for dependent and endogenous variables. The newer models exhibit much broader distributions, producing some highly accurate descriptions alongside many poorly matched strings. For instrumental variables, the distributions overlap more closely, but substantial variation remains within every measurement system. These results reinforce the distinction between successfully identifying the presence of a concept and accurately describing what that concept is. For example, GPT-6.1 slightly outperforms SFT in identifying whether a paper uses rainfall as an IV, yet its mean semantic similarity in naming those rainfall instruments is approximately 0.54, compared with 0.67 for SFT. Moreover, because semantic similarity is evaluated conditional on successful binary identification, these distributions do not capture the additional loss of information arising when models fail to identify a relevant concept altogether.
	
	\subsection{Paper-Level Comparison}
	
\begin{figure}[!htbp] 
		\begin{minipage}{\linewidth}		
			\begin{center}
				\caption{Frontier Models: Evidence, Performance, and Semantic Stability  \label{fig:frnt_eval}}
				\includegraphics[width=.84\linewidth,keepaspectratio]{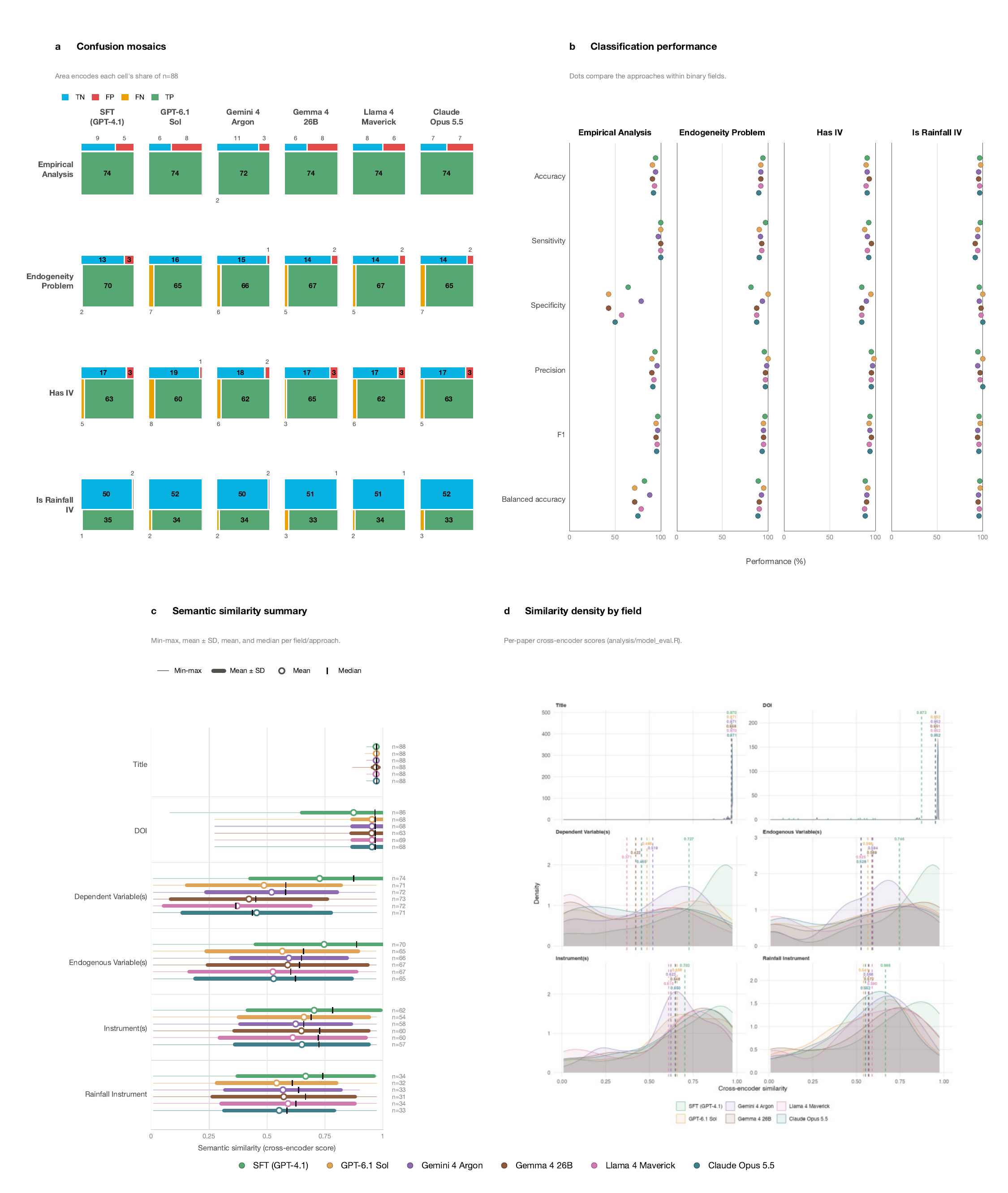}
			\end{center}
			\footnotesize  \textit{Note}: The figure compares the supervised fine-tuning (SFT) implementation of GPT-4.1 with zero-shot implementations of four newer models: GPT-6.1 Sol, Gemma 4 26B, Llama 4 Maverick, and Claude Opus 5.5. All models are evaluated against the same human-labeled evaluation set of 88 papers. Panel A presents confusion mosaics for the four binary classification tasks, reporting true negatives (TN), false positives (FP), false negatives (FN), and true positives (TP). Panel B reports accuracy, sensitivity, specificity, precision, F1 score, and balanced accuracy for each task and model. Panels C and D evaluate string extraction using semantic similarity between model-generated and human-labeled strings. Panel C reports the minimum--maximum range, mean $\pm$ one standard deviation, mean, and median, with the corresponding sample size. Panel D presents the distributions of semantic similarity scores, with dashed vertical lines indicating model-specific means. Semantic similarity is calculated conditional on correct positive classification, so sample sizes vary across fields and models.
		\end{minipage}
	\end{figure}

	Whether the differences in performance documented in Figure~\ref{fig:frnt_eval} matter for our description of the rainfall-IV literature is a separate question. Table~\ref{tab:frontier_agree} replicates the analysis in Table~\ref{tab:gpt_agree}, using SFT as the basis for comparison with each of the four newer models. Interestingly, SFT identifies more papers as using a rainfall IV (180) than either GPT-6.1 Sol (146), Gemma (137), Llama (167), or Claude (137). Thus, despite the relatively minor differences in accuracy on the held-out evaluation set, there are meaningful differences in the estimated size of the rainfall-IV literature. Of the $222$ papers identified as using a rainfall IV by at least one model, all five models agree on $118$ papers ($53\%$), while $59$ papers ($27\%$) are identified by only one model. For comparison, the corresponding shares across the three implementations of GPT-4.1 are $62\%$ and $27\%$, respectively.
	
	Conditional on all five models identifying a paper as using a rainfall IV, we again examine the degree to which models agree on what the rainfall IV is and what it instruments. Across pairwise comparisons with SFT, exact agreement on both the rainfall IV and endogenous variable lists ranges from $4\%$ to $6\%$. When we group variables into categories, agreement on both sets of categories ranges from $24\%$ to $36\%$. The corresponding rates of agreement when comparing SFT to either zero-shot or RAG in Table~\ref{tab:gpt_agree} range from $8\%$ to $11\%$ for exact strings and from $38\%$ to $39\%$ for categories. While these comparisons are conditional on different sets of papers, the differences suggest that varying the model itself introduces additional measurement-system dependence beyond that which we document when varying the implementation of a single model.
	
	\subsection{Relationship-Level Comparison}
	
\begin{figure}[!htbp] 
		\begin{minipage}{\linewidth}		
			\begin{center}
				\caption{SFT: Rainfall Instruments and Endogenous Variables \label{fig:gpt_sft_app}}
				\includegraphics[width=\linewidth,keepaspectratio]{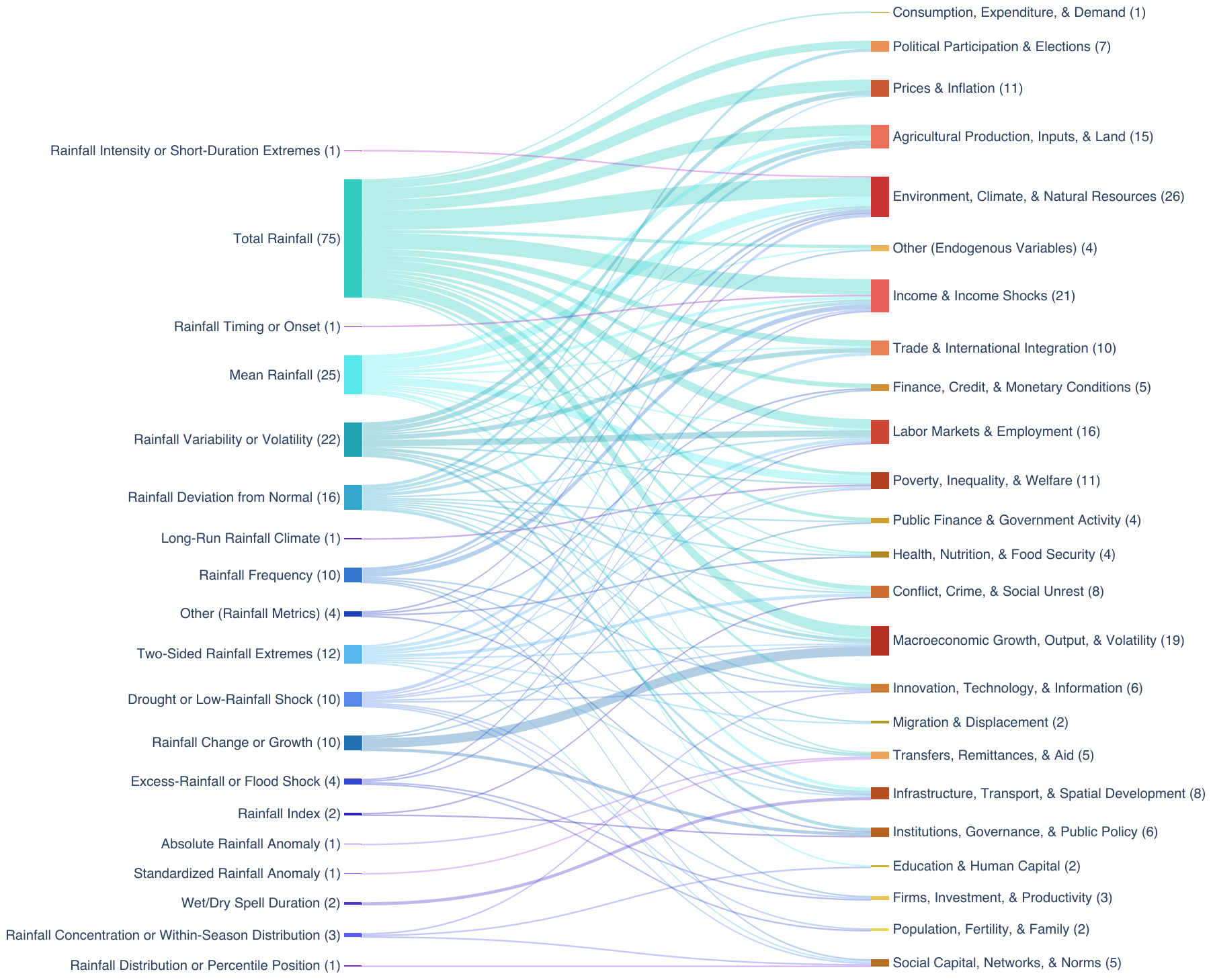}
			\end{center}
			\footnotesize  \textit{Note}: The Sankey diagram maps harmonized categories of rainfall instruments identified by the supervised fine-tuning (SFT) implementation (left) to the corresponding harmonized categories of endogenous variables (right). Flow width is proportional to the number of extracted rainfall instrument–endogenous variable linkages, with flows colored according to the rainfall-instrument category. Values in parentheses report the total number of extracted linkages associated with each category. Counts refer to model-generated linkages and need not correspond to unique papers.
		\end{minipage}
	\end{figure}

\begin{figure}[!htbp] 
		\begin{minipage}{\linewidth}		
			\begin{center}
				\caption{GPT-6.1 Sol: Rainfall Instruments and Endogenous Variables\label{fig:gpt_sol}}
				\includegraphics[width=\linewidth,keepaspectratio]{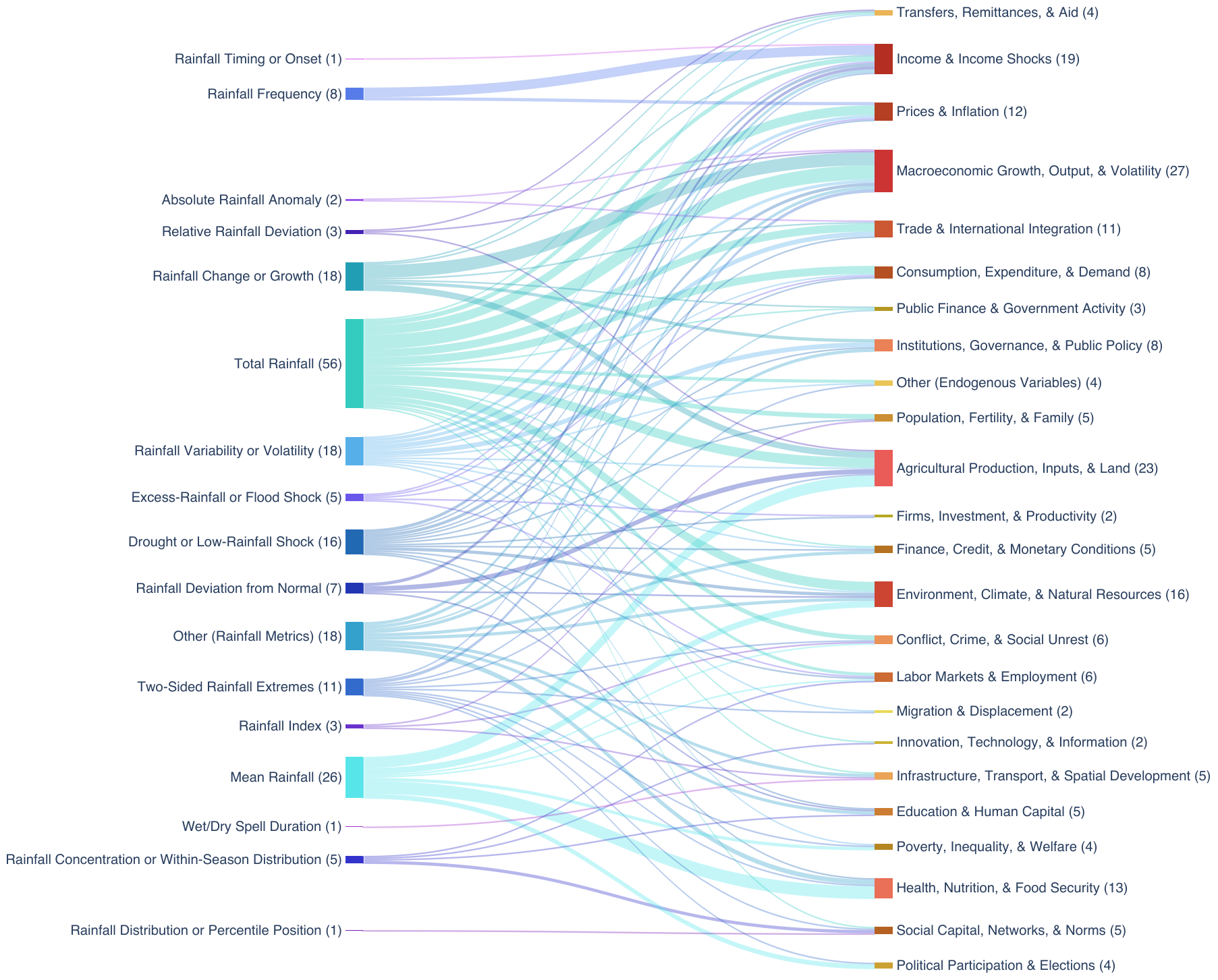}
			\end{center}
			\footnotesize  \textit{Note}: The Sankey diagram maps harmonized categories of rainfall instruments identified by the zero-shot implementation (left) to the corresponding harmonized categories of endogenous variables (right). Flow width is proportional to the number of extracted rainfall instrument–endogenous variable linkages, with flows colored according to the rainfall-instrument category. Values in parentheses report the total number of extracted linkages associated with each category. Counts refer to model-generated linkages and need not correspond to unique papers.
		\end{minipage}
	\end{figure}

\begin{figure}[!htbp] 
		\begin{minipage}{\linewidth}		
			\begin{center}
				\caption{Gemma 4 26B: Rainfall Instruments and Endogenous Variables\label{fig:gpt_gemma}}
				\includegraphics[width=\linewidth,keepaspectratio]{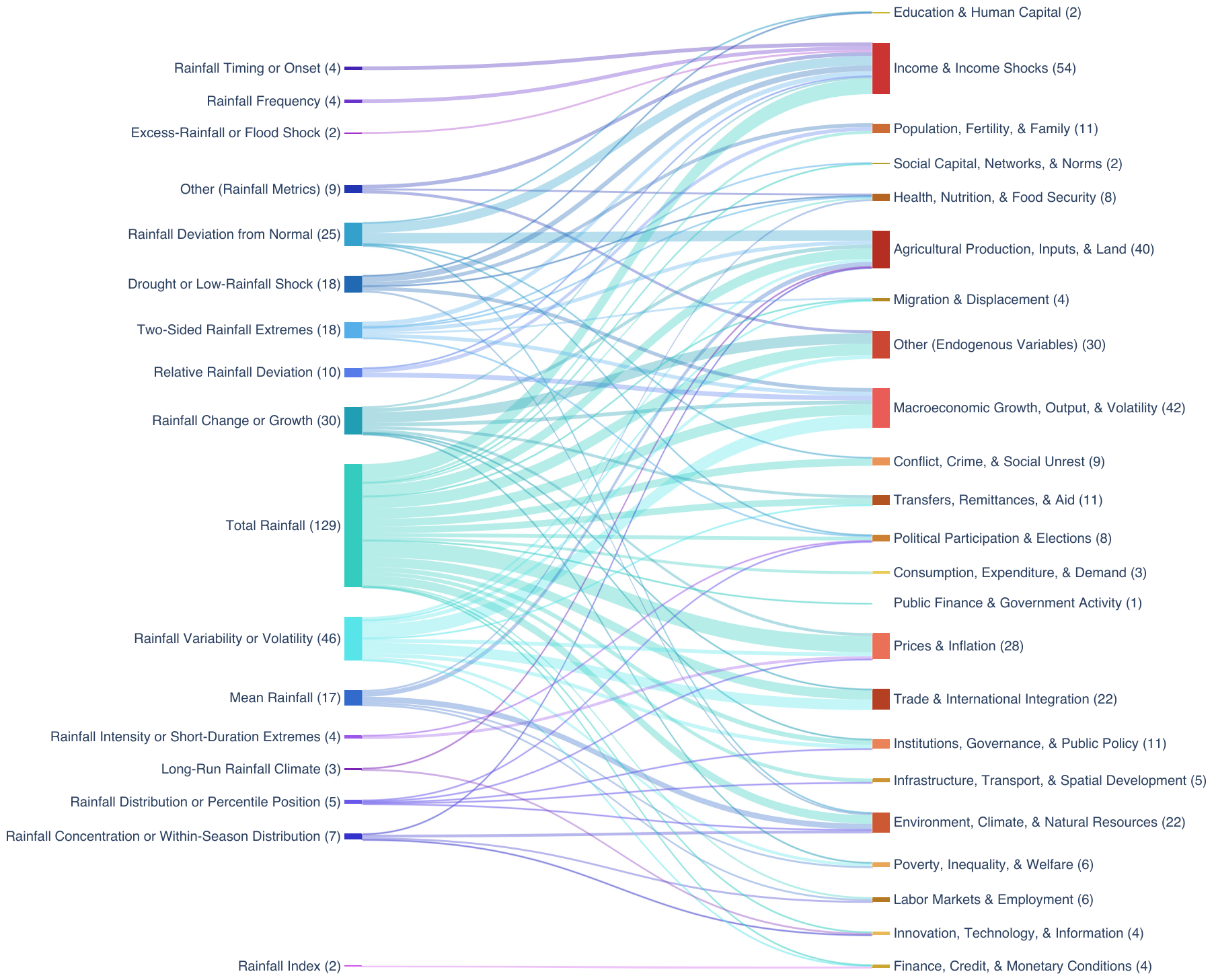}
			\end{center}
			\footnotesize  \textit{Note}: The Sankey diagram maps harmonized categories of rainfall instruments identified by the zero-shot implementation (left) to the corresponding harmonized categories of endogenous variables (right). Flow width is proportional to the number of extracted rainfall instrument–endogenous variable linkages, with flows colored according to the rainfall-instrument category. Values in parentheses report the total number of extracted linkages associated with each category. Counts refer to model-generated linkages and need not correspond to unique papers.
		\end{minipage}
	\end{figure}

\begin{figure}[!htbp] 
		\begin{minipage}{\linewidth}		
			\begin{center}
				\caption{Llama 4 Maverick: Rainfall Instruments and Endogenous Variables\label{fig:gpt_llama}}
				\includegraphics[width=\linewidth,keepaspectratio]{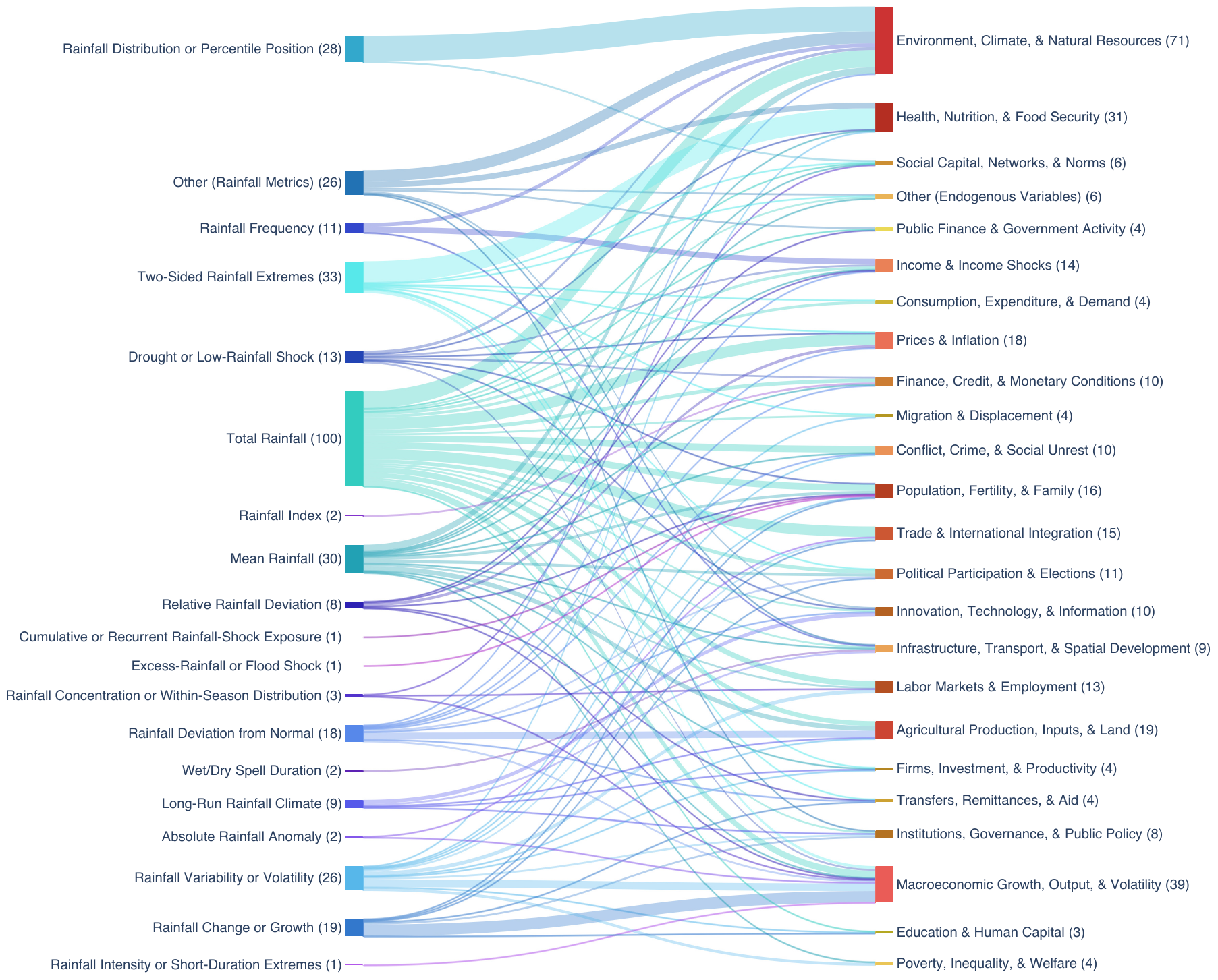}
			\end{center}
			\footnotesize  \textit{Note}: The Sankey diagram maps harmonized categories of rainfall instruments identified by the zero-shot implementation (left) to the corresponding harmonized categories of endogenous variables (right). Flow width is proportional to the number of extracted rainfall instrument–endogenous variable linkages, with flows colored according to the rainfall-instrument category. Values in parentheses report the total number of extracted linkages associated with each category. Counts refer to model-generated linkages and need not correspond to unique papers.
		\end{minipage}
	\end{figure}

\begin{figure}[!htbp] 
		\begin{minipage}{\linewidth}		
			\begin{center}
				\caption{Claude Opus 5.5: Rainfall Instruments and Endogenous Variables\label{fig:gpt_claude}}
				\includegraphics[width=\linewidth,keepaspectratio]{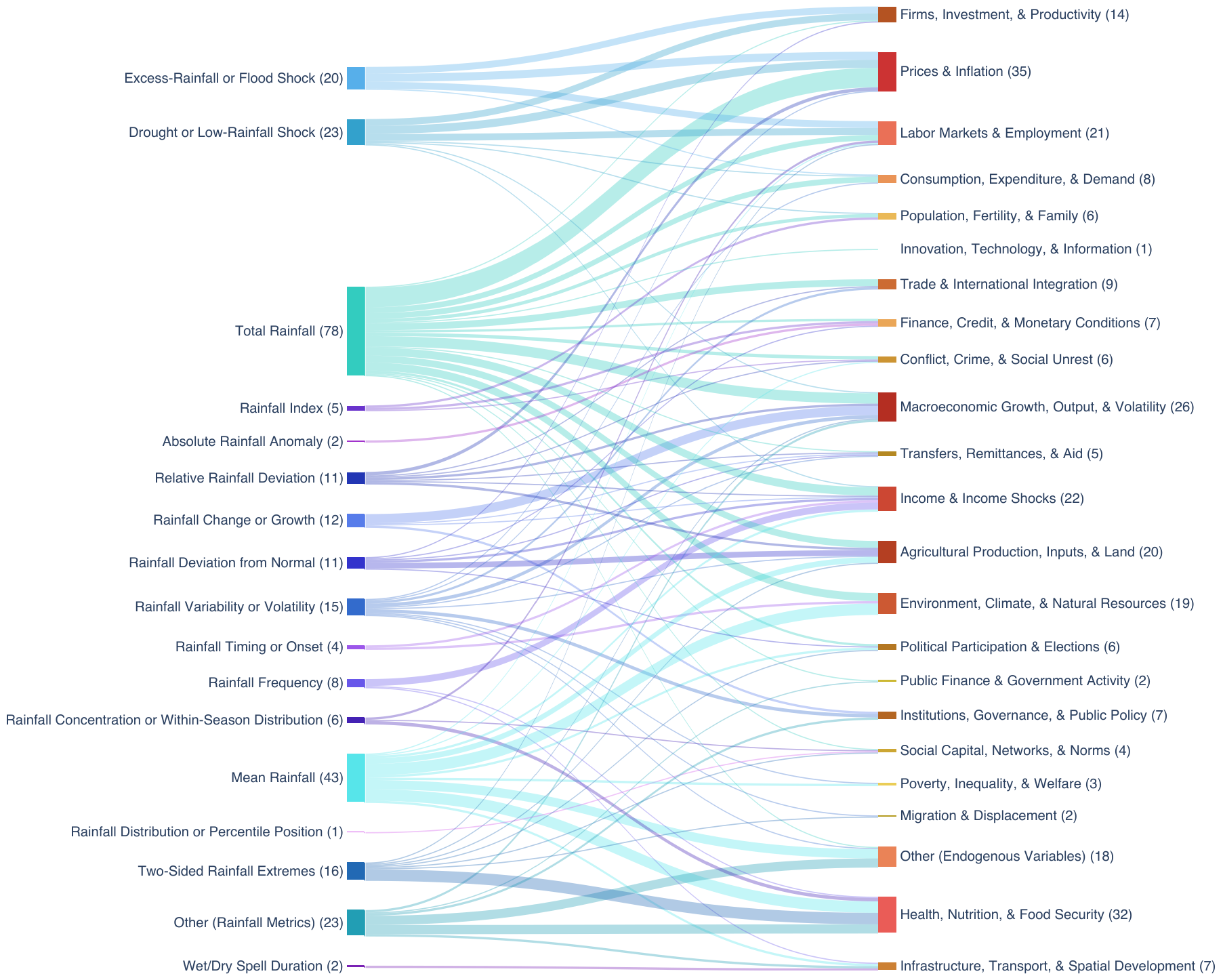}
			\end{center}
			\footnotesize  \textit{Note}: The Sankey diagram maps harmonized categories of rainfall instruments identified by the zero-shot implementation (left) to the corresponding harmonized categories of endogenous variables (right). Flow width is proportional to the number of extracted rainfall instrument–endogenous variable linkages, with flows colored according to the rainfall-instrument category. Values in parentheses report the total number of extracted linkages associated with each category. Counts refer to model-generated linkages and need not correspond to unique papers.
		\end{minipage}
	\end{figure}

	We next examine whether the broader characterization of the rainfall-IV literature is robust to our choice of model. One of the more interesting differences across the Sankey diagrams in Figures~\ref{fig:gpt_sft_app}--\ref{fig:gpt_claude} is in the number of rainfall-IV/endogenous-variable relationships the models identify. While GPT-6.1 Sol identifies roughly the same number of relationships as SFT (199 versus 201), Gemma, Llama, and Claude identify considerably more (333, 333, and 280, respectively). Recall that Gemma and Claude identify the same number of papers as using rainfall IVs (137) and substantially fewer papers than SFT (180). Yet Gemma identifies $66\%$ more instrumenting relationships than SFT. This difference is even more pronounced when we account for the number of papers contributing to the diagrams. Gemma identifies an average of 3.7 relationships per contributing paper compared with 1.5 for SFT. Thus, differences in our characterization of the literature are a function not only of which papers the models identify as being part of the literature but also of how much information the models generate about those papers.
	
	There are also considerably larger differences in the relative importance of particular economic concepts than we observe across the three implementations of GPT-4.1. For example, Income \& Income Shocks accounts for $16\%$ of the relationships identified by Gemma but only $4\%$ of those identified by Llama. In contrast, Environment, Climate, \& Natural Resources accounts for $21\%$ of the relationships identified by Llama but only $7\%$ of those identified by Gemma. Claude identifies a relatively large number of relationships involving Health, Nutrition, \& Food Security (32 versus four for SFT) and Prices \& Inflation (35 versus 11 for SFT). There are similarly large differences in the ways in which rainfall is quantified. For example, Llama identifies 28 relationships involving Rainfall Distribution or Percentile Position while both SFT and Claude identify only one. Thus, if our estimand of interest is how frequently economists use a particular measure of rainfall or instrument for a particular economic concept, our estimates can be quite sensitive to the choice of model.
	
	Despite this additional measurement-system dispersion, we recover much the same picture of the rainfall-IV literature as we do in Section~\ref{sec:rel}. Total Rainfall remains the most frequently used rainfall IV across the models examined and is used to instrument for a wide range of economic concepts. Similarly, there continues to be considerable variation in the measures of rainfall used to instrument for any one economic concept. What changes as we vary the model is the relative importance of particular nodes and edges in the network, sometimes quite substantially.

	
	\newpage
	
	\FloatBarrier
	
	\begin{landscape}
		\input{tables/frontier_agree}   
	\end{landscape}


\end{document}

%% file: tables/gpt_agree.tex
\begin{table}[!htbp]
	\centering
	\caption{Cross-Implementation Agreement in Model-Labeled Rainfall-IV Records}
	\label{tab:gpt_agree}
	\begin{threeparttable}
		\footnotesize
		\setlength{\tabcolsep}{5pt}
		\renewcommand{\arraystretch}{1.18}

		\begin{tabular}{lrrrr}
			\toprule

			\multicolumn{5}{l}{\textit{Panel A: Rainfall-IV identification}} \\
			\addlinespace[2pt]
				& Zero-shot & RAG & SFT & \\
			\midrule
			Rainfall-IV papers & 151 & 160 & 180 & \\
			Share of common model-labeled corpus (\%) & 3.95 & 4.19 & 4.71 & \\
			Unique to implementation & 8 & 14 & 35 & \\

			\addlinespace[8pt]
			\multicolumn{5}{l}{\textit{Panel B: Cross-implementation overlap in rainfall-IV identification}} \\
			\addlinespace[2pt]
				& \multicolumn{2}{c}{Papers} & \multicolumn{2}{c}{Share of union (\%)} \\
			\midrule
			Identified by at least one implementation & \multicolumn{2}{c}{209} & \multicolumn{2}{c}{100.0} \\
			Identified by all three & \multicolumn{2}{c}{130} & \multicolumn{2}{c}{62.2} \\
			Identified by exactly two & \multicolumn{2}{c}{22} & \multicolumn{2}{c}{10.5} \\
			Identified by exactly one & \multicolumn{2}{c}{57} & \multicolumn{2}{c}{27.3} \\
			Any disagreement across implementations & \multicolumn{2}{c}{79} & \multicolumn{2}{c}{37.8} \\

			\addlinespace[8pt]
			\multicolumn{5}{l}{\textit{Panel C: Exact agreement conditional on rainfall-IV identification}} \\
			\addlinespace[2pt]
				& Zero-shot--RAG & Zero-shot--SFT & RAG--SFT & All three \\
			\midrule
			Dependent variable(s) & 56 (43.1) & 6 (4.6) & 7 (5.4) & 5 (3.8) \\
			Endogenous variable(s) & 64 (49.2) & 42 (32.3) & 46 (35.4) & 35 (26.9) \\
			Rainfall instrument(s) & 64 (49.2) & 28 (21.5) & 26 (20.0) & 23 (17.7) \\
			Rainfall instrument and endogenous variable lists & 35 (26.9) & 14 (10.8) & 10 (7.7) & 8 (6.2) \\

			\addlinespace[8pt]
			\multicolumn{5}{l}{\textit{Panel D: Category agreement conditional on rainfall-IV identification}} \\
			\addlinespace[2pt]
				& Zero-shot--RAG & Zero-shot--SFT & RAG--SFT & All three \\
			\midrule
			Rainfall instrument categories & 84 (64.6) & 61 (46.9) & 61 (46.9) & 52 (40.0) \\
			Endogenous variable categories & 102 (78.5) & 96 (73.8) & 91 (70.0) & 84 (64.6) \\
			Both category sets & 68 (52.3) & 51 (39.2) & 49 (37.7) & 40 (30.8) \\

			\bottomrule
		\end{tabular}

		\begin{tablenotes}[flushleft]
			\footnotesize
			\item \textit{Note:} The sample contains 3,821 unique papers with model-generated labels from all three implementations: the full-corpus extraction plus the papers of the common held-out evaluation set. The non-held-out human-reference records are not included. Panel A counts explicit rainfall-IV labels of 1; all other codes are not counted as positive. Panel B reports overlap among the 209 papers labeled positive by at least one implementation. Panels C and D report agreement among the 130 papers labeled positive by all three, with percentages of that sample in parentheses. In Panel C, strings are lowercased, whitespace is standardized, and semicolon-separated lists are sorted without removing repeated items. Missing-value strings remain literal and count as agreement when identical. Exact-string disagreement need not imply disagreement in meaning. Panel D reports agreement after assigning variables to the categories defined in Appendix~\ref{app:class}. Agreement requires identical sets of categories, including the residual categories Other and Unclassifiable. The last row of each panel requires agreement on both rainfall instruments and endogenous variables.
		\end{tablenotes}
	\end{threeparttable}
\end{table}

%% file: tables/econ_tax.tex
\newpage
\begingroup
\footnotesize
\setlength{\tabcolsep}{4pt}
\renewcommand{\arraystretch}{1.18}

\begin{ThreePartTable}
\begin{TableNotes}[flushleft]
    \footnotesize
    \item \textit{Note:} The table reports 23 economic concepts and two residual categories used to classify extracted endogenous variables, along with their definitions and related JEL codes. Other denotes interpretable variables outside the predefined concepts; Unclassifiable denotes entries whose meaning cannot be identified from the available information.
\end{TableNotes}

\begin{longtable}{
    @{} >{\raggedright\arraybackslash}p{0.27\textwidth}
    >{\raggedright\arraybackslash}p{0.51\textwidth}
    >{\raggedright\arraybackslash}p{0.15\textwidth} @{} }

    \caption{Taxonomy of Economic Concepts}
    \label{tab:econ_tax} \\

    \\[-1.8ex]\hline
    \noalign{\vskip 2pt}
    \hline \\[-1.8ex]
    Concept & Description & Related JEL codes \\
    \midrule
    \endfirsthead

    \multicolumn{3}{c}{\footnotesize\textit{Table \ref{tab:econ_tax} continued}} \\
    \\[-1.8ex]\hline
    \noalign{\vskip 2pt}
    \hline \\[-1.8ex]
    Concept & Description & Related JEL codes \\
    \midrule
    \endhead

    \midrule
    \multicolumn{3}{r}{\footnotesize\textit{Continued on next page}} \\
    \endfoot

    \\[-1.8ex]\hline
    \noalign{\vskip 2pt}
    \hline
    \insertTableNotes \\
    \endlastfoot

    Income \& income shocks
    & Individual, household, farm, or other income levels, changes, seasonality, and shocks. Excludes wages and aggregate GDP.
    & D31; O15 \\

    Consumption, expenditure \& demand
    & Individual or household consumption, expenditure, purchases, or demand for goods and services.
    & D12; E21 \\

    Poverty, inequality \& welfare
    & Poverty, deprivation, wealth, inequality, distributional outcomes, and general measures of material or subjective welfare.
    & D31; D63; I31; I32 \\

    Agricultural production, inputs \& land
    & Agricultural production and productivity, crop outcomes, agricultural inputs, farming practices, land use and tenure, irrigation, and agricultural technology.
    & Q11; Q12; Q15; Q16 \\

    Health, nutrition \& food security
    & Physical or mental health, disease, mortality, anthropometrics, nutrition, food security, health behavior, and utilization or quality of healthcare.
    & I10--I15; Q18 \\

    Education \& human capital
    & Schooling, educational attainment, skills, cognitive development, training, and other measures of human-capital accumulation.
    & I21--I26; J24 \\

    Labor markets \& employment
    & Employment, unemployment, labor-force participation, labor supply, wages, occupations, working conditions, and work arrangements.
    & J21--J33; J64 \\

    Population, fertility \& family
    & Population size and composition, fertility, marriage, family formation, household structure, and demographic change other than migration.
    & J11--J14 \\

    Migration \& displacement
    & Migration, geographic mobility, displacement, resettlement, migrant stocks, and movement between locations.
    & F22; J61; O15; R23 \\

    Social capital, networks, norms \& attitudes
    & Social connections, peer effects, social capital, trust, cultural or social norms, attitudes, empowerment, and socially transmitted behavior.
    & D85; Z10--Z13 \\

    Political participation \& elections
    & Voting, turnout, electoral behavior and outcomes, political representation, candidacy, and election-related political participation.
    & D72 \\

    Institutions, governance \& public policy
    & Institutional quality, state capacity, corruption, regulation, property rights, legal institutions, governance, policy regimes, and policy uncertainty. Excludes government expenditure, taxation, and transfers.
    & D02; D73; H11; H83; P16 \\

    Conflict, crime \& social unrest
    & Armed conflict, political violence, criminal activity, riots, protests, dissent, repression, organized crime, and related social unrest.
    & D74; F51; K14; K42 \\

    Macroeconomic growth, output \& volatility
    & Economy-wide output, GDP, economic growth, output gaps, aggregate productivity, and macroeconomic shocks or volatility.
    & E23; E32; O40; O47 \\

    Prices \& inflation
    & Prices of goods, commodities, outputs, or inputs, as well as aggregate price indices and inflation.
    & D40; E31 \\

    Trade \& international integration
    & Imports, exports, trade openness, exposure to international markets, trade policy, and cross-border productive capital flows.
    & F10--F21 \\

    Finance, credit \& monetary conditions
    & Credit, borrowing, banking, savings, insurance, financial access, asset markets, interest rates, exchange rates, money, and monetary conditions.
    & E41--E52; G10--G29; O16 \\

    Firms, investment \& productivity
    & Nonagricultural firm outcomes, productive investment, capital accumulation, profitability, productivity, business formation, and firm growth or survival.
    & D21--D25; G31--G34; L20--L26 \\

    Innovation, technology \& information
    & Innovation, research and development, technological adoption outside agriculture, knowledge, digital technologies, information access, and media.
    & D83; L82; L86; O30--O33 \\

    Environment, climate \& natural resources
    & Environmental quality and pollution, emissions, conservation, ecosystems, climate or weather variables when themselves endogenous, natural disasters, energy, and the stock, extraction, or rents of natural resources.
    & Q31--Q43; Q50--Q57 \\

    Infrastructure, transport \& urban/spatial development
    & Physical infrastructure, transportation, accessibility, housing, urbanization, spatial concentration, and location-based or regional development.
    & O18; R11--R42 \\

    Public finance \& government activity
    & Taxation, government expenditure, public debt and deficits, procurement, subsidies, intergovernmental finance, public investment, and other fiscal activity.
    & H20--H72 \\

    Transfers, remittances \& aid
    & Private transfers, remittances, cash or in-kind transfers, pensions, foreign assistance, development aid, and other external transfers when the transfer itself is the endogenous economic object.
    & F24; F35; H53; H55; I38 \\

    Other
    & Clearly identified economic variables whose substantive concept does not fit any preceding category. The meaning of the variable must be sufficiently specified to assess category fit. Excludes entries whose meaning is missing, ambiguous, corrupted, or uninterpretable.
    & None \\

    Unclassifiable
    & Extracted strings that do not provide enough information to identify a specific economic variable, including gibberish, uninterpretable or corrupted text, unexplained variable codes or abbreviations, generic labels, and incomplete fragments. Examples include \texttt{x1}, \texttt{vul}, and \texttt{Driving factor} when no supplied definition or context resolves their meaning.
    & None \\

\end{longtable}
\end{ThreePartTable}
\endgroup

%% file: tables/rain_tax.tex
\newpage
\begingroup
\footnotesize
\setlength{\tabcolsep}{4pt}
\renewcommand{\arraystretch}{1.18}

\begin{ThreePartTable}
\begin{TableNotes}[flushleft]
    \footnotesize
    \item \textit{Note:} The table reports 21 rainfall categories and two residual categories used to classify extracted rainfall instruments, along with their definitions and representative examples. Other denotes sufficiently specified measures outside the predefined categories, including identifiable non-rainfall measures; Unclassifiable denotes entries whose meaning or construction cannot be determined from the available information.
\end{TableNotes}

\begin{longtable}{
    @{} >{\raggedright\arraybackslash}p{0.21\textwidth}
    >{\raggedright\arraybackslash}p{0.36\textwidth}
    >{\raggedright\arraybackslash}p{0.36\textwidth} @{} }

    \caption{Taxonomy of Rainfall Measures}
    \label{tab:rain_tax} \\

    \\[-1.8ex]\hline
    \noalign{\vskip 2pt}
    \hline \\[-1.8ex]
    Category & What belongs & Examples \\
    \midrule
    \endfirsthead

    \multicolumn{3}{c}{\footnotesize\textit{Table \ref{tab:rain_tax} continued}} \\
    \\[-1.8ex]\hline
    \noalign{\vskip 2pt}
    \hline \\[-1.8ex]
    Category & What belongs & Examples \\
    \midrule
    \endhead

    \midrule
    \multicolumn{3}{r}{\footnotesize\textit{Continued on next page}} \\
    \endfoot

    \\[-1.8ex]\hline
    \noalign{\vskip 2pt}
    \hline
    \insertTableNotes \\
    \endlastfoot

    Total rainfall
    & Raw accumulated precipitation over a specified period, rather than an average across observations or a historical reference period
    & Total annual rainfall; total rainy-season precipitation; total growing-season precipitation; monthly precipitation total; accumulated rainfall \\

    Mean rainfall
    & Average precipitation across observations within a specified period, rather than accumulated precipitation or a long-run climatological mean
    & Mean daily rainfall during the growing season; average monthly rainfall within a year; mean weekly precipitation; average rainfall during the current season \\

    Rainfall change or growth
    & Change in rainfall between periods rather than the contemporaneous rainfall level or departure from a climatological norm
    & Rainfall growth; annual rainfall change; year-on-year change in seasonal rainfall; proportional change from the previous year \\

    Rainfall deviation from normal
    & Signed difference between realized rainfall and a historical or reference amount, expressed in rainfall units rather than percentages or standardized units
    & Rainfall minus long-run mean; annual deviation from average; deviation from historical mean; deviation from normal rainfall; rainfall deviation from a fitted trend \\

    Relative rainfall deviation
    & Departure from normal expressed relative to the reference rainfall level rather than in rainfall units
    & Percentage deviation from historical rainfall; rainfall-to-normal ratio; log rainfall minus log historical rainfall; fractional deviation from average \\

    Standardized rainfall anomaly
    & Continuous signed departure from normal expressed in units of the historical standard deviation, excluding separately defined named rainfall indices
    & Seasonal rainfall \(z\)-score; standardized annual rainfall deviation; standardized monsoon rainfall; standardized precipitation anomaly \\

    Absolute rainfall anomaly
    & Magnitude of departure from normal where the direction of the anomaly is intentionally removed, using absolute values or squared deviations
    & Absolute standardized anomaly; \(\lvert z\rvert\); squared standardized deviation; absolute rainfall deviation; squared rainfall deviation from normal \\

    Rainfall distribution or percentile position
    & Position of realized rainfall within its historical distribution, without necessarily identifying an extreme event
    & Rainfall percentile; fitted gamma-distribution percentile; rainfall quartile; percentile-discretized precipitation; rainfall-distribution bins \\

    Two-sided rainfall extremes
    & Measures that jointly identify or categorize unusually wet and unusually dry realizations, rather than retaining only one tail of the rainfall distribution
    & Above or below one standard deviation; below the 20th or above the 80th percentile; high/normal/low rainfall classification; negative/normal/positive rainfall shock classification \\

    Drought or low-rainfall shock
    & Measures retaining only the lower tail of the rainfall distribution or rainfall deficits, including indicators and measures of deficit severity
    & Drought indicator; rainfall below the 15th percentile; negative rainfall shock indicator; rainfall shortfall; one-sided drought-intensity measure \\

    Excess-rainfall or flood shock
    & Measures retaining only the upper tail of the rainfall distribution or rainfall surpluses, including indicators and measures of excess-rainfall severity
    & Positive rainfall shock indicator; excessive rainfall; high-rainfall shock; rainfall surplus; flood rainfall shock \\

    Rainfall frequency
    & Number or frequency of days, months, or other observations experiencing rainfall, dryness, or a specified rainfall state, without measuring consecutive duration
    & Number of rainy days; number of wet days; number of dry days; number of drought months in a year; share of days with precipitation \\

    Wet/dry spell duration
    & Persistence or consecutive duration of wet or dry conditions, rather than the total frequency of such conditions
    & Longest dry spell; consecutive dry days; drought duration; consecutive drought months; longest spell of rainfall days \\

    Rainfall timing or onset
    & Timing of the beginning, end, or arrival of rainfall or the rainy season, rather than the amount or frequency of precipitation
    & Monsoon onset; rainy-season onset; delay in onset; start or end date of the rainy season; timing of first rainfall \\

    Rainfall intensity or short-duration extremes
    & Amount or maximum intensity of precipitation during a short event or interval, rather than an indicator of whether an extreme event occurred
    & Maximum 24-hour rainfall; maximum consecutive five-day precipitation; extreme daily precipitation amount; maximum hourly rainfall; peak rainfall intensity \\

    Rainfall concentration or within-season distribution
    & Temporal concentration or uneven distribution of rainfall within a season or year, measured through the relative allocation of rainfall across intervals rather than its total amount
    & Wettest-month rainfall relative to average rainfall; ratio of maximum monthly rainfall to average rainfall; share of seasonal rainfall occurring in the wettest month; temporal concentration index \\

    Rainfall variability or volatility
    & Statistical dispersion of rainfall across repeated observations or periods, rather than the temporal concentration of rainfall within a particular season
    & Standard deviation of rainfall; rainfall variance; coefficient of variation; standard deviation of monthly rainfall; rolling rainfall CV \\

    Rainfall index
    & Named or composite precipitation-derived indices that transform or combine rainfall information using a specified index construction, including indices incorporating other hydroclimatic variables
    & Standardized Precipitation Index (SPI); Standardized Precipitation-Evapotranspiration Index (SPEI); Palmer Drought Severity Index (PDSI); Weighted Anomaly of Standardized Precipitation (WASP) \\

    Cumulative or recurrent rainfall-shock exposure
    & Measures summarizing repeated rainfall shocks or accumulated exposure over multiple prior periods rather than a single rainfall realization from a particular period
    & Number of droughts in the previous five or ten years; fraction of prior years classified as dry; cumulative drought count; recurrent rainfall shocks \\

    Long-run rainfall climate
    & Historical or climatological rainfall conditions used directly as the measure of interest rather than solely as a reference for contemporaneous weather
    & Long-run mean precipitation; historical median rainfall; precipitation normal used as an instrument; long-run average rainfall \\

    Rainfall trend or long-run change
    & Persistent changes in rainfall conditions, including estimated trends and long-run shifts, rather than contemporaneous anomalies or year-to-year fluctuations
    & Long-run precipitation trend; change in precipitation trend; secular change in rainfall; historical rainfall trend coefficient; multi-decade change in mean rainfall \\

    Other
    & Clearly identified and sufficiently specified measures that do not fit any preceding rainfall category, including identifiable non-rainfall measures. Excludes entries whose meaning or construction is too incomplete or ambiguous to assess category fit.
    & Covariance between monthly rainfall and monthly temperature; Wind speed; Air temperature \\

    Unclassifiable
    & Extracted strings that do not identify an interpretable measure, or that lack enough information about its construction to distinguish among the preceding rainfall categories. Includes gibberish, uninterpretable or corrupted text, unexplained codes or abbreviations, generic labels, and incomplete fragments when no supplied definition or context resolves their meaning.
    & Uninterpretable or corrupted mathematical text; \texttt{x1}; \texttt{vul}; \texttt{Driving factor}; rainfall shock with unspecified construction; rainfall with unspecified measurement construction \\

\end{longtable}
\end{ThreePartTable}
\endgroup

%% file: tables/frontier_agree.tex
\begin{table}[!htbp]
    \centering
    \caption{Cross-Model Agreement in Model-Labeled Rainfall-IV Records}
    \label{tab:frontier_agree}
    \begin{minipage}{\linewidth}
        \footnotesize
        \setlength{\tabcolsep}{4pt}
        \renewcommand{\arraystretch}{1.00}
        \setlength{\parskip}{0pt}

        \noindent\begin{tabular}{@{}p{0.40\linewidth}*{5}{@{}>{\centering\arraybackslash}p{0.12\linewidth}}@{}}
            \toprule

            \multicolumn{6}{@{}l@{}}{\textit{Panel A: Rainfall-IV identification}} \\
            \addlinespace[1pt]
                & \shortstack[c]{GPT-4.1\\SFT}
                & \shortstack[c]{GPT-6.1\\Sol}
                & \shortstack[c]{Gemma 4\\26B}
                & \shortstack[c]{Llama 4\\Maverick}
                & \shortstack[c]{Claude\\Opus 5.5} \\
            \midrule
            Rainfall-IV papers & 180 & 146 & 137 & 167 & 137 \\
            Share of common model-labeled corpus (\%) & 4.71 & 3.82 & 3.59 & 4.37 & 3.59 \\
            Unique to model & 31 & 10 & 3 & 14 & 1 \\

            \addlinespace[4pt]
            \multicolumn{6}{@{}l@{}}{\textit{Panel B: Cross-model overlap in rainfall-IV identification}} \\
            \addlinespace[1pt]
                & \multicolumn{2}{@{}c@{}}{Papers} & \multicolumn{3}{@{}c@{}}{Share of union (\%)} \\
            \midrule
            Identified by at least one model & \multicolumn{2}{@{}c@{}}{222} & \multicolumn{3}{@{}c@{}}{100.0} \\
            Identified by all five & \multicolumn{2}{@{}c@{}}{118} & \multicolumn{3}{@{}c@{}}{53.2} \\
            Identified by exactly four & \multicolumn{2}{@{}c@{}}{7} & \multicolumn{3}{@{}c@{}}{3.2} \\
            Identified by exactly three & \multicolumn{2}{@{}c@{}}{14} & \multicolumn{3}{@{}c@{}}{6.3} \\
            Identified by exactly two & \multicolumn{2}{@{}c@{}}{24} & \multicolumn{3}{@{}c@{}}{10.8} \\
            Identified by exactly one & \multicolumn{2}{@{}c@{}}{59} & \multicolumn{3}{@{}c@{}}{26.6} \\
            Any disagreement across models & \multicolumn{2}{@{}c@{}}{104} & \multicolumn{3}{@{}c@{}}{46.8} \\

        \end{tabular}\par\vspace{4pt}
        \noindent\begin{tabular}{@{}p{0.40\linewidth}*{4}{@{}>{\centering\arraybackslash}p{0.15\linewidth}}@{}}
            \multicolumn{5}{@{}l@{}}{\textit{Panel C: Exact agreement conditional on rainfall-IV identification}} \\
            \addlinespace[1pt]
                & SFT--Sol & SFT--Gemma & SFT--Llama & SFT--Claude \\
            \midrule
            Dependent variable(s) & 7 (5.9) & 3 (2.5) & 4 (3.4) & 3 (2.5) \\
            Endogenous variable(s) & 29 (24.6) & 20 (16.9) & 16 (13.6) & 24 (20.3) \\
            Rainfall instrument(s) & 22 (18.6) & 17 (14.4) & 19 (16.1) & 18 (15.3) \\
            Rainfall instrument and endogenous variable lists & 7 (5.9) & 6 (5.1) & 5 (4.2) & 6 (5.1) \\

            \addlinespace[4pt]
            \multicolumn{5}{@{}l@{}}{\textit{Panel D: Category agreement conditional on rainfall-IV identification}} \\
            \addlinespace[1pt]
                & SFT--Sol & SFT--Gemma & SFT--Llama & SFT--Claude \\
            \midrule
            Rainfall instrument categories & 58 (49.2) & 46 (39.0) & 46 (39.0) & 60 (50.8) \\
            Endogenous variable categories & 84 (71.2) & 70 (59.3) & 63 (53.4) & 82 (69.5) \\
            Both category sets & 42 (35.6) & 28 (23.7) & 28 (23.7) & 42 (35.6) \\

            \bottomrule
        \end{tabular}

        \par\smallskip
        \footnotesize
        \raggedright
        \noindent\textit{Note:} The sample contains 3,821 unique papers with model-generated labels from all five models: the full-corpus extraction plus the papers of the common held-out evaluation set. The non-held-out human-reference records are not included. Panel A counts explicit rainfall-IV labels of 1; all other codes are not counted as positive. Panel B reports overlap among the 222 papers labeled positive by at least one model. Panels C and D compare GPT-4.1 SFT to each of the four frontier models among the 118 papers labeled positive by all five, with percentages of that sample in parentheses. In Panel C, strings are lowercased, whitespace is standardized, and semicolon-separated lists are sorted without removing repeated items. Missing-value strings remain literal and count as agreement when identical. Exact-string disagreement need not imply disagreement in meaning. Panel D reports agreement after assigning variables to the categories defined in Appendix~\ref{app:class}. Agreement requires identical sets of categories, including the residual categories Other and Unclassifiable. The last row of Panels C and D requires agreement on both rainfall instruments and endogenous variables.
    \end{minipage}
\end{table}